\documentclass[twocolumn]{aastex631}

\newcommand{\cmtbk}[1]{\textcolor{black}{#1}}

\newcommand{\revision}[1]{#1}

\usepackage{dcolumn}

\shorttitle{eDisk: IRAS 04169+2702}
\shortauthors{Han et al.}

\begin{document}

\title{Early Planet Formation in Embedded Disks (eDisk) XVII: A Compact but Structured Keplerian Disk and Large-scale Streamers Revealed in the Class I Protostellar System IRAS 04169$+$2702}

\author[0000-0002-9143-1433]{Ilseung Han}
\affiliation{Institut de Ci\`encies de l'Espai (ICE-CSIC), Campus UAB, Can Magrans S/N, E-08193 Cerdanyola del Vall\`es, Catalonia, Spain}
\affiliation{Department of Earth Science Education, Seoul National University, 1 Gwanak-ro, Gwanak-gu, Seoul 08826, Republic of Korea}
\affiliation{Department of Astronomy and Space Science, University of Science and Technology, 217 Gajeong-ro, Yuseong-gu, Daejeon 34113, Republic of Korea}
\affiliation{Korea Astronomy and Space Science Institute, 776 Daedeok-daero, Yuseong-gu, Daejeon 34055, Republic of Korea}

\author[0000-0003-4022-4132]{Woojin Kwon}
\affiliation{Department of Earth Science Education, Seoul National University, 1 Gwanak-ro, Gwanak-gu, Seoul 08826, Republic of Korea}
\affiliation{SNU Astronomy Research Center, Seoul National University, 1 Gwanak-ro, Gwanak-gu, Seoul 08826, Republic of Korea}
\affiliation{\cmtbk{The Center for Educational Research, Seoul National University, 1 Gwanak-ro, Gwanak-gu, Seoul 08826, Republic of Korea}}
\correspondingauthor{Woojin Kwon}
\email{wkwon@snu.ac.kr}

\author[0000-0002-8238-7709]{Yusuke Aso}
\affiliation{Korea Astronomy and Space Science Institute, 776 Daedeok-daero, Yuseong-gu, Daejeon 34055, Republic of Korea}
\affiliation{Department of Astronomy and Space Science, University of Science and Technology, 217 Gajeong-ro, Yuseong-gu, Daejeon 34113, Republic of Korea}

\author[0000-0003-0998-5064]{Nagayoshi Ohashi}
\affiliation{Academia Sinica Institute of Astronomy $\&$ Astrophysics, 11F of Astronomy-Mathematics Building, AS/NTU, No. 1, Sec. 4, Roosevelt Rd, Taipei 10617, Taiwan, R.O.C.}

\author[0000-0002-6195-0152]{John J. Tobin}
\affiliation{National Radio Astronomy Observatory, 520 Edgemont Rd., Charlottesville, VA 22903 USA}

\author[0000-0001-9133-8047]{Jes K. J{\o}rgensen}
\affiliation{Niels Bohr Institute, University of Copenhagen, Jagtvej 155A, DK-2200 Copenhagen N., Denmark}

\author[0000-0003-0845-128X]{Shigehisa Takakuwa}
\affiliation{Department of Physics and Astronomy, Graduate School of Science and Engineering, Kagoshima University, 1-21-35 Korimoto, Kagoshima, Kagoshima 890-0065, Japan}
\affiliation{Academia Sinica Institute of Astronomy $\&$ Astrophysics, 11F of Astronomy-Mathematics Building, AS/NTU, No. 1, Sec. 4, Roosevelt Rd, Taipei 10617, Taiwan, R.O.C.}

\author[0000-0002-4540-6587]{Leslie W. Looney}
\affiliation{Department of Astronomy, University of Illinois, 1002 West Green St, Urbana, IL 61801, USA}
\affiliation{National Radio Astronomy Observatory, 520 Edgemont Rd., Charlottesville, VA 22903 USA}

\author[0000-0003-3283-6884]{Yuri Aikawa}
\affiliation{Department of Astronomy, Graduate School of Science, The University of Tokyo, 7-3-1 Hongo, Bunkyo-ku, Tokyo 113-0033, Japan}

\author[0000-0002-8591-472X]{Christian Flores}
\affiliation{Academia Sinica Institute of Astronomy $\&$ Astrophysics, 11F of Astronomy-Mathematics Building, AS/NTU, No. 1, Sec. 4, Roosevelt Rd, Taipei 10617, Taiwan, R.O.C.}

\author[0000-0003-4518-407X]{Itziar de Gregorio-Monsalvo}
\affiliation{European Southern Observatory, Alonso de Cordova 3107, Casilla 19, Vitacura, Santiago, Chile}

\author[0000-0003-2777-5861]{Patrick M. Koch}
\affiliation{Academia Sinica Institute of Astronomy $\&$ Astrophysics, 11F of Astronomy-Mathematics Building, AS/NTU, No. 1, Sec. 4, Roosevelt Rd, Taipei 10617, Taiwan, R.O.C.}

\author[0000-0002-3179-6334]{Chang Won Lee}
\affiliation{Department of Astronomy and Space Science, University of Science and Technology, 217 Gajeong-ro, Yuseong-gu, Daejeon 34113, Republic of Korea}
\affiliation{Korea Astronomy and Space Science Institute, 776 Daedeok-daero, Yuseong-gu, Daejeon 34055, Republic of Korea}

\author[0000-0003-3119-2087]{Jeong-Eun Lee}
\affiliation{Department of Physics and Astronomy, Seoul National University, 1 Gwanak-ro, Gwanak-gu, Seoul 08826, Republic of Korea}
\affiliation{SNU Astronomy Research Center, Seoul National University, 1 Gwanak-ro, Gwanak-gu, Seoul 08826, Republic of Korea}

\author[0000-0002-7402-6487]{Zhi-Yun Li}
\affiliation{University of Virginia, 530 McCormick Rd., Charlottesville, Virginia 22904, USA}

\author[0000-0001-7233-4171]{Zhe-Yu Daniel Lin}
\affiliation{University of Virginia, 530 McCormick Rd., Charlottesville, Virginia 22904, USA}

\author[0000-0003-4361-5577]{Jinshi Sai}
\affiliation{Academia Sinica Institute of Astronomy $\&$ Astrophysics, 11F of Astronomy-Mathematics Building, AS/NTU, No. 1, Sec. 4, Roosevelt Rd, Taipei 10617, Taiwan, R.O.C.}

\author[0000-0003-0334-1583]{Travis J. Thieme}
\affiliation{Institute of Astronomy, National Tsing Hua University, No. 101, Section 2, Kuang-Fu Road, Hsinchu 30013, Taiwan}
\affiliation{Center for Informatics and Computation in Astronomy, National Tsing Hua University, No. 101, Section 2, Kuang-Fu Road, Hsinchu 30013, Taiwan}
\affiliation{Department of Physics, National Tsing Hua University, No. 101, Section 2, Kuang-Fu Road, Hsinchu 30013, Taiwan}

\author[0000-0001-5058-695X]{Jonathan P. Williams}
\affiliation{Institute for Astronomy, University of Hawai‘i at Mānoa, 2680 Woodlawn Dr., Honolulu, HI 96822, USA}

\author[0000-0001-5782-915X]{Sacha Gavino}
\affiliation{Niels Bohr Institute, University of Copenhagen, {\O}ster Voldgade 5-7, 1350, Copenhagen K, Denmark}

\author[0000-0002-2902-4239]{Miyu Kido}
\affiliation{Department of Physics and Astronomy, Graduate School of Science and Engineering, Kagoshima University, 1-21-35 Korimoto, Kagoshima, Kagoshima 890-0065, Japan}
\affiliation{Academia Sinica Institute of Astronomy $\&$ Astrophysics, 11F of Astronomy-Mathematics Building, AS/NTU, No. 1, Sec. 4, Roosevelt Rd, Taipei 10617, Taiwan, R.O.C.}

\author[0000-0001-5522-486X]{Shih-Ping Lai}
\affiliation{Institute of Astronomy, National Tsing Hua University, No. 101, Section 2, Kuang-Fu Road, Hsinchu 30013, Taiwan}
\affiliation{Center for Informatics and Computation in Astronomy, National Tsing Hua University, No. 101, Section 2, Kuang-Fu Road, Hsinchu 30013, Taiwan}
\affiliation{Department of Physics, National Tsing Hua University, No. 101, Section 2, Kuang-Fu Road, Hsinchu 30013, Taiwan}
\affiliation{Academia Sinica Institute of Astronomy $\&$ Astrophysics, 11F of Astronomy-Mathematics Building, AS/NTU, No. 1, Sec. 4, Roosevelt Rd, Taipei 10617, Taiwan, R.O.C.}

\author[0000-0002-4372-5509]{Nguyen Thi Phuong}
\affiliation{Vietnam National Space Center, Vietnam Academy of Science and Techonology, 18 Hoang Quoc Viet, Cau Giay, Hanoi, Vietnam}
\affiliation{Korea Astronomy and Space Science Institute, 776 Daedeok-daero, Yuseong-gu, Daejeon 34055, Republic of Korea}

\author[0000-0001-6267-2820]{Alejandro Santamaría-Miranda}
\affiliation{European Southern Observatory, Alonso de Cordova 3107, Casilla 19, Vitacura, Santiago, Chile}

\author[0000-0003-1412-893X]{Hsi-Wei Yen}
\affiliation{Academia Sinica Institute of Astronomy $\&$ Astrophysics, 11F of Astronomy-Mathematics Building, AS/NTU, No. 1, Sec. 4, Roosevelt Rd, Taipei 10617, Taiwan, R.O.C.}



\begin{abstract}
    We present high-resolution ($\sim$0$\farcs$05; 8 au) dust continuum and molecular line observations toward the Class I protostellar system IRAS 04169$+$2702 in the Taurus B213 region, as part of the ALMA Large Program Early Planet Formation in Embedded Disks (eDisk).
    The 1.3-mm dust continuum emission traces a circumstellar disk with a central depression toward the protostar.
    Our VLA observations of the same target reveal a single central peak dominated by the free-free emission, which coincides with the depression of the thermal dust emission.
    The mean spectral index of the thermal dust emission from 1.3 mm to 1.4 cm is approximately 2.8, suggestive of the presence of grains grown to millimeter or centimeter sizes in the disk.
    Velocity gradients along the disk major axis are seen in emission from $^{12}$CO (2--1), $^{13}$CO (2--1), and C$^{18}$O (2--1) molecular lines.
    The position-velocity diagrams of these lines unveil a Keplerian-rotating disk with a radius of $\sim$21 au around a 1.3 $M_{\odot}$ protostar, as well as an infalling and rotating envelope with the angular momentum conserved.
    In addition to the compact disk, large-scale infalling spiral structures \revision{extending up to approximately 1400 au}, streamers, are discovered in C$^{18}$O (2--1), SO (6$_5-$5$_4$), and H$_2$CO (3$_{0, 3}-$2$_{0, 2}$) as well as in the 1.3-mm continuum emission.
    Notably, in the region closer to the protostar, the spatial coincidence of C$^{18}$O and SO may indicate the presence of a shock related to accretion through the \revision{spiral} arms.
\end{abstract}



\section{Introduction}
\label{sec:introduction}

Stars form through the gravitational collapse of dense cores in molecular clouds, and circumstellar disks form due to the conservation of angular momentum during collapse \citep[e.g.,][]{2011ARA&A..49...67W, 2024ARA&A..62..203T}.
Interferometric observations from the Atacama Large Millimeter/submillimeter Array (ALMA) and its predecessors have demonstrated that Class 0/I young stellar objects (YSOs) embedded in envelopes are surrounded by rotationally supported disks, whose velocity profile is Keplerian: the rotational velocity $V$ is proportional to $R^{-0.5}$, where $R$ denotes the disk radius
\citep[e.g.,][]{2009A&A...507..861J, 2012Natur.492...83T, 2013ApJ...772...22Y, 2013A&A...560A.103M, 2014A&A...562A..77H, 2014ApJ...796..131O, 2015ApJ...812...27A, 2017ApJ...849...56A}.
Furthermore, a few such embedded disks have recently been revealed to have substructures \citep[e.g.,][]{2020A&A...634L..12D, 2020Natur.586..228S, 2020ApJ...902..141S}.
These substructures, such as rings, cavities, and spirals, which have been commonly observed in the Class II phase \citep[e.g.,][]{2018ApJ...869L..41A, 2021MNRAS.501.2934C}, are expected to result from planet-disk interaction (e.g., \citealt{2023ASPC..534..423B} and references therein).
Indeed, protoplanets were directly imaged in the central cavity of a disk around the Class II protostar PDS 70 at (sub)millimeter wavelengths \citep[e.g.,][]{2019ApJ...879L..25I, 2021ApJ...916L...2B}.
The presence of substructures in younger Class 0/I protostellar disks thus suggests that planet formation may begin much earlier.
Grain growth, which is the first step in the planet-forming process, has already been reported in the Class 0 phase
\citep[e.g.,][]{2007ApJ...659..479J, 2009ApJ...696..841K, 2012ApJ...756..168C, 2013ApJ...771...48T, 2019A&A...632A...5G}.

Previously, material was thought to be infalling from a rotationally flattened envelope onto a disk \citep[e.g.,][]{1976ApJ...210..377U, 1981Icar...48..353C, 1984ApJ...286..529T}.
While these theoretical models assumed axisymmetry, actual observations have revealed that dense cores and infalling envelopes have elongated, filamentary, or non-axisymmetric shapes \citep[e.g.,][]{1989ApJS...71...89B, 1991ApJ...376..561M, 1997ApJ...488..317O, 2007ApJ...670L.131L, 2010ApJ...712.1010T, 2011ApJ...740...45T, 2019ApJ...879...25K}.
Numerical simulations predicted that material is non-axisymmetrically accreted along narrow filamentary structures during the early evolutionary phase \citep[e.g.,][]{2015MNRAS.446.2776S, 2017ApJ...846....7K, 2019A&A...628A.112K}.
Such filamentary structures, called streamers, were indeed detected toward several Class 0/I YSOs by molecular line observations at (sub)millimeter wavelengths \cmtbk{\citep[e.g.,][]{2014ApJ...793....1Y, 2019ApJ...880...69Y, 2020NatAs...4.1158P}}.
These structures were kinematically interpreted as channels funneling material from a cloud to an envelope or from an envelope onto a disk.
Since a growing number of streamers ranging in size from hundreds to thousands of au have recently been observed in Class 0/I YSOs \citep[e.g.,][]{2022ApJ...925...32T, 2022A&A...667A..12V, 2024A&A...687A..71V, 2023A&A...669A.137H, 2023ApJ...953...82L}, they may be common in the accretion phase of low-mass star formation.

To systematically investigate the nature of disks in Class 0/I protostellar systems and ultimately to study when and how planet formation begins, we initiated the ALMA Large Program ``Early Planet Formation in Embedded Disks (eDisk)'' \citep{2023ApJ...951....8O}.
This program surveys 19 protostars, 12 Class 0 and 7 Class I sources, which were observed at high angular resolutions of up to 0$\farcs$04 in the 1.3-mm dust continuum and various molecular lines, including CO (2--1) isotopologues.
The primary goal of the program is to (1) examine whether disks surrounding the targeted sources have substructures and (2) identify Keplerian motion to constrain their central protostellar masses.
\citet{2023ApJ...951....8O} found that our targets exhibit overall less substructures than Class II protoplanetary disks reported in previous surveys \citep[e.g.,][]{2018ApJ...869L..41A, 2021MNRAS.501.2934C}.
The initial observations for individual targets have been published in a series of papers.
For example, in the edge-on Class I disk IRAS 04302+2247, dust grains have not yet settled significantly in the vertical direction \citep{2023ApJ...951....9L}.
Three sources\revision{, R CrA IRS 7B, Ced 110 IRS 4, and R CrA IRAS 32,} were newly identified as binary systems, with projected separations between roughly 100 and 250 au \citep{2023ApJ...951....8O, 2023ApJ...954...67S, 2024ApJ...966...32E}.
Other four sources\revision{, IRAS 16253$-$2429, Oph IRS 63, IRAS 16544$-$1604, and L1489 IRS,} exhibit prominent accretion streamers \citep{2023ApJ...954..101A, 2023ApJ...958...98F, 2023ApJ...953..190K, 2025arXiv250400495K, 2023ApJ...951...11Y}.

IRAS 04169+2702 (hereafter I04169), one of the eDisk targets, is a protostar located in the B213 region of the Taurus molecular cloud \citep[e.g.,][]{1996A&A...311..858B, 2013A&A...554A..55H}.
The distance to I04169 is not directly measured, but the mean distance to the B213 region has been calculated to be about 160 pc by parallax observations of other optically visible members therein \citep[e.g.,][]{2019A&A...630A.137G, 2020A&A...638A..85R}.
Recently, \citet{2021AJ....162..110K} reported the mean distance to B213 as 155.94 pc, using the early third data release of \textit{Gaia}
\citep[Gaia EDR3;][]{2021A&A...649A...1G}.
In this paper, we adopt 156 pc as the distance to I04169 (see also Table 1 in \citealt{2023ApJ...951....8O}).
Note that after the third Gaia data was fully released \citep[Gaia DR3;][]{2023A&A...674A...1G}, the mean distance to B213 was measured as 158 pc \citep{2023AJ....165...37L}.
However, this small difference of a few pc does not affect the results that are presented here.

The physical properties of I04169 have been actively studied from near-infrared to millimeter wavelengths.
Previously, this protostar was classified as a Class I YSO based on an infrared spectral index of 0.53 \citep[e.g.,][]{2007AJ....133.1528C}, a bolometric luminosity of 0.7--1.4 $L_{\odot}$ \citep[e.g.,][]{1990AJ.....99..869K, 1993ApJ...414..676K, 1995ApJS..101..117K, 2003ApJS..145..111Y, 2007AJ....133.1528C}, and a bolometric temperature of 133--170 K \citep[e.g.,][]{2001A&A...365..440M, 2003ApJS..145..111Y}.
Utilizing the spectral energy distribution (SED) for this source, \citet{2023ApJ...951....8O} re-calculated its bolometric luminosity and temperature to be 1.4 $L_{\odot}$ and 161 K, respectively; thus, it is still a Class I object.
Furthermore, \citet{2008AJ....135.2496C} reported that this target was identified as a close binary with a separation of 0$\farcs$18 (28 au) through $L^{'}$-band observations with a median angular resolution of 0$\farcs$3.
However, it was not clearly resolved at the angular resolution (see Figure 6 from \citealt{2008AJ....135.2496C}).
Its status as a binary is revisited in this paper.

In addition to the protostar, its surrounding environment has been actively studied at (sub)millimeter wavelengths.
A molecular outflow was first detected in the $^{12}$CO (3--2) spectrum of I04169 \citep{1992ApJ...400..260M}.
Subsequent observations in $^{12}$CO and $^{13}$CO emission further resolved its structure, revealing an elongation along the northeast--southwest direction with a position angle (P.A.) of 64$\arcdeg$ \citep[e.g.,][]{1996A&A...311..858B, 1997ApJ...488..317O, 2018ApJ...865...51T}.
A cometary structure extending over roughly 4,700 au toward the southwest was discovered by several near-IR $K$-band and H$_2$ observations \citep[e.g.,][]{1991ApJ...374L..25T, 1993ApJ...414..773K, 1997ApJ...485..703W, 1997AJ....114.1138G, 2007AJ....133.1528C}.
This structure coincides with the blueshifted lobe of the outflow, indicating that it is a near-IR nebula reflected by the outflow cavity wall.
Observations of C$^{18}$O (1--0) and (2--1) emission reveal a rotationally flattened envelope elongated along the northwest--southeast direction with a P.A. of 154$\arcdeg$, perpendicular to the outflow axis: the outer envelope extends to 2200 au \citep{1997ApJ...488..317O}, while the inner envelope, with a radius of 1000 au, exhibits infall motion \citep{2018ApJ...865...51T}.
Furthermore, \citet{2018ApJ...865...51T} suggested the presence of a Keplerian disk with a radius of 200 au and a central protostellar mass of 0.1 $M_{\odot}$; however, these parameters were not well-constrained due to the limited angular resolution.
\citet{2017ApJ...851...45S} observed 1.3-mm dust continuum emission at a higher angular resolution, revealing a compact, dusty disk with a radius of 39 au embedded in a massive envelope.

This paper is organized as follows.
Section \ref{sec:observations} describes the observations, data reduction, and imaging procedures for I01469.
In Section \ref{sec:results}, we present the overall results of the 1.3-mm dust continuum and molecular line observations.
Based on these results, in Sections \ref{sec:substructures} and \ref{sec:kinematics}, we discuss the binarity/substructure and kinematics of this protostellar system, respectively.
Furthermore, in Section \ref{sec:streamer}, we discuss the nature and possible origins of the streamers first revealed in the system.
Finally, our conclusions are given in Section \ref{sec:conclusions}.

\section{Observations and Data Reduction}
\label{sec:observations}

\begin{deluxetable*}{c@{ }c@{ }c@{ }l@{:}rcccccccc}[htbp]
    \tablenum{1}
    \tablecaption{Summary of the ALMA Observations\label{tab:observations}}
    \tablehead{
        \multicolumn{5}{c}{Date} &
        \colhead{Config.} &
        \colhead{$N_{\rm ant}$} &
        \colhead{Baseline} &
        \colhead{\revision{$\theta_{\rm MRS}$$\tablenotemark{a}$}} &
        \colhead{$t_{\rm on-source}$} &
        \multicolumn2c{Calibrators} &
        \colhead{Memo} \\
        \multicolumn{5}{c}{(UTC)} &
        \colhead{} &
        \colhead{} &
        \colhead{(m)} &
        \colhead{\revision{($\arcsec$)}} &
        \colhead{(min)} &
        \colhead{Flux/Bandpass} &
        \colhead{Phase} &
        \colhead{}}
    \startdata
        2021 & Sep & 30 & 05 & 20 & C43-8 & 44 & 70--11886 & \revision{2.30}  & 24.79 & J0238$+$1636 & J0433$+$2905 & LB \\
        2021 & Oct & 01 & 05 & 22 & C43-8 & 45 & 70--10803 & \revision{2.30}  & 24.79 & J0238$+$1636 & J0438$+$3004 & LB \\
        2021 & Oct & 18 & 04 & 32 & C43-8 & 43 & 46--8983  & \revision{3.50}  & 24.79 & J0238$+$1636 & J0438$+$3004 & LB \\
        2021 & Oct & 24 & 07 & 20 & C43-8 & 45 & 46--8983  & \revision{3.50}  & 24.79 & J0510$+$1800 & J0438$+$3004 & LB \\
        2022 & Jul & 03 & 13 & 15 & C43-5 & 41 & 15--1301  & \revision{10.73} & 20.16 & J0510$+$1800 & J0438$+$3004 & SB \\
        2022 & Jul & 03 & 14 & 26 & C43-5 & 41 & 15--1301  & \revision{10.73} & 20.16 & J0510$+$1800 & J0438$+$3004 & SB
    \enddata
    \revision{\tablenotetext{a}{$\theta_{\rm MRS}$ indicates the maximum recoverable scale (MRS): $\theta_{\rm MRS}$ $\approx$ 0.6$\lambda$/$L_{\rm min}$, where $\lambda$ is the observing wavelength, and $L_{\rm min}$ is the minimum baseline length in a given array configuration \citep{cortes_2025_14933753}.}}
\end{deluxetable*}

\begin{deluxetable*}{cccccc@{ }rccr@{ $\pm$ }l}[htbp]
    \tablenum{2}
    \tablecaption{ALMA and VLA Continuum Image Properties}\label{tab:continuum}
    \tablehead{
        \colhead{Image} &
        \colhead{Data} &
        \colhead{$\nu$} &
        \colhead{$\lambda$} &
        \colhead{Robust} &
        \multicolumn2c{Beam (P.A.)} &
        \colhead{$\sigma_I$\tablenotemark{a}} &
        \colhead{$I_{\rm peak}$} &
        \multicolumn2c{$F_{3\sigma}$\tablenotemark{b}} \\
        \colhead{} &
        \colhead{} &
        \colhead{(GHz)} &
        \colhead{(mm)} &
        \colhead{} &
        \multicolumn2c{($\arcsec$ $\times$ $\arcsec$ ($^{\circ}$))} &
        \colhead{($\mu$Jy beam$^{-1}$)} &
        \colhead{($\mu$Jy beam$^{-1}$)} &
        \multicolumn2c{(mJy)}
    }
    \startdata
        Band 6          & SB   & 225 & 1.3 & 2.0 & 0.463 $\times$ 0.331 & ($-$2.5)  & 30.0 & 76.98 & 92.93 & 0.12 \\
                        & SBLB &     &     & 0.5 & 0.071 $\times$ 0.052 & (28.0)    & 13.1 & 8.40  & 92.19 & 1.64 \\
                        & LB   &     &     & 0.0 & 0.044 $\times$ 0.034 & (19.1)    & 18.8 & 3.71  & 91.71 & 2.02 \\
        Q               & -    & 44  & 7   & 1.0 & 0.056 $\times$ 0.051 & ($-$89.4) & 10.6 & 0.29  & 1.02  & 0.04 \\
        K               & -    & 22  & 14  & 0.0 & 0.083 $\times$ 0.071 & ($-$79.5) & 6.8  & 0.21  & 0.33  & 0.02 \\
        C$_{\rm high}$  & -    & 7   & 43  & 1.0 & 0.337 $\times$ 0.289 & ($-$52.9) & 8.5  & 0.16  & 0.22  & 0.02 \\
        C$_{\rm low}$   & -    & 5   & 60  & 0.0 & 0.334 $\times$ 0.297 & ($-$56.0) & 11.2 & 0.14  & 0.17  & 0.02
    \enddata
    \tablenotetext{a}{$\sigma_I$ denotes the rms noise level of the specific intensity.}
    \tablenotetext{b}{$F_{3\sigma}$ indicates the flux density measured within the 3$\sigma_I$ region of each image.
                      The uncertainty in the flux density represents a statistical uncertainty only.
                      The flux measurement of the SB-only image is confined within a radius of 1$\arcsec$ from the image center.}
\end{deluxetable*}

    \begin{deluxetable*}{ccccc@{ }rcccc}
        \tablenum{3}
        \tablecaption{ALMA Molecular Line Image Properties\label{tab:line}}
        \tablehead{
            \colhead{Image} &
            \colhead{$\nu$} &
            \colhead{$\Delta v$} &
            \colhead{Data} &
            \multicolumn2c{Beam (P.A.)} &
            \colhead{$\sigma_I$} &
            \colhead{$V_{\rm start}$\tablenotemark{a}} &
            \colhead{$V_{\rm end}$\tablenotemark{a}} &
            \colhead{$\sigma_{\rm M0}$\tablenotemark{b}} \\
            \colhead{} &
            \colhead{(GHz)} &
            \colhead{(km s$^{-1}$)} &
            \colhead{} &
            \multicolumn2c{($\arcsec$ $\times$ $\arcsec$ ($^{\circ}$))} &
            \colhead{(mJy beam$^{-1}$)} &
            \colhead{(km s$^{-1}$)} &
            \colhead{(km s$^{-1}$)} &
            \colhead{(mJy beam$^{-1}$ km s$^{-1}$)}
        }
        \decimals
        \startdata
            $^{12}$CO (2--1)                 & 230.538000 & 0.635 & SB   & 0.455 $\times$ 0.323 & ($-$1.4) & 1.88 & $-$12.37 & 18.75 & 11.08 \\
                                             &            &       & SBLB & 0.182 $\times$ 0.147 & (6.5)    & 0.95 & $-$13.01 & 18.75 & 5.61  \\
                                             &            &       & LB   & 0.129 $\times$ 0.109 & (12.9)   & 1.07 & $-$7.93  & 18.11 & 5.75  \\
            $^{13}$CO (2--1)                 & 220.398684 & 0.167 & SB   & 0.478 $\times$ 0.343 & ($-$1.9) & 4.16 & 1.68     & 11.37 & 10.40 \\
                                             &            &       & SBLB & 0.189 $\times$ 0.151 & (4.0)    & 2.09 & 1.01     & 11.37 & 5.27  \\
                                             &            &       & LB   & 0.132 $\times$ 0.111 & (10.8)   & 2.35 & 2.18     & 10.20 & 5.06  \\
            C$^{18}$O (2--1)                 & 219.560354 & 0.167 & SB   & 0.482 $\times$ 0.343 & ($-$2.0) & 3.17 & 3.18     & 9.86  & 4.72  \\
                                             &            &       & SBLB & 0.188 $\times$ 0.149 & (6.6)    & 1.57 & 3.18     & 10.20 & 2.44  \\
                                             &            &       & LB   & 0.133 $\times$ 0.111 & (14.2)   & 1.75 & 3.35     & 9.86  & 2.63  \\
            SO (6$_5$--5$_4$)                & 219.949442 & 0.167 & SB   & 0.475 $\times$ 0.344 & ($-$2.3) & 3.76 & 3.35     & 9.36  & 5.34  \\
                                             &            &       & SBLB & 0.188 $\times$ 0.152 & (8.6)    & 1.88 & 2.68     & 9.70  & 2.93  \\
                                             &            &       & LB   & 0.133 $\times$ 0.112 & (17.0)   & 2.11 & 3.52     & 9.70  & 3.08  \\
            H$_2$CO (3$_{0, 3}$--2$_{0, 2}$) & 218.222192 & 1.340 & SB   & 0.480 $\times$ 0.346 & ($-$4.0) & 1.08 & 4.74     & 10.1  & 3.97  \\
                                             &            &       & SBLB & 0.186 $\times$ 0.148 & (5.5)    & 0.52 & 3.40     & 8.76  & 1.91  \\
                                             &            &       & LB   & 0.132 $\times$ 0.110 & (12.4)   & 0.58 & 3.40     & 8.76  & 2.10  \\
        \enddata
        \tablenotetext{a}{$V_{\rm start}$ and $V_{\rm end}$ are the starting and ending channel velocities of each emission range detected above the 3$\sigma_I$ level, respectively.}
        \tablenotetext{b}{$\sigma_{\rm M0}$ indicates the rms noise level of the intensity map integrated over the determined velocity range (moment 0) for each molecular line.}
    \end{deluxetable*}

\subsection{ALMA}
\label{sec:almaobservations}

We observed I04169 using ALMA in Band 6, as part of the ALMA Large Program eDisk (2019.1.00261.L; PI: Nagayoshi Ohashi).
The correlator was configured to observe the 1.3-mm dust continuum and various molecular lines simultaneously, including \revision{$^{12}$CO, $^{13}$CO, C$^{18}$O,} SO, \revision{and} H$_2$CO.
To cover a wide range of physical scales, i.e., from a small embedded disk and a surrounding envelope up to a large bipolar outflow, the observations were made in the C43-8 and C43-5 configurations, from 2021 September 30 to October 24 and on 2022 July 3, respectively.
The observations with the C43-8 configuration, which we will refer to as Long-Baseline (LB) observations, were executed four times, whereas those with the C43-5 configuration, Short-Baseline (SB) observations, were executed twice.
Details of the observations for both configurations, including the number of 12-m antennas used, projected baseline length, on-source integration time, and calibrators, are summarized in Table \ref{tab:observations}.
Further details of the observations, such as the correlator setup and visibility calibration, should be referred to \citet{2023ApJ...951....8O}.

The standard calibration was carried out using the ALMA calibration pipeline included in the Common Astronomy Software Application \citep[CASA;][]{2007ASPC..376..127M, 2022PASP..134k4501C} version 6.2.1, and subsequent imaging and self-calibration were also carried out with the same CASA version.
Details of the overall procedures are explained in \citet{2023ApJ...951....8O}.
Here, we briefly describe the imaging procedures for I04169, and the imaging scripts for this target can be found in \citet{2023zndo...7986682T}\footnote{\url{http://github.com/jjtobin/edisk}}.
The self-calibration was performed on the SB-only data, and then the self-calibrated SB data were combined with the LB data to self-calibrate the combined data set (hereafter SBLB).
The self-calibration was conducted until the signal-to-noise ratio (S/N) began decreasing with shorter solution intervals for the CASA task \textit{gaincal}.
We used the Briggs weighting with a robust parameter of 0.5 for all images created in the self-calibration process.
For the SB-only data, we did six iterations of phase-only self-calibration and two iterations of phase and amplitude self-calibration.
The self-calibration for the SBLB data was performed in the following two steps: using the SB-only data first, we did six iterations of phase-only self-calibration and then an iteration of phase and amplitude self-calibration; second, for the combined data, we did five iterations of phase-only self-calibration and two iterations of phase and amplitude self-calibration.
The self-calibration solutions obtained from the continuum data were applied to the molecular line data.

The final SB-only and SBLB images of the dust continuum and molecular line emission were created from the self-calibrated SB and SBLB data, respectively, using the CASA task \textit{tclean}.
The final LB-only images were created by selecting only the LB measurement sets of the self-calibrated SBLB data.
All the images were corrected for the primary beam attenuation.
Regarding the dust continuum, we individually made three images from the SB-only, SBLB, and LB-only visibilities.
Different Briggs parameters were chosen for each of the three images to better represent both small- and large-scale structures: SB-only (robust = 2.0), SBLB (0.5), and LB-only (0.0).
Among the observed molecular lines (see also Table 2 from \citealt{2023ApJ...951....8O}), only five lines were clearly detected toward I04169: $^{12}$CO (2--1), $^{13}$CO (2--1), C$^{18}$O (2--1), SO (6$_5$--5$_4$), and H$_2$CO (3$_{0, 3}$--2$_{0, 2}$).
Likewise, the images of the molecular lines were individually made from the SB-only, SBLB, and LB-only visibilities.
In contrast to the dust continuum, all these images have the same robust parameter of 2.0.
These molecular-line images of the SBLB and LB-only data are uv-tapered by a circular Gaussian with a full width at half maximum (FWHM) of 2000 $k\lambda$ to recover emission from large-scale structures.
The properties of the final images, such as synthesized beam sizes and rms noise levels, are listed in Tables \ref{tab:continuum} and \ref{tab:line}.

\subsection{VLA}
\label{sec:vlaobservations}

Three eDisk targets in Taurus\revision{, including I04169,} were observed with the Karl G. Jansky Very Large Array (VLA) in the A configuration at the Q- (0.7 cm), K- (1.4 cm), and C-bands (5.0 cm) (Program 20B-322).
Details of the observations, data reduction, and imaging procedures are described in \citet{2022ApJ...934...95S}, and here, we briefly introduce the imaging procedures of I04169.
The observations of I04169 were conducted three times: on 2020 December 10, 24, and 30.
We imaged the observed Q- and K-band data with robust parameters of 1.0 and 0.0, respectively.
The angular resolutions of the Q- and K-band observations are comparable to that of the ALMA LB-only image, and thus, the direct comparison between the millimeter- and centimeter-wavelength images is feasible, which is discussed in Section \ref{sec:substructures}.
The flux density within the 3$\sigma_I$ region of the Q-band image is almost the same as the natural-weighted image (robust $=$ 2.0), where $\sigma_I$ is the rms noise level of the specific intensity, while the K-band image also has a comparable flux density with the natural-weighted within 25$\%$.
This means that the emission arising from the scales of 0$\farcs$06 and 0$\farcs$13, corresponding to the synthesized beam sizes of the natural-weighted images, is not severely resolved out in the presented Q- and K-band images.
Regarding the C-band, we divided it into two sub-bands for imaging, named C$_{\rm high}$ (4.3 cm) and C$_{\rm low}$ (6.0 cm), which are the high- and low-frequency ends of this band.
These two sub-bands provide us with better constraints on a free-free spectral slope of the target (see Section \ref{sec:substructures}).
We chose robust parameters of 1.0 and 0.0 for the C$_{\rm high}$- and C$_{\rm low}$-band images to achieve comparable synthesized beam sizes each other.
Details of these four-band images are summarized in Table \ref{tab:continuum}.

We corrected for the proper motion of I04169 to directly compare the ALMA LB-only and VLA images.
The observations of the two telescopes were taken approximately 10 months apart: December 2020 for VLA and October 2021 for ALMA LB.
However, given that the ALMA LB-only and VLA images of this target show significantly different appearances, it is not straightforward to use them to derive the proper motion.
Instead, we computed the proper motion of the nearby protostar \revision{IRAS 04166$+$2706 (hereafter I04166)} because it was observed by ALMA simultaneously with I04169 \citep{2023ApJ...951....8O} and was observed by VLA on similar dates to when I04169 was observed.
I04166 exhibits a single central peak within its disk in the ALMA LB-only and VLA Q-band observations, and the synthesized beam sizes of both images are comparable: 0$\farcs$047 $\times$ 0$\farcs$035 and 0$\farcs$046 $\times$ 0$\farcs$040 for ALMA and VLA, respectively.
Using the CASA task \textit{imfit}, we obtained the elliptical Gaussian centers of the two images of I04166 separately.
The Gaussian center of the ALMA image is at $\alpha_{\rm ICRS}$ $=$ 4$^{\rm h}$19$^{\rm m}$42$\fs$50512 and $\delta_{\rm ICRS}$ $=$ 27$\arcdeg$13$\arcmin$35$\farcs$82036, while that of the VLA Q-band image is at $\alpha_{\rm ICRS}$ $=$ 4$^{\rm h}$19$^{\rm m}$42$\fs$50430 and $\delta_{\rm ICRS}$ $=$ 27$\arcdeg$13$\arcmin$35$\farcs$85001.
The proper motion of I04166 is finally derived as (14.76 mas yr$^{-1}$, $-$35.64 mas yr$^{-1}$).
We applied this proper motion to I04169.

\section{Results}
\label{sec:results}

\subsection{Continuum}
\label{sec:cont}

\subsubsection{ALMA}
\label{sec:alma}

\begin{figure*}[t]
    \gridline{
        \leftfig{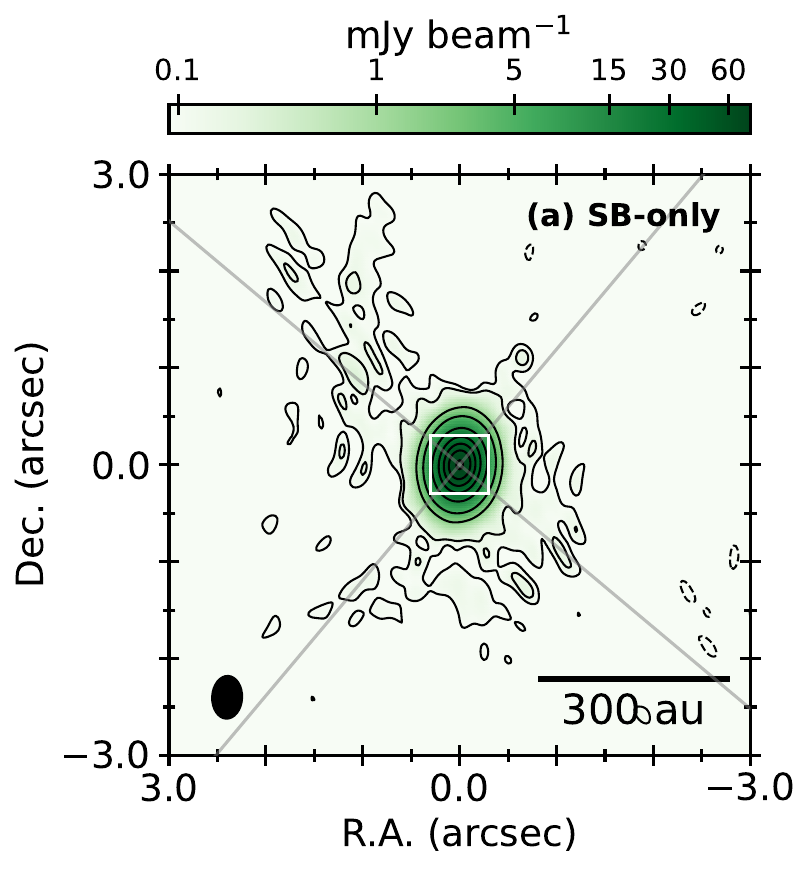}{0.3\textwidth}{}
        \fig{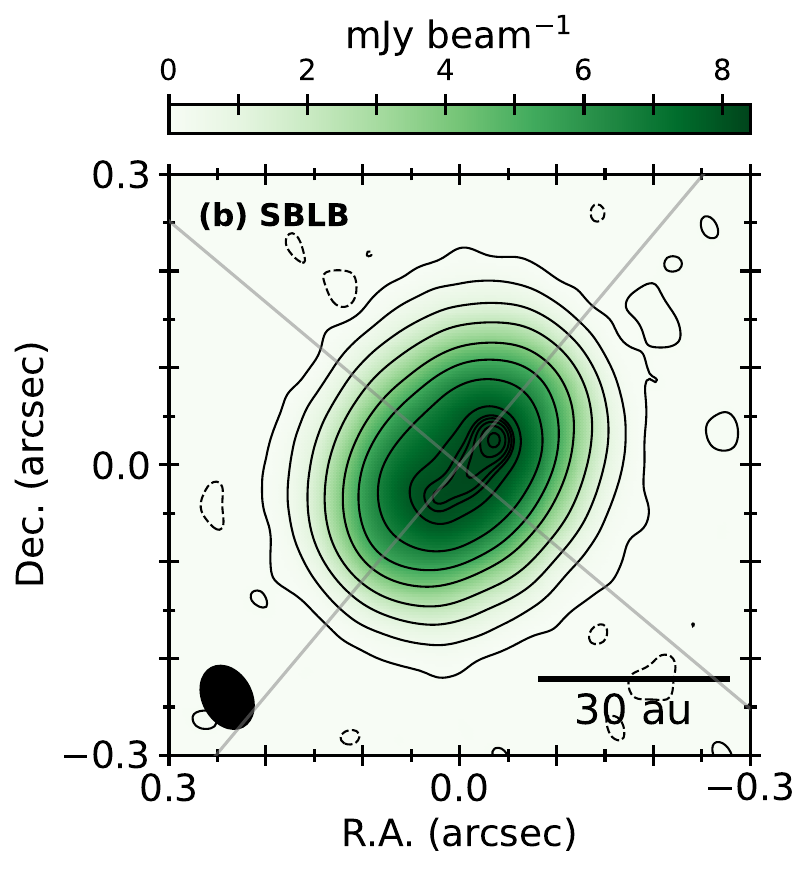}{0.3\textwidth}{}
        \rightfig{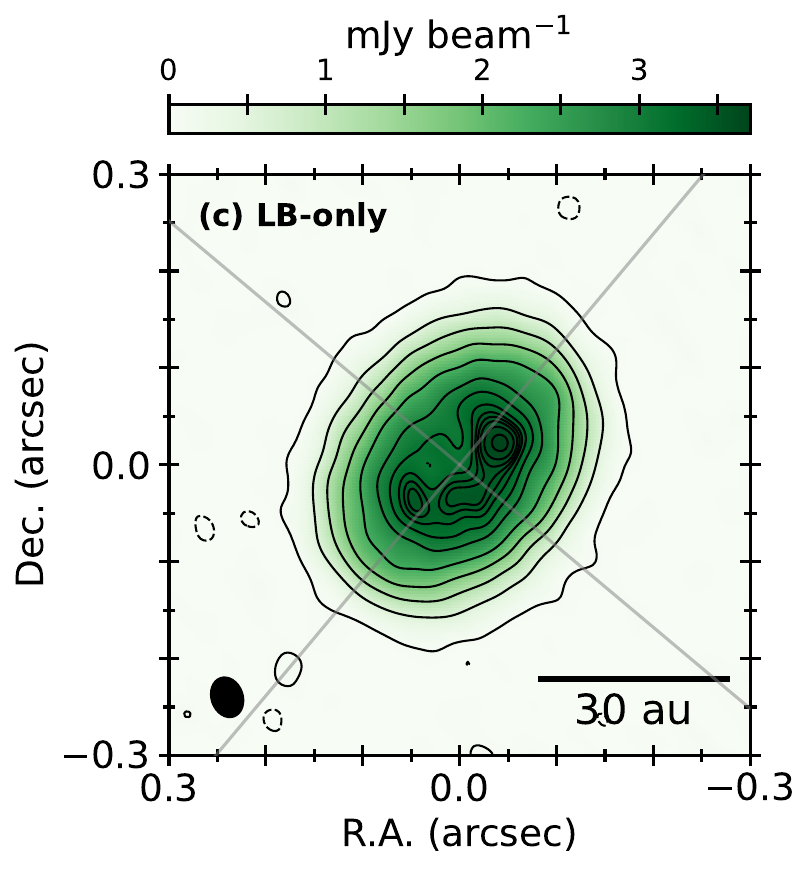}{0.3\textwidth}{}
    }
    \vspace{-0.5cm}
    \caption{ALMA Band 6 (1.3 mm) dust continuum images of I04169 obtained from the three different data sets: SB-only, SBLB, and LB-only. \textbf{(a)} SB-only continuum image with a robust parameter of 2.0. The color scale is stretched by the asinh function to cover a wide dynamic range from 0.09 to 77.0 mJy beam$^{-1}$. The contour levels are \{$-$3, 3, 5, 50, 150, 500, 1000, 1500, 2000, 2500\} $\times$ $\sigma_{\rm SB}$, where 1$\sigma_{\rm SB}$ $=$ 30.0 $\mu$Jy beam$^{-1}$. The white box indicates the entire region of the middle and right panels. The black ellipse shown in the lower left denotes the synthesized beam with an FWHM of 0$\farcs$463 $\times$ 0$\farcs$331 (P.A. $=$ $-$2.5$^{\circ}$). \textbf{(b)} SBLB continuum image with a robust parameter of 0.5. The color is linearly scaled from 0 mJy beam$^{-1}$ to a peak intensity of 8.40 mJy beam$^{-1}$. The contour levels are \{$-$3, 3, 30, 90, 180, 300, 400, 500, 600, 620, 625, 630, 635, 640\} $\times$ $\sigma_{\rm SBLB}$, where 1$\sigma_{\rm SBLB}$ $=$ 13.1 $\mu$Jy beam$^{-1}$. The black ellipse in the lower left denotes the synthesized beam of 0$\farcs$071 $\times$ 0$\farcs$052 (P.A. $=$ 28.0$^{\circ}$). \textbf{(c)} LB-only continuum image with a robust parameter of 0.0. The color is also linearly scaled from 0 mJy beam$^{-1}$ to a peak intensity of 3.71 mJy beam$^{-1}$. The contour levels are \{$-$3, 3, 30, 60, 90, 120, 150, 165, 175, 180, 183, 186, 190, 195\} $\times$ $\sigma_{\rm LB}$, where 1$\sigma_{\rm LB}$ $=$ 18.8 $\mu$Jy beam$^{-1}$. The black ellipse in the lower left denotes the synthesized beam of 0$\farcs$044 $\times$ 0$\farcs$034 (P.A. $=$ 19.1$^{\circ}$). The two grey lines in these three panels indicate the directions of the disk major (P.A. $=$ 140$\arcdeg$) and minor (P.A. $=$ 50$\arcdeg$) axes, respectively, of the LB-only image.} 
    \label{fig:cont}
\end{figure*}

\begin{figure*}[t]
    \gridline{
        \fig{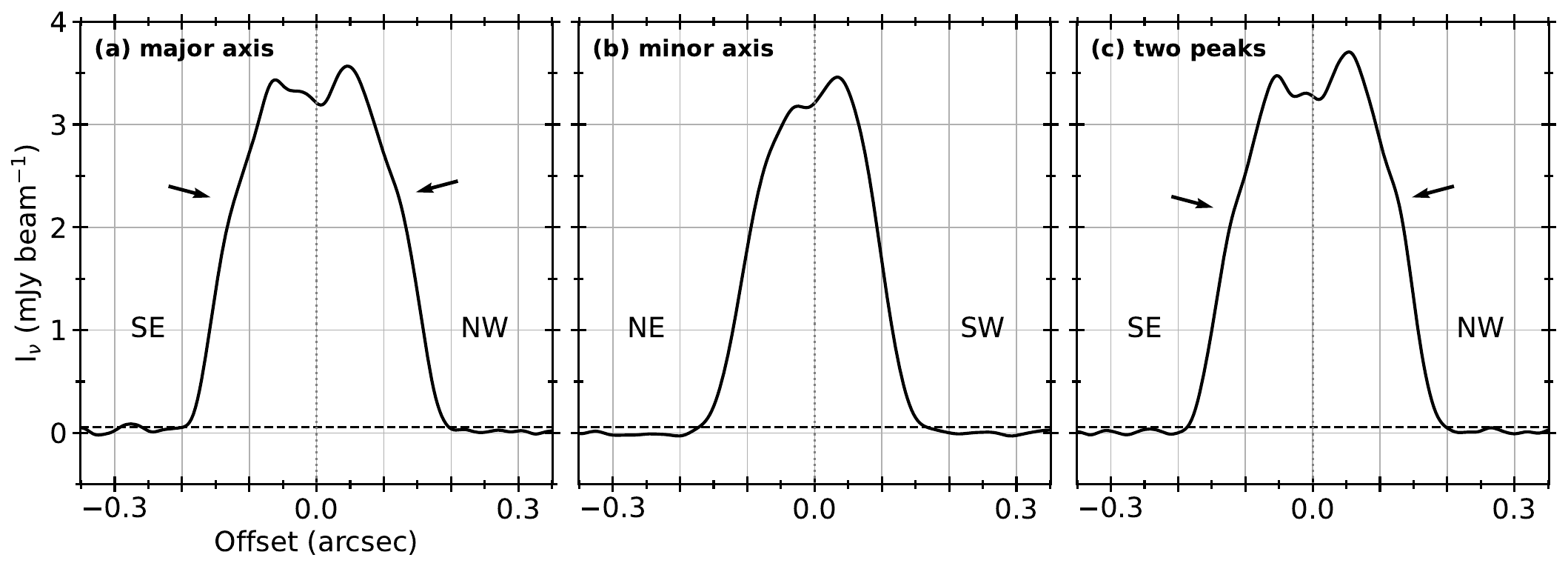}{0.9\textwidth}{}
    }
    \vspace{-0.8cm}
    \caption{Radial intensity cuts of the LB-only dust continuum image, shown in Figure \ref{fig:cont}(c), along three directions. The black horizontal dashed lines in the three panels mark the 3$\sigma_{\rm LB}$ level, where 1$\sigma_{\rm LB}$ $=$ 18.8 $\mu$Jy beam$^{-1}$. \textbf{(a)} Intensity cut along the major axis (P.A. $=$ 140$\arcdeg$). The left and right indicate the southeastern (SE) and northwestern (NW) parts of the disk, respectively, along the major axis. \textbf{(b)} Intensity cut along the minor axis (P.A. $=$ 50$\arcdeg$). The left and right indicate the northeastern (NE) and southwestern (SW) of the disk, respectively, along the minor axis. \textbf{(c)} Intensity cut along the two peaks (P.A. $=$ 124$\arcdeg$). The zero-offset position in this panel is the same as those in the left and middle panels, which is the Gaussian center of the LB-only dust continuum image (Section \ref{sec:cont}), and only the position angle is changed from 140$\arcdeg$ of panel (a) to 124$\arcdeg$. Note that there are small bumps in the outer region, located 0$\farcs$13 (20 au) from the center, which are denoted by the black arrows in panels (a) and (c).}
    \label{fig:cut}
\end{figure*}

Figures \ref{fig:cont}a, \ref{fig:cont}b, and \ref{fig:cont}c show the 1.3-mm continuum images made from the SB-only, SBLB, and LB-only data, respectively, to represent a wide range of physical scales from the overall structure of the dusty disk and envelope to hidden substructures of the disk.
The SB-only image exhibits the compact component at the center\revision{, interpreted as the circumstellar disk,} and the surrounding extended emission,
\revision{likely from part of the inner envelope.}
\revision{The compact component, when fitted with an elliptical Gaussian,} yields the deconvolved full width at half maximum (FWHM) of 0$\farcs$19 $\times$ 0$\farcs$14 (30 au $\times$ 22 au) with a P.A. of 140$\arcdeg$\revision{, which is comparable to the previously reported disk radius ($R_{\rm disk}$ $=$ 39 au; \citealt{2017ApJ...851...45S})}.
\revision{The extended emission}
shows features stretching out along the northeast--southwest axis.
In particular, the northeastern arm-like one appears to curve northward.
A similar feature was previously detected at the 3--5$\sigma_I$ level in the Submillimeter Array (SMA) 1.3-mm observations, albeit with a much coarser angular resolution and lower sensitivity (see Figure 1c from \citealt{2018ApJ...865...51T}).

The SBLB and LB-only images show details of the compact protostellar disk, such as its overall shape, which is elongated along the northwest--southeast direction.
The Gaussian fitting to the disk emission gives quite similar results: for the SBLB image, the deconvolved FWHM is 0$\farcs$21 $\times$ 0$\farcs$15 (33 au $\times$ 23 au), and the P.A. is 139$\arcdeg$, while for the LB-only image, they are 0$\farcs$22 $\times$ 0$\farcs$16 (34 au $\times$ 25 au) and 140$\arcdeg$, respectively.
Furthermore, these two images unveil substructures at the center of the disk.
The SBLB image exhibits a bean-shaped asymmetry, and the higher resolution of the LB-only image enables resolving this feature into two distinct peaks located in the northwest (primary) and southeast (secondary) along the disk major axis.
There is a central depression between the two peaks, and an arc-shaped component in the southwest connects them.

We measure the flux densities within the 3$\sigma_I$ areas of the SBLB and LB-only images to be 92.19 $\pm$ 1.64 and 91.71 $\pm$ 2.02 mJy, respectively.
Note that these flux densities are slightly smaller than the one measured by the elliptical Gaussian fitting presented in \citet{2023ApJ...951....8O}.
Since the two values calculated in this paper are almost identical, we derive the physical parameters of the disk from the higher-resolution LB-only image.
The results of the Gaussian fitting to this image enable us to estimate the disk inclination angle to be 43$\arcdeg$ by assuming that the disk is geometrically thin and using the ratio of the major axis to the minor axis of the FWHM: $i$ $=$ cos$^{-1}$($\theta_{\rm min}$/$\theta_{\rm maj}$).
The Gaussian center of this image is at $\alpha_{\rm ICRS}$ $=$ 04$^{\rm h}$19$^{\rm m}$58$\fs$475 and $\delta_{\rm ICRS}$ $=$ 27$\arcdeg$09$\arcmin$56$\farcs$81.
We also performed Gaussian fitting to the primary and secondary peaks individually to derive their central positions and separation.
To minimize the contribution of the surrounding emission, the fitting region \cmtbk{for both the primary and secondary peaks} is restricted to the single synthesized beam size.
The Gaussian center of the primary peak is derived at $\alpha_{\rm ICRS}$ $=$ 04$^{\rm h}$19$^{\rm m}$58$\fs$472 and $\delta_{\rm ICRS}$ $=$ 27${\arcdeg}$09${\arcmin}$56$\farcs$84, while the secondary center is at $\alpha_{\rm ICRS}$ $=$ 04$^{\rm h}$19$^{\rm m}$58$\fs$478 and $\delta_{\rm ICRS}$ $=$ 27${\arcdeg}$09${\arcmin}$56$\fs$78.
The intensities of the primary and secondary peaks are 3.71 and 3.48 mJy beam$^{-1}$, respectively.
The projected separation of the two peaks is 0$\farcs$11 (17 au).
This is smaller than the separation of the possible binary reported from the previous near-IR observations by \citet{2008AJ....135.2496C}, which is 0$\farcs$18 (28 au).

To further investigate the nature of the substructure in the LB-only image, we compare three radial intensity cuts, which are along the major and minor axes and along the two peaks, as presented in Figure \ref{fig:cut}.
First, the intensity cut of the major axis (P.A. $=$ 140$\arcdeg$) shown in Figure \ref{fig:cut}a has the central two peaks.
The right and left peaks correspond to the primary (northwestern) and secondary (southeastern) peaks, respectively.
Small bumps appear in the outer region, located 0$\farcs$13 (20 au) from the Gaussian center of the image, where the intensity drops off sharply.
In Figure \ref{fig:cut}b, the radial cut along the minor axis (P.A. $=$ 50$\arcdeg$) shows \cmtbk{an asymmetric} brightness distribution in the central region: the southwestern part is brighter than the northeastern part, probably because of the bridge-like feature connecting the primary and secondary peaks.
In contrast to the cut along the major axis, there are no bumps in the outer region.
Lastly, Figure \ref{fig:cut}c shows the intensity cut along the two peaks (P.A. $=$ 124$\arcdeg$).
The projected separation between the two peaks is 0$\farcs$11 (17 au), as measured in the image plane above, and the contrast in their intensities is around 12$\sigma_{\rm LB}$, where 1$\sigma_{\rm LB}$ $=$ 18.8 $\mu$Jy beam$^{-1}$.
Small outer bumps also appear on both sides, located 0$\farcs$13 (20 au) from the center, the same as those seen along the major axis.

The dust mass of the disk is estimated from the 1.3-mm continuum flux density of the LB-only image, assuming that the continuum emission originates from isothermal dust grains and is optically thin \citep[e.g.,][]{1983QJRAS..24..267H}:
\begin{equation}
    M_{\rm dust} = \frac{F_{\nu} d^2}{\kappa_{\nu} B_{\nu}(T_{\rm dust})},
\end{equation}
where $M_{\rm dust}$ is the dust mass, $F_{\nu}$ is the flux density at the given frequency $\nu$, $d$ is the distance to the source, $\kappa_{\nu}$ is the dust mass absorption coefficient at the given frequency $\nu$, and $B_{\nu}$($T_{\rm dust}$) is the Planck function at the dust temperature $T_{\rm dust}$.
The central frequency of the observations is 225 GHz.
The dust opacity adopted at this frequency is 2.25 cm$^{2}$ g$^{-1}$ based on the equation proposed by \citet{1990AJ.....99..924B}: $\kappa_{\nu}$ $=$ ($\nu$/100 GHz)$^{\beta}$ and $\beta$ $=$ 1, where $\beta$ is the dust opacity spectral index.
Note that this opacity is comparable to those of large grain populations with a maximum grain size $a_{\rm max}$ of 1 mm \citep[e.g.,][]{2009ApJ...700.1502A, 2011ApJ...732...42A, 2018ApJ...869L..45B, 2023ApJ...956....9H}.
Regarding the dust temperature, we assume a typical value of 20 K, which was
previously used to calculate dust masses of YSOs in Taurus, including I04169 \citep[e.g.,][]{2005ApJ...631.1134A, 2017ApJ...851...45S}.
In addition, \citet{2020ApJ...890..130T} recently derived a simple equation for the mean dust temperature of protostellar disks, which is scaled as follows: $T_{\rm dust}$ $=$ 43 K ($L_{\rm bol}$/$L_{\sun}$)$^{0.25}$, where $L_{\rm bol}$ is the bolometric luminosity.
Note that \citet{2020ApJ...890..130T} also suggested a temperature distribution depending on the disk radius ($T_{\rm dust}$ $\propto$ $R^{-0.46}$).
\citet{2022ApJ...929...76S} developed this equation through radiative transfer modeling so that the mean dust temperature of embedded protostellar disks can be better constrained when considering the disk size as well as the bolometric luminosity.
The above equation is thus developed into the following:
\begin{equation}
    T_{\rm dust} = 43\ {\rm K}\ \left(\frac{L_{\rm bol}}{L_{\odot}}\right)^{0.25} \left(\frac{R_{\rm disk}}{50\ {\rm au}}\right)^{-0.5},
    \label{eq:temperature}
\end{equation}
where $R_{\rm disk}$ is the disk radius.
\citet{2020ApJ...890..130T} assumed the disk radius to be
\revision{twice the standard deviation (hereafter $\sigma$) obtained by fitting a Gaussian to the deconvolved major axis.}
\revision{For I04169, the major axis of the deconvolved FWHM from our LB-only image is 0$\farcs$22 (34 au), resulting in the 2$\sigma$ disk radius of 29 au.}
\revision{Thus,} according to Equation \ref{eq:temperature}, the dust temperature of I04169 is derived to be 61 K.
Adopting the two temperatures of 20 and 61 K, the dust masses of the protostellar disk are derived to be 67 and 18 $M_{\earth}$, respectively.
Assuming a typical gas-to-dust mass ratio of 100, the total disk mass is 0.020 and 0.005 $M_{\sun}$.
Note that our mass estimates would be lower limits if the 1.3-mm continuum emission is optically thick.

\subsubsection{VLA}
\label{sec:vla}

\begin{figure*}[t]
    \gridline{
        \leftfig{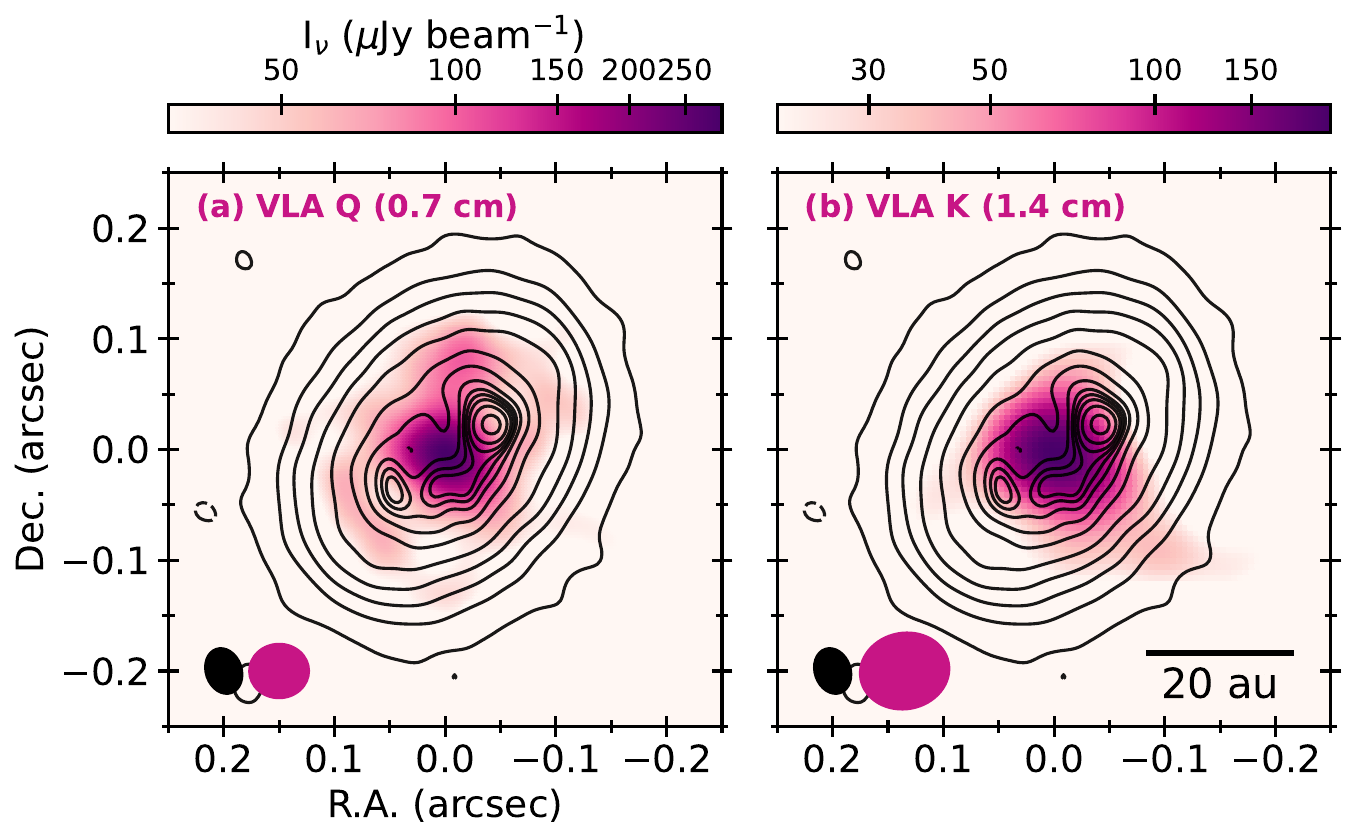}{0.46\textwidth}{}
        \rightfig{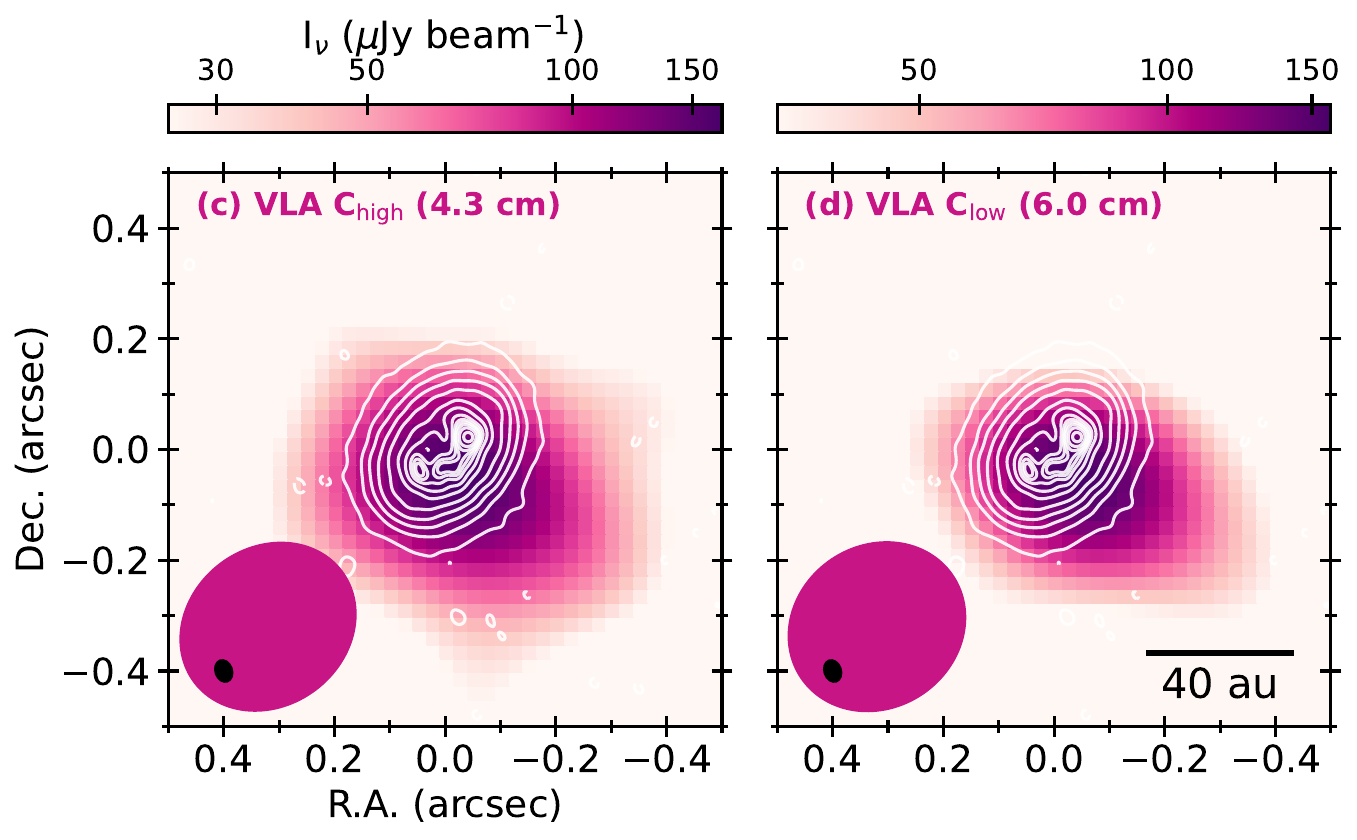}{0.46\textwidth}{}
    }
    \vspace{-0.5cm}
    \caption{VLA Q- (0.7 cm), K- (1.4 cm), C$_{\rm high}$- (4.3 cm), and C$_{\rm low}$-band (6 cm) continuum images of I04169. The black or white contours represent the ALMA Band 6 (1.3 mm) LB-only continuum image with a robust parameter of 0.0, shown in Figure \ref{fig:cont}c. The contour level and the synthesized beams denoted by the black ellipse at the lower left are the same as in Figure \ref{fig:cont}c. The color maps indicate the VLA continuum images, corrected for the proper motion between the ALMA and VLA observations. In all the panels, the color scale is logarithmically scaled from the 3$\sigma_I$ level. The red-violet ellipses in the lower left denote the synthesized beams of the VLA images. The two left panels have an identical spatial scale of 0$\farcs$5 $\times$ 0$\farcs$5, while the two right panels show a four times larger area of 1$\farcs$0 $\times$ 1$\farcs$0 to cover the whole emission of the C$_{\rm high}$ and C$_{\rm low}$ bands. \textbf{(a)} Q-band (7 mm) image with a robust parameter of 1.0. Its synthesized beam and rms noise level are 0$\farcs$056 $\times$ 0$\farcs$051 (P.A. $=$ $-$89.4$^{\circ}$) and $\sigma_{\rm Q}$ $=$ 10.6 $\mu$Jy beam$^{-1}$, respectively. \textbf{(b)} K-band (1.4 cm) image with a robust parameter of 0.0. The synthesized beam is 0$\farcs$083 $\times$ 0$\farcs$071 (P.A. $=$ $-$79.5$^{\circ}$), and the rms noise level is $\sigma_{\rm K}$ $=$ 6.8 $\mu$Jy beam$^{-1}$. Note that this image shows a single central peak extending southwest, located between the two peaks of the ALMA image. \textbf{(c)} C$_{\rm high}$-band (4.3 cm) image with a robust parameter of 1.0. The synthesized beam is 0$\farcs$337 $\times$ 0$\farcs$289 (P.A. $=$ $-$52.9$^{\circ}$), and the rms noise level is 1$\sigma_{\rm C_{high}}$ $=$ 8.5 $\mu$Jy beam$^{-1}$. \textbf{(d)} C$_{\rm low}$-band (6.0 cm) image with a robust parameter of 0.0 and a synthesized beam of 0$\farcs$334 $\times$ 0$\farcs$297 (P.A. $=$ $-$56.0$^{\circ}$), where 1$\sigma_{\rm C_{low}}$ $=$ 11.2 $\mu$Jy beam$^{-1}$. In addition to the K-band image, the emission in these two C-band images is also elongated southwest, implying a jet extending toward this direction.}
    \label{fig:vla}
\end{figure*}

Figure \ref{fig:vla} shows the proper motion-corrected VLA images with the ALMA LB-only image overlaid as contours.
For each image, the flux density of the 3$\sigma_I$ region and the peak intensity are listed in Table \ref{tab:continuum}.
The Q-band image shown in Figure \ref{fig:vla}a is centrally peaked between the two ALMA peaks, slightly elongated toward the southwest.
This suggests that the two ALMA peaks do not represent two distinct protostars.
The source multiplicity is discussed further in Section \ref{sec:substructures}.
In addition, the emission is spread across the disk region, with low surface brightness, and interestingly, it is faint around the two ALMA peaks.
This anti-correlation suggests that the compact Q-band emission may be dominated by free-free emission rather than thermal dust emission traced in the 1.3-mm continuum emission.
The K-band image in Figure \ref{fig:vla}b shows a central peak extending toward the southwest, the same as the outflow direction, and this peak is also located between the two ALMA peaks.
Lastly, the C$_{\rm high}$- and C$_{\rm low}$-band images in Figures \ref{fig:vla}c and \ref{fig:vla}d also show emission elongated in the southwest direction.
Note that it is difficult to directly compare the features with those of the Q- and K-bands due to the large difference in the synthesized beam sizes.
The southwest extension in the K-, C$_{\rm high}$-, and C$_{\rm low}$-band images suggests that these three bands trace part of the protostellar jet.

\subsection{Molecular Lines}
\label{sec:lines}

\begin{figure*}[t]
    \gridline{
        \rightfig{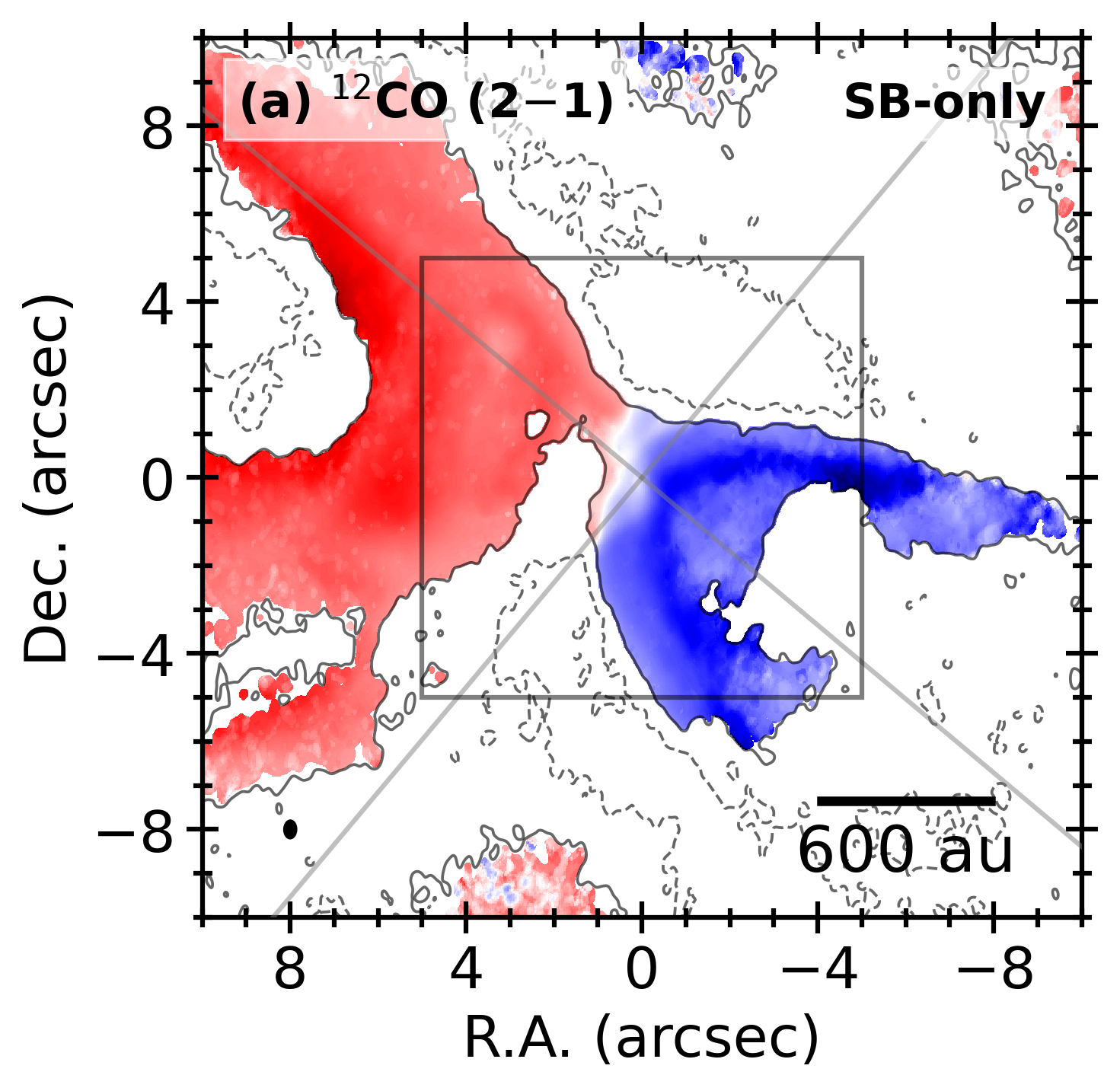}{0.24\textwidth}{}
        \hspace{-1.0cm}
        \fig{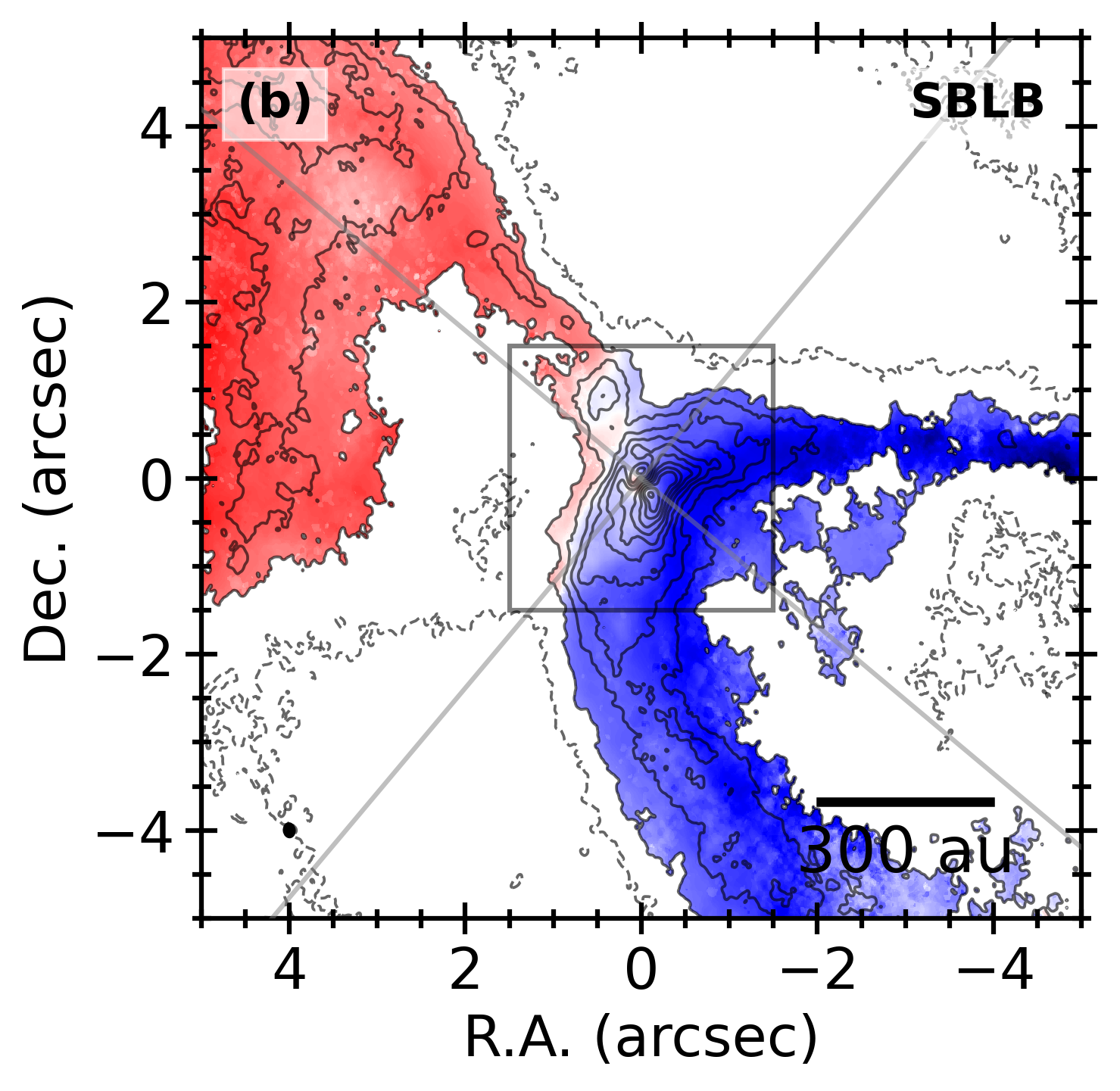}{0.24\textwidth}{}
        \hspace{-1.0cm}
        \leftfig{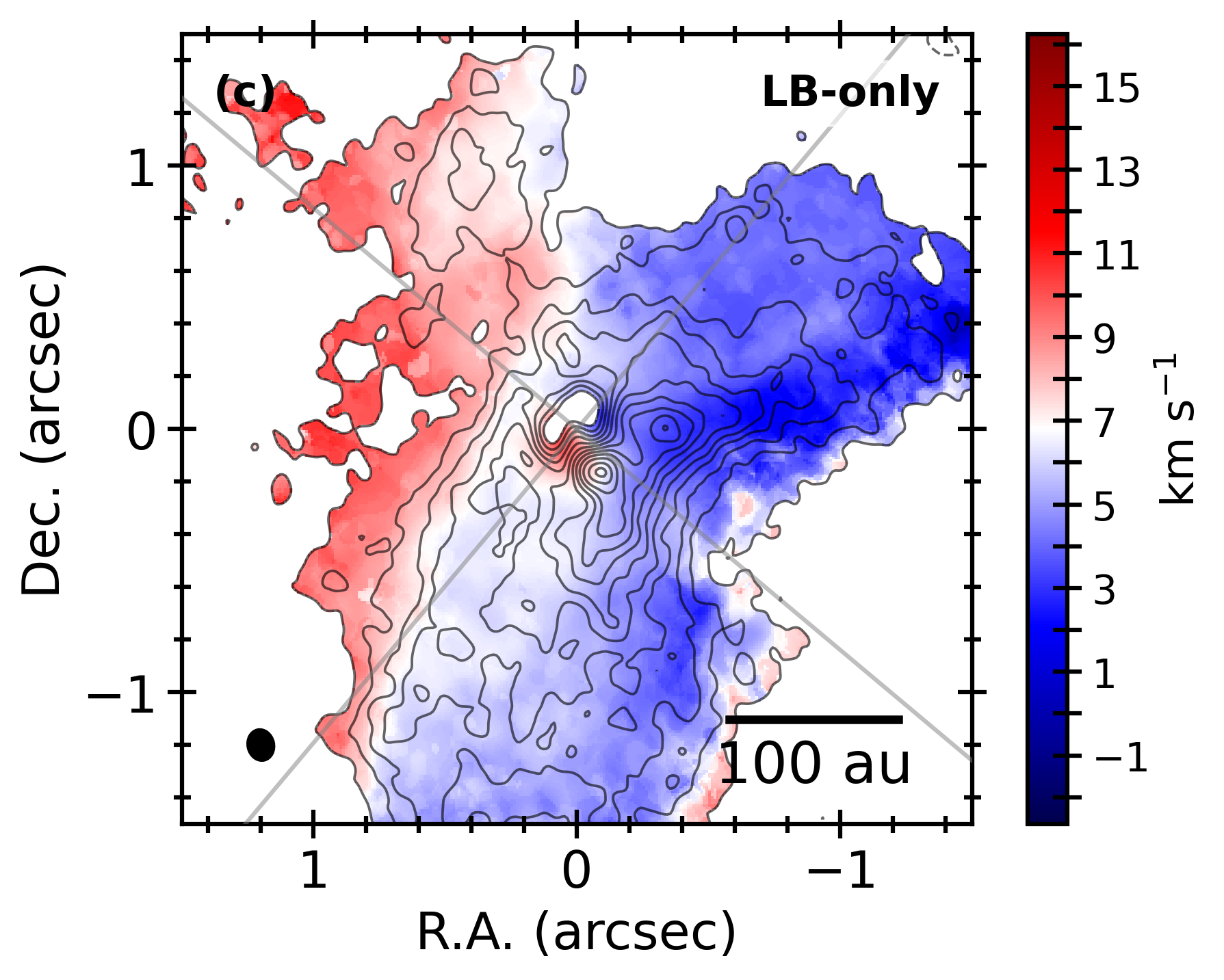}{0.2919\textwidth}{}
    }
    \vspace{-1.55cm}
    \gridline{
        \rightfig{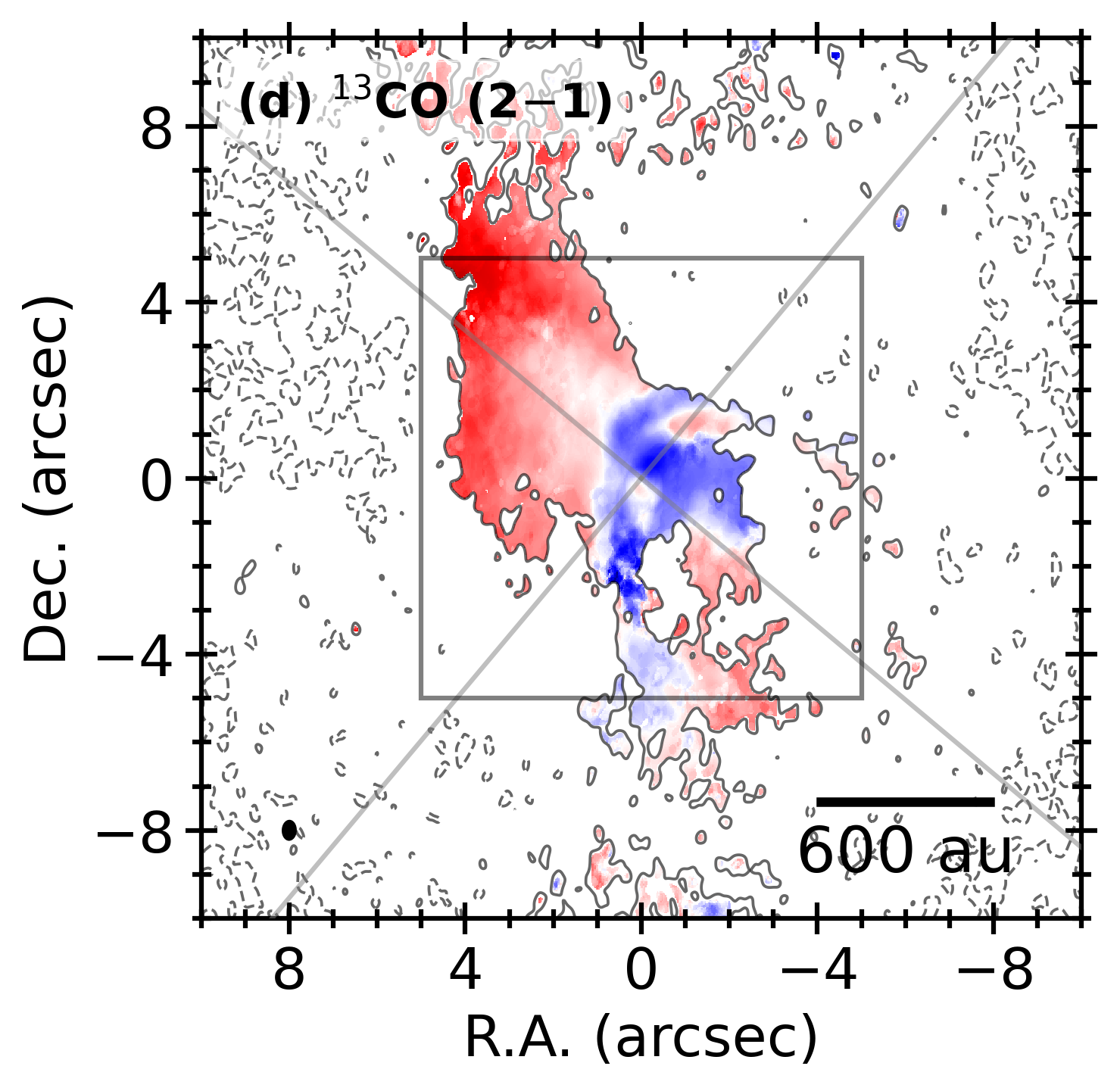}{0.24\textwidth}{}
        \hspace{-1.0cm}
        \fig{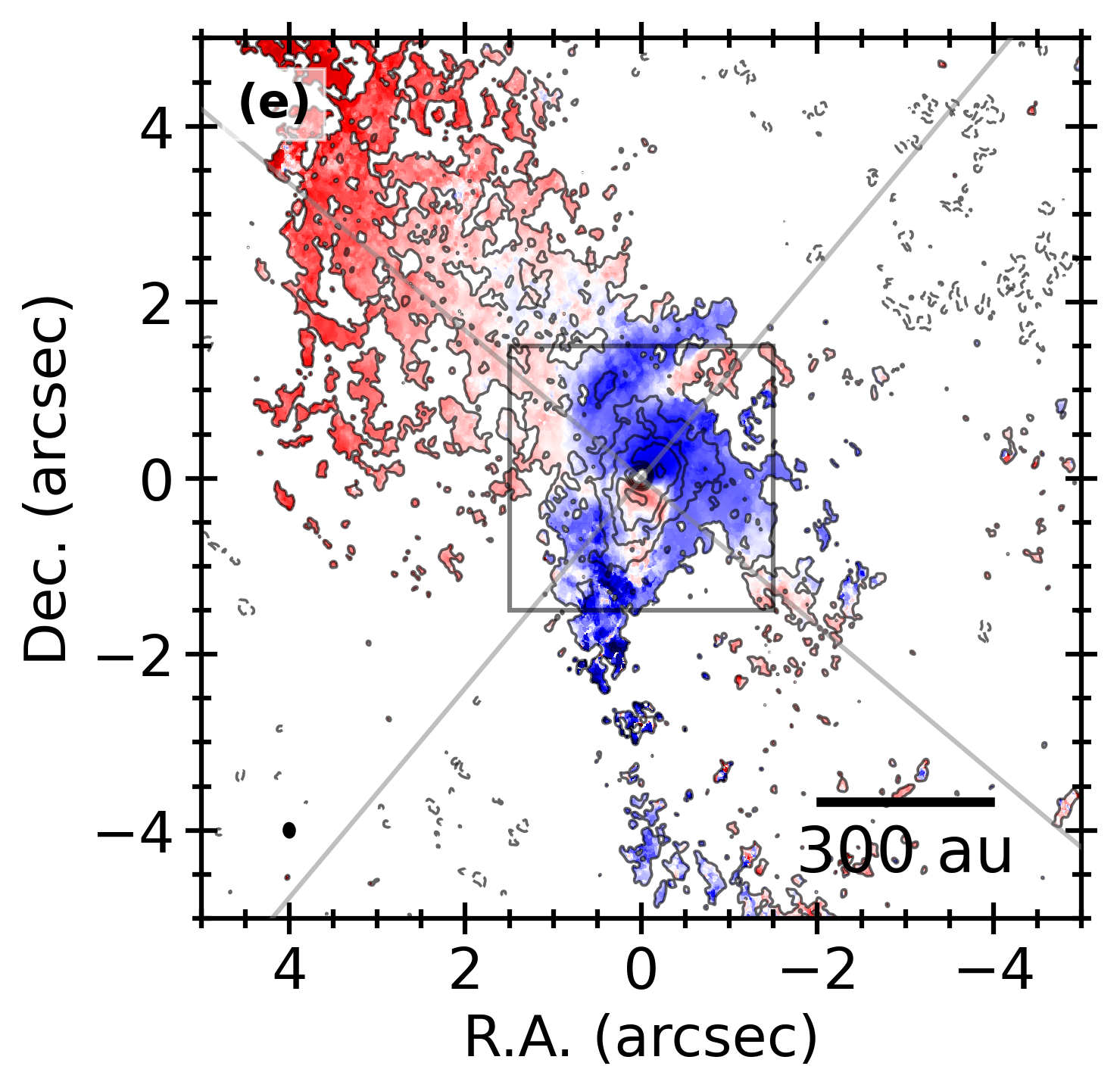}{0.24\textwidth}{}
        \hspace{-1.0cm}
        \leftfig{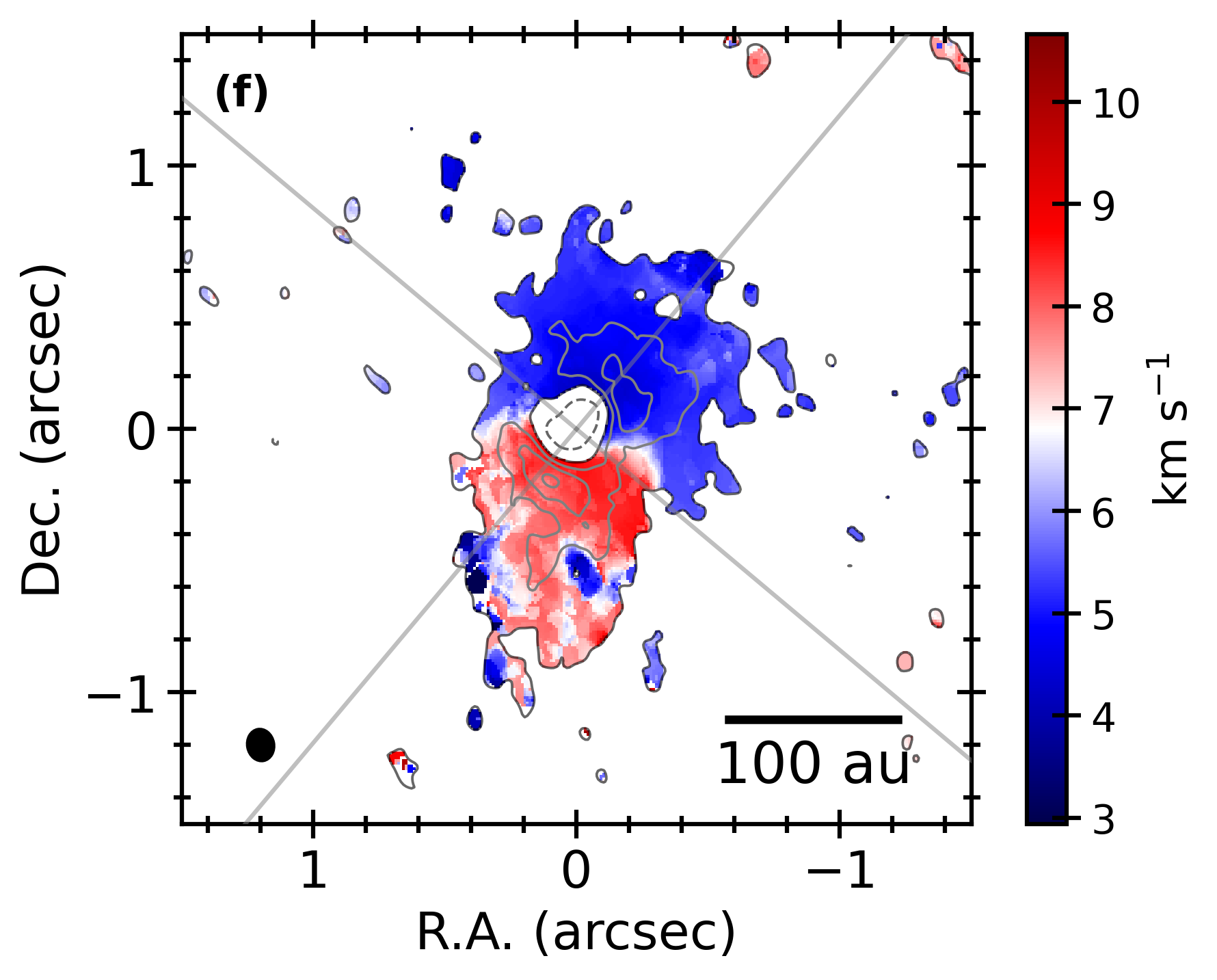}{0.2919\textwidth}{}
    }
    \vspace{-1.55cm}
    \gridline{
        \rightfig{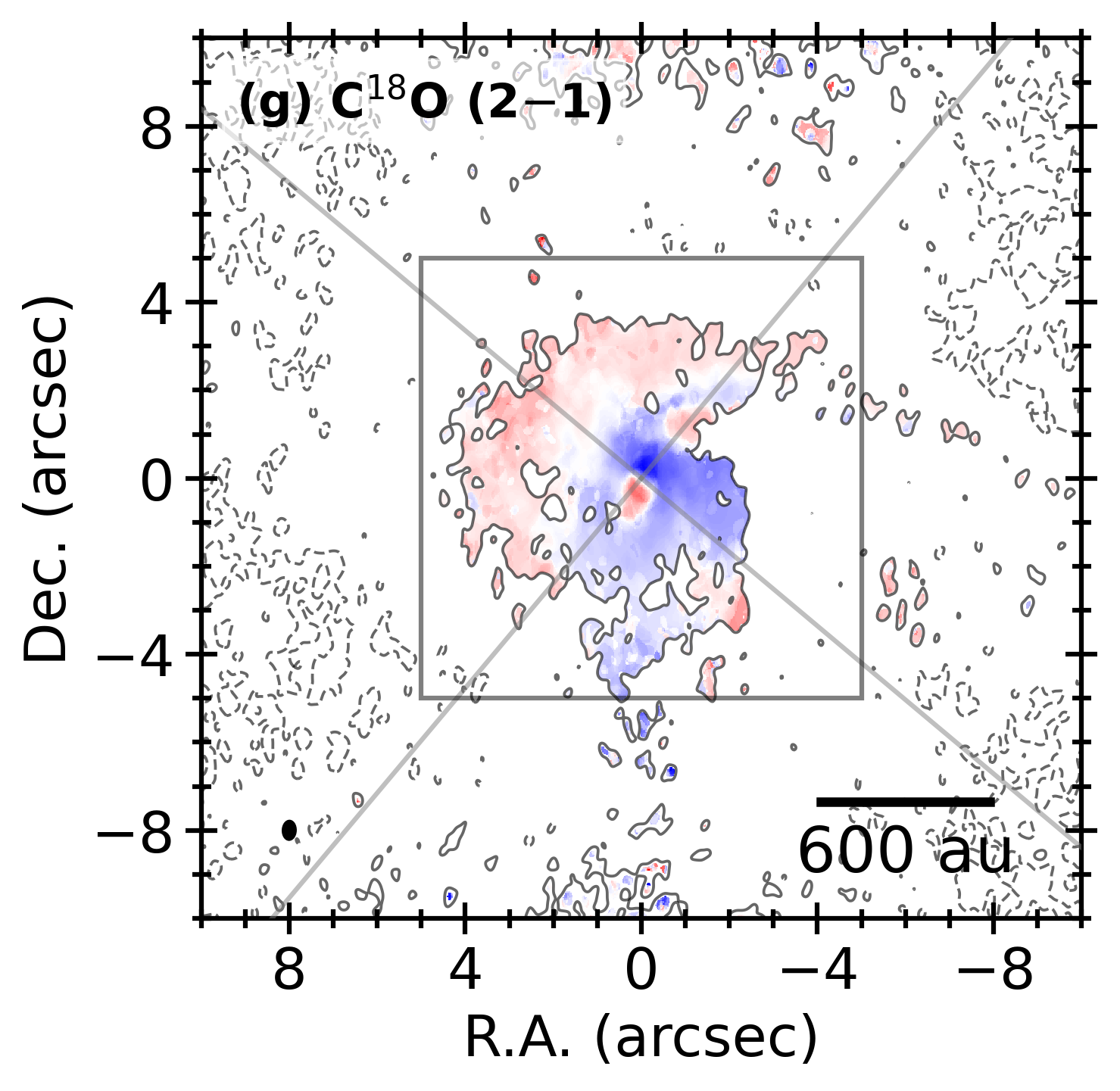}{0.24\textwidth}{}
        \hspace{-1.0cm}
        \fig{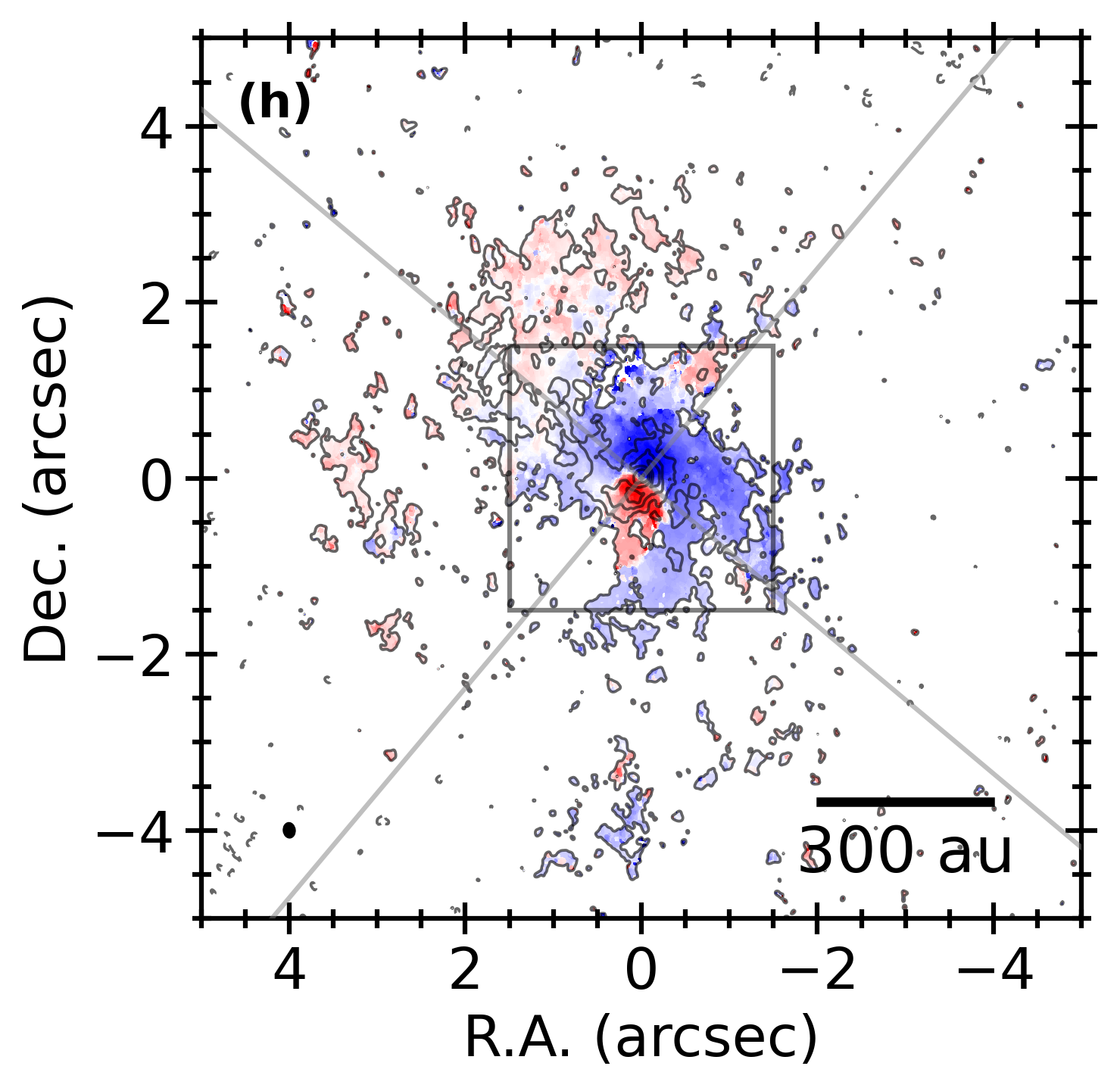}{0.24\textwidth}{}
        \hspace{-1.0cm}
        \leftfig{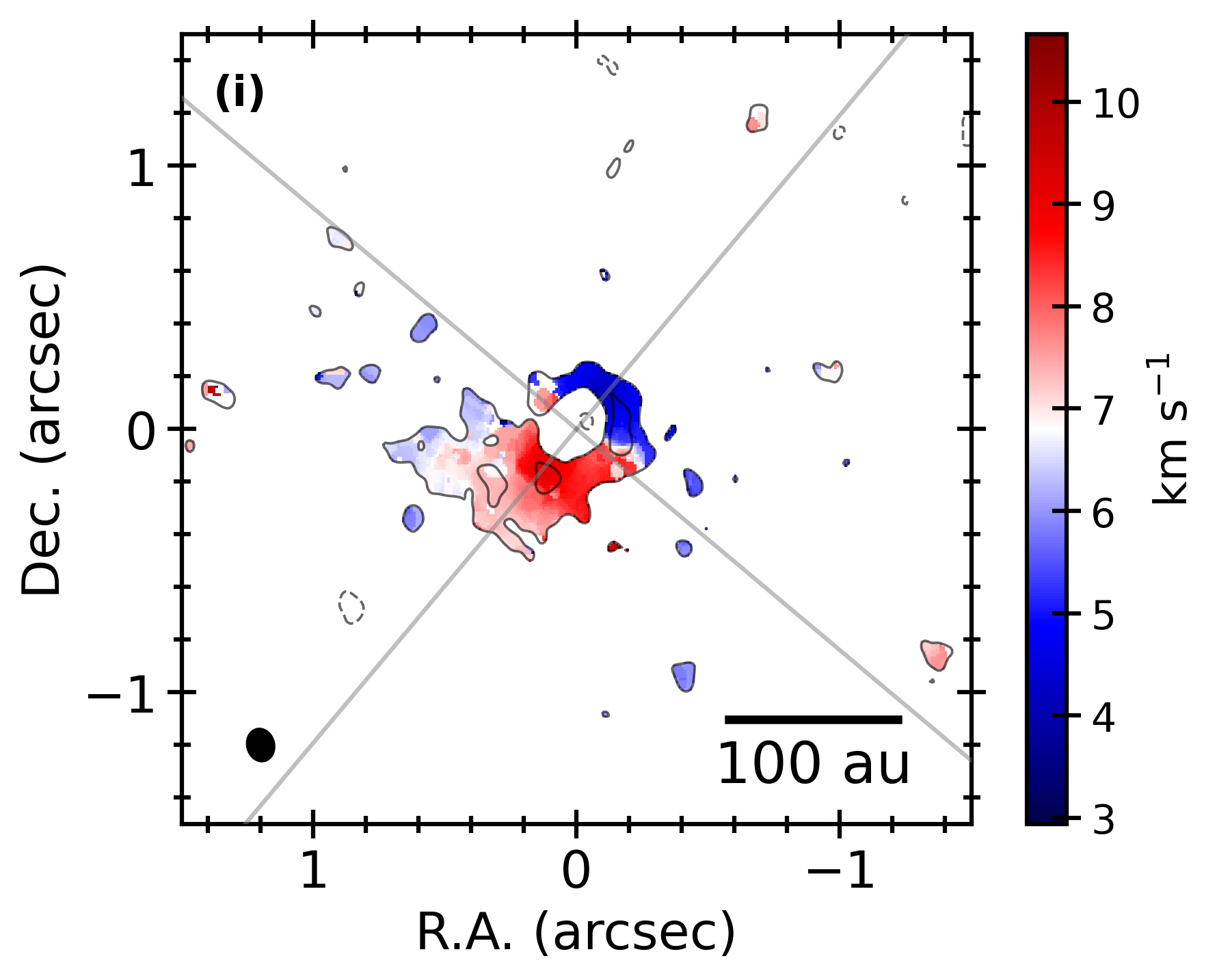}{0.2919\textwidth}{}
    }
    \vspace{-1.55cm}
    \gridline{
        \rightfig{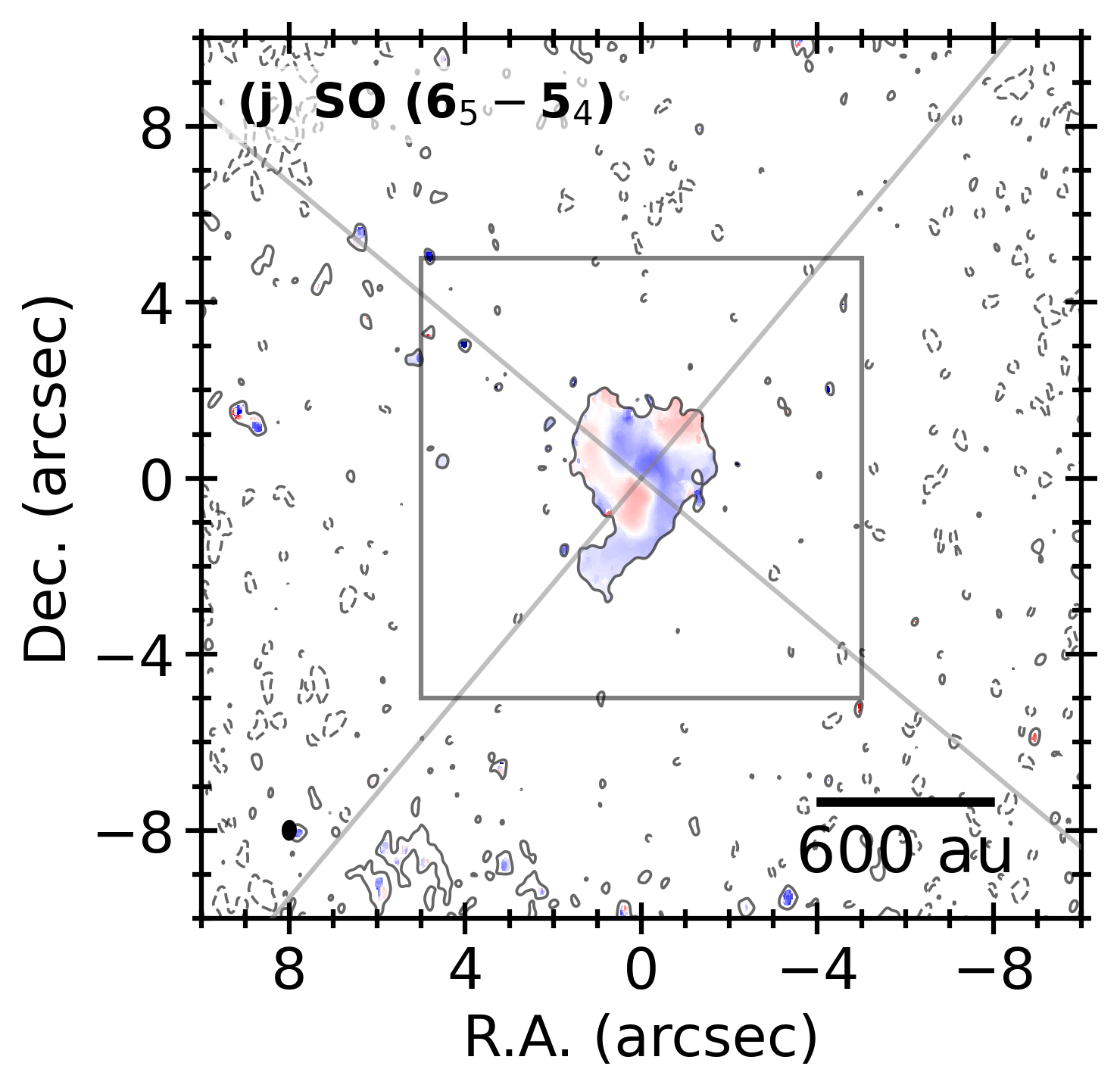}{0.24\textwidth}{}
        \hspace{-1.0cm}
        \fig{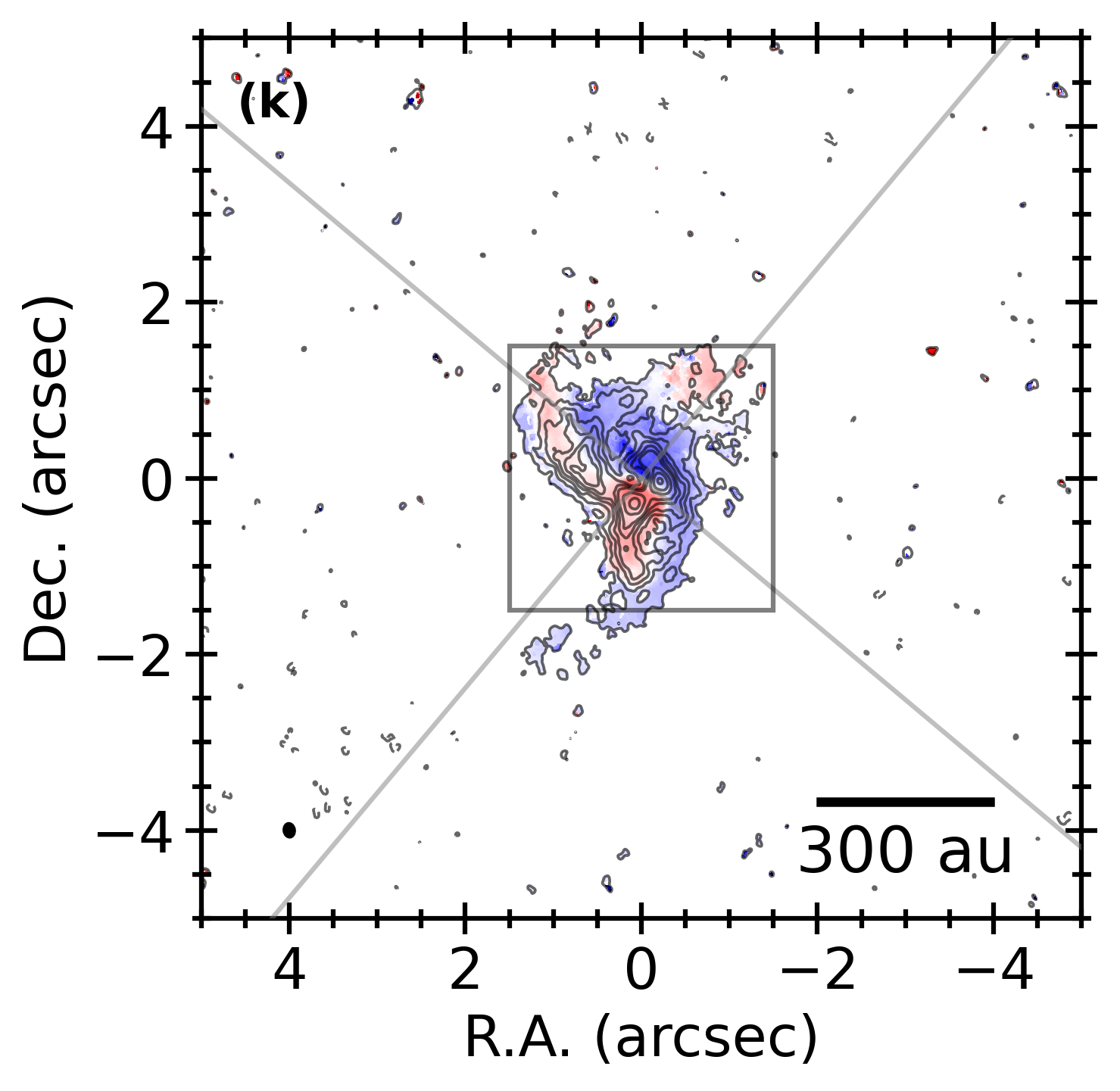}{0.24\textwidth}{}
        \hspace{-1.0cm}
        \leftfig{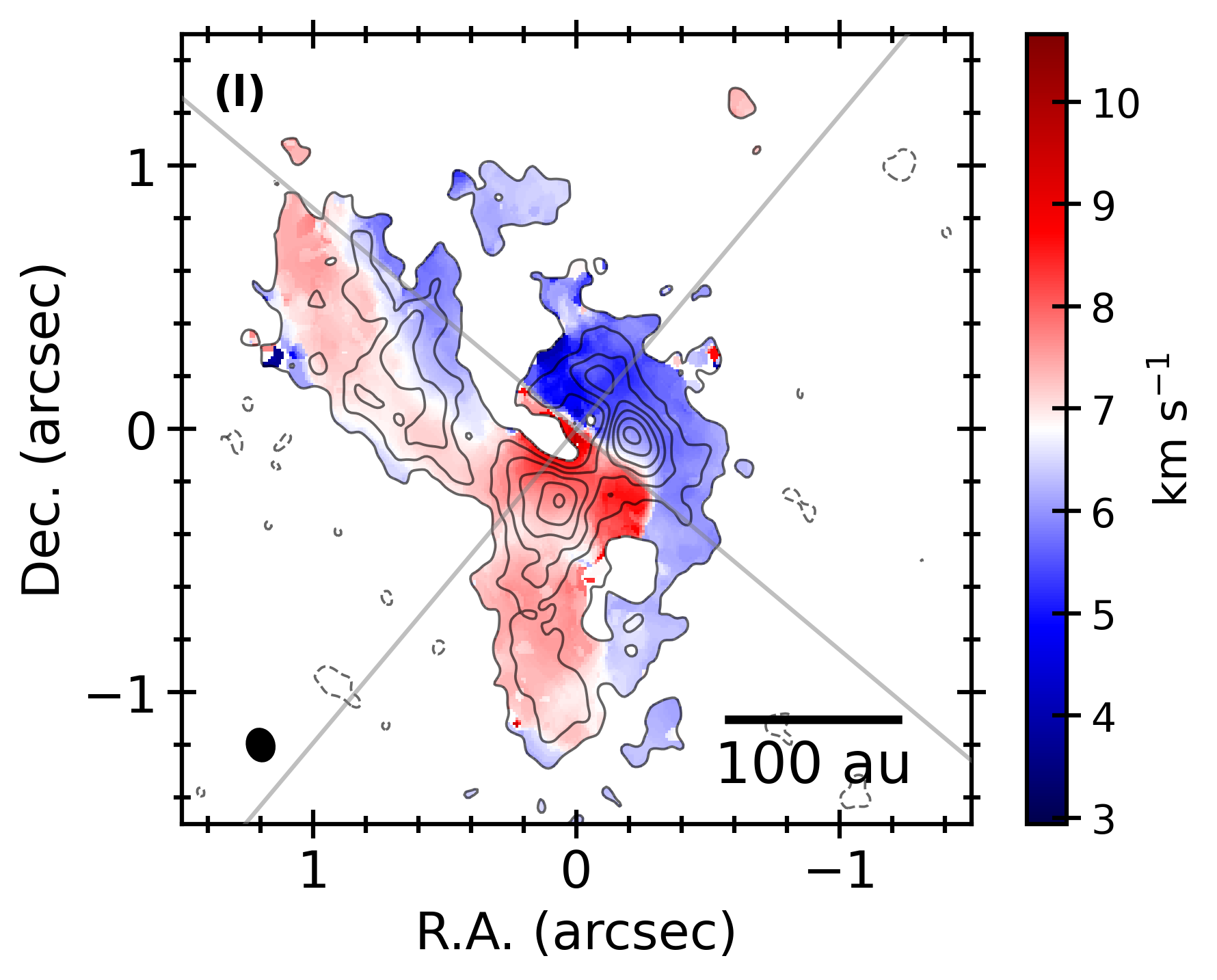}{0.2919\textwidth}{}
    }
    \vspace{-1.55cm}
    \gridline{
        \rightfig{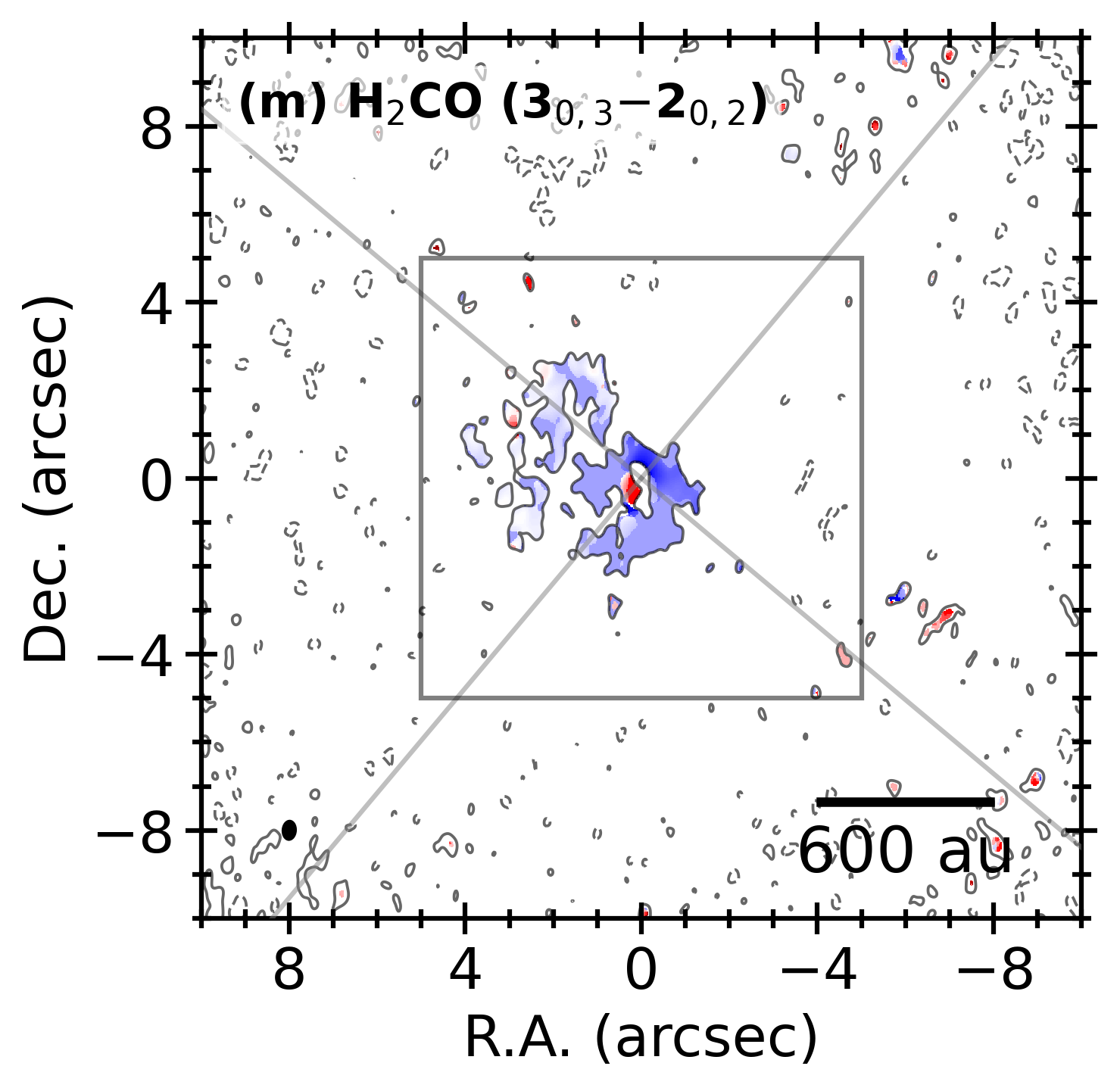}{0.24\textwidth}{}
        \hspace{-1.0cm}
        \fig{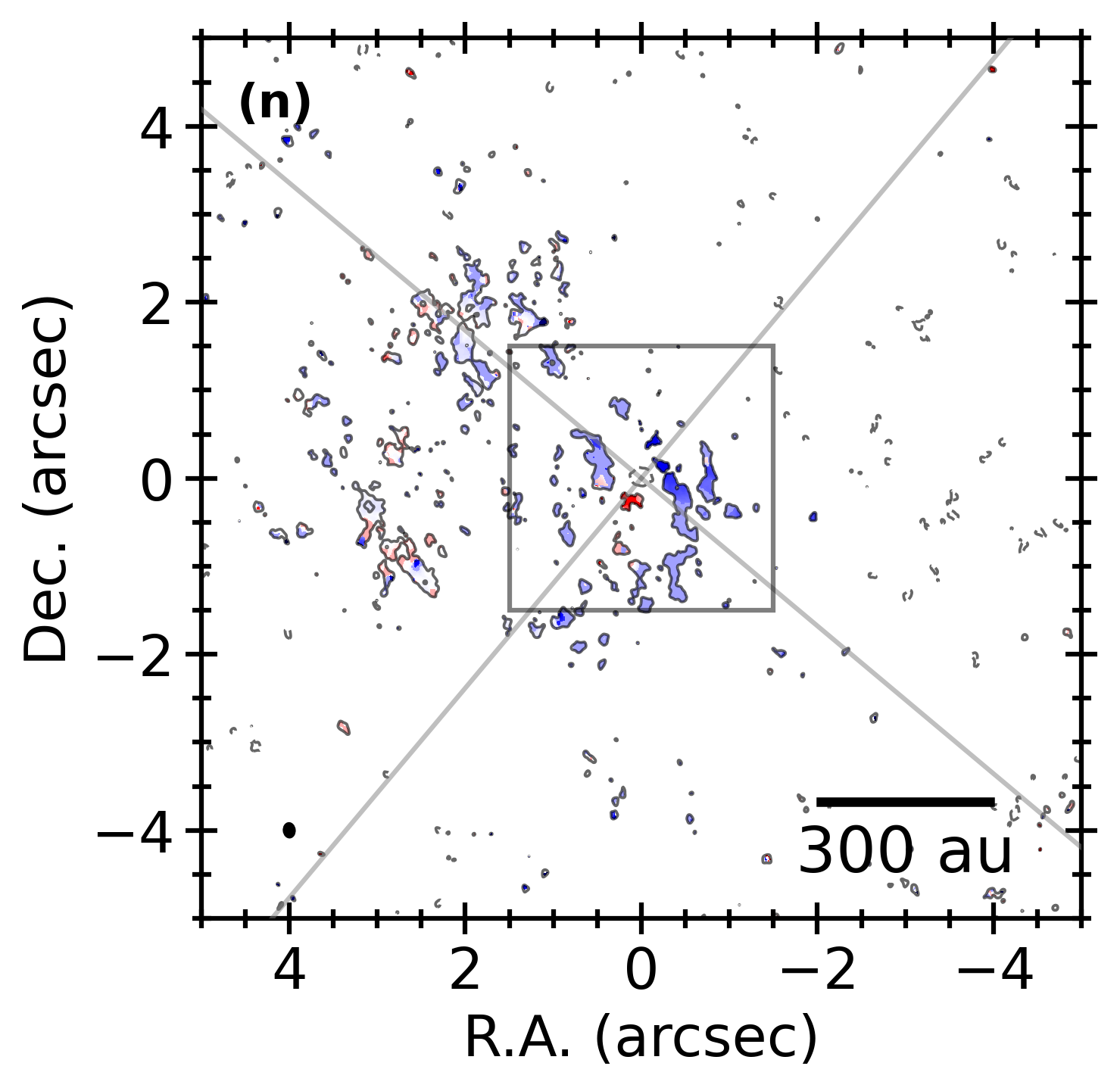}{0.24\textwidth}{}
        \hspace{-1.0cm}
        \leftfig{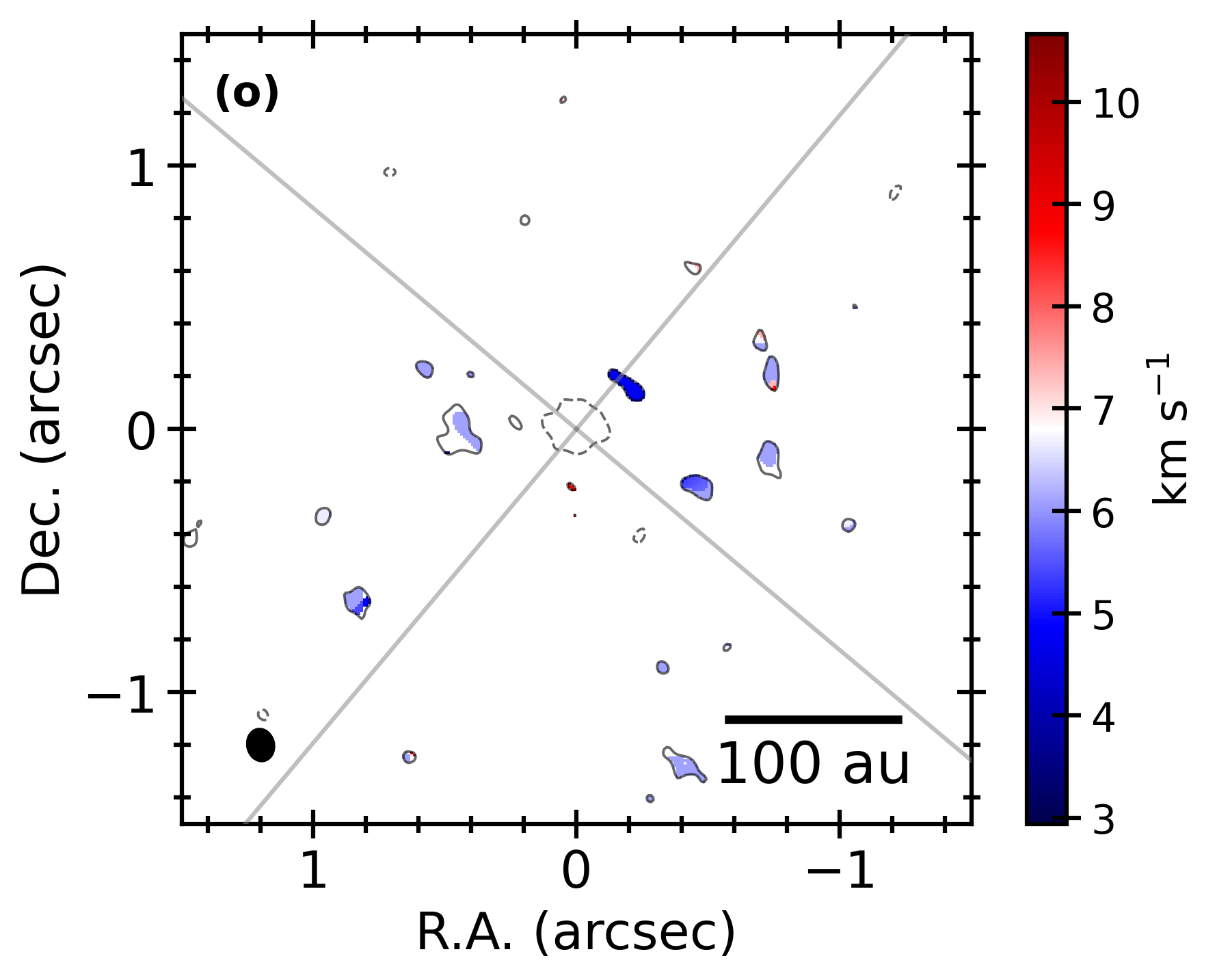}{0.2919\textwidth}{}
    }
    \vspace{-0.5cm}
    \caption{Overview of the moment 0/1 maps for the five molecular lines detected in I04169. The five rows (top to bottom) show the maps of $^{12}$CO (2--1), $^{13}$CO (2--1), C$^{18}$O (2$-$1), SO (6$_5$--5$_4$), and H$_2$CO (3$_{0, 3}$--2$_{0, 2}$). The left, middle, and right columns indicate the SB-only, SBLB, and LB-only data sets, respectively. To compare the spatial distributions of the five lines together, each column has an identical spatial scale: 20$\arcsec$ $\times$ 20$\arcsec$ (SB-only), 10$\arcsec$ $\times$ 10$\arcsec$ (SBLB), and 3$\arcsec$ $\times$ 3$\arcsec$ (LB-only). The velocity range of panels (a), (b), and (c) is from $-$2.64 to 16.24 km s$^{-1}$, and the other panels share the same velocity range as panel (d): 2.94 to 10.66 km s$^{-1}$. The solid contours start at 6$\sigma_{\rm M0}$ for the top three panels and at 3$\sigma_{\rm M0}$ for the remaining panels, where the 1$\sigma_{\rm M0}$ level for each panel is listed in Table \ref{tab:line}. The contour intervals are 6$\sigma_{\rm M0}$ in panel (b) and 3$\sigma_{\rm M0}$ in the other panels of the middle and right columns. The level of the dashed contours is $-$3$\sigma_{\rm M0}$ in all the panels. The dark grey boxes in the left and middle columns indicate the shown regions for the middle and right columns, respectively. The two light grey lines orthogonal to each other indicate the directions of the disk major (P.A. $=$ 140$\arcdeg$) and minor (P.A. $=$ 50$\arcdeg$) axes of the LB-only dust continuum image (Figure \ref{fig:cont}c). The black ellipses at the lower left denote the synthesized beams listed in Table \ref{tab:line}.}
    \label{fig:lines}
\end{figure*}

Figure \ref{fig:lines} shows the integrated intensity (moment 0) and mean velocity (moment 1) maps of I04169 in $^{12}$CO (2--1), $^{13}$CO (2--1), C$^{18}$O (2$-$1), SO (6$_5$--5$_4$), and H$_2$CO (3$_{0, 3}$--2$_{0, 2}$), made from the SB-only, SBLB, and LB-only visibility data individually.
The image parameters are summarized in Table \ref{tab:line}.
The moment 1 maps are obtained by integrating only the emission above the 3$\sigma_I$ level within the velocity range in which each line is detected.
Regarding the systemic velocity, we adopt 6.9 km s$^{-1}$ derived from our detailed position-velocity (PV) diagram analysis (see Section \ref{sec:kinematics}), which is comparable to the former measurement of 6.8 km s$^{-1}$ \citep[e.g.,][]{1997ApJ...488..317O, 2001ApJ...547..840S, 2018ApJ...865...51T}.

\subsubsection{$^{12}$CO (2--1)}
\label{sec:12co}

The SB-only and SBLB moment 0/1 maps of the $^{12}$CO (2--1) emission presented in Figures \ref{fig:lines}a and \ref{fig:lines}b trace a bipolar outflow, which is perpendicular to the disk major axis (P.A. $=$ 50$\arcdeg$).
The northeastern part of the outflow is redshifted, while the southwestern part is blueshifted.
Both parts exhibit U-shaped morphologies, but their detailed structures are different.
\revision{As shown in the SB-only map (Figure \ref{fig:lines}a), which covers a wider region,} the redshifted lobe is larger than the blueshifted one, and the cavity wall of this redshifted lobe appears asymmetric.
The northeastern wall starts at the protostellar position and extends over 9$\arcsec$ ($\sim$1,400 au).
On the other hand, the eastern wall does not start from the center but branches out the northeastern wall.
Furthermore, the emission on the northeastern side is stronger than on the eastern side, which is more clearly seen in the channel maps of the SBLB data (Figure \ref{fig:chan12cosblb} in Appendix).
In contrast to the redshifted lobe, the blueshifted one appears less extended, with most of the emission enhanced along the cavity wall.
The position angles of the symmetric axes of the red- and blue-shifted lobes are different from each other\revision{: roughly 75$\arcdeg$ and 240$\arcdeg$, respectively.}
\revision{The} redshifted
\revision{lobe}
is more misaligned from the disk minor axis \revision{(P.A. $=$ 50$\arcdeg$) than the blueshifted one}.
The overall morphology of the outflow appears consistent with the previous SMA $^{12}$CO (2--1) observations reported by \citet{2018ApJ...865...51T}.

The moment 0/1 maps of the LB-only data shown in Figure \ref{fig:lines}c reveal the spatial and velocity structures of the base of the outflows.
In addition to the outflow, another redshifted emission peak is found \revision{near} the center of the maps, which coincides with the continuum emission region.
More specifically, this redshifted feature is located in the southeastern part of the disk, and the blueshifted one is in the northwestern part.
This velocity gradient is more clearly seen in the channel maps of the LB-only data (Figure \ref{fig:chan12colb} in Appendix), which suggests that these components may trace a rotationally supported disk.
The kinematics of this possible disk will be discussed in Section \ref{sec:kinematics}.
In addition, the maps exhibit an intensity dip at the center (see also the channel maps).
This deficit is likely due to absorption of the continuum by optically thick emission near the source velocity, and continuum subtraction results in the negative.

\subsubsection{$^{13}$CO (2--1)}
\label{sec:13co}

Figures \ref{fig:lines}d and \ref{fig:lines}e present the moment 0/1 maps of the $^{13}$CO (2--1) emission made using the SB-only and SBLB data, respectively.
The emission is elongated along the disk minor axis, and a velocity gradient is clearly seen along this axis: the northeastern part is redshifted, and the southwestern part is blueshifted.
The velocity of the redshifted component increases from the center toward the northeast up to around 8$\arcsec$ ($\sim$1200 au), indicating that this emission likely traces part of the bipolar outflow.
The blueshifted lobe is concentrated in the center and also slightly extends toward the southwest and south.
These two extensions correspond to the blueshifted cavity wall of $^{12}$CO.
In addition, as exhibited in Figure \ref{fig:lines}d, another redshifted component appears to the southwest, which is also seen in the SMA image of the $^{13}$CO (2--1) emission \citep{2018ApJ...865...51T}.

The moment 0/1 maps of the LB-only data in Figure \ref{fig:lines}f show elongated emission along the major axis of the disk, with a clear velocity gradient from the blueshifted part in the northwest through the redshifted part in the southeast.
This velocity gradient, in contrast to that of the outflow, likely traces the rotational motion of the disk \revision{and/}or envelope.
Note that \citet{2018ApJ...865...51T} also reported a similar velocity gradient through the SMA $^{13}$CO (3--2) observations.
\revision{They interpreted this gradient as indicative of a Keplerian-rotating disk, although its radius and central mass were not well-constrained due to the limited angular resolution of their data (see also Section \ref{sec:kinematics} for further discussion).}
Additionally, there is a negative region in the center of the map at the location of the continuum emission, as in the case of the $^{12}$CO emission.

\subsubsection{C$^{18}$O (2--1)}
\label{sec:c18o}

Figures \ref{fig:lines}g and \ref{fig:lines}h show the moment 0/1 maps of the C$^{18}$O (2--1) emission with the SB-only and SBLB data, respectively.
This emission forms an S-shaped structure along the minor axis of the disk, which is significantly different from the shape of the $^{12}$CO and $^{13}$CO emission.
The northeastern arm is redshifted, while the southwestern arm is blueshifted.
The spiral arms making up this \revision{S-shaped} structure are more clearly seen in a few individual velocity channels near the systemic velocity of 6.9 km s$^{-1}$ (see Figures \ref{fig:chanc18osb} and \ref{fig:chanc18osblb} of Appendix).
Even though the trend of the velocity gradient is the same as that of the outflow detected in $^{12}$CO and $^{13}$CO, the velocity range of the gradient is much narrower, within $|V-V_{\rm sys}|$ $\lesssim$ 0.5 km s$^{-1}$.
Thus, the C$^{18}$O emission likely traces a distinct component rather than the outflow traced by the $^{12}$CO and $^{13}$CO emission.

The moment 0/1 maps of the LB-only data in Figure \ref{fig:lines}i show that the morphology of the emission is different from that shown in the maps of the SB-only and SBLB data.
The emission is elongated along the disk major axis, similar to $^{13}$CO (Figure \ref{fig:lines}f), but its spatial distribution is more compact than $^{13}$CO, with a length of 1$\arcsec$ (156 au).
The northwestern part is blueshifted, while the southeastern part is redshifted.
These features indicate that C$^{18}$O likely traces the rotating envelope or disk, whose kinematics is discussed in Section \ref{sec:kinematics}.
As shown in $^{12}$CO and $^{13}$CO, an emission deficiency toward the central continuum peak is also seen in C$^{18}$O.

\subsubsection{SO (6$_5-$5$_4$) and H$_2$CO (3$_{0, 3}-$2$_{0, 2}$)}
\label{sec:so+h2co}

Figures \ref{fig:lines}j, \ref{fig:lines}k, and \ref{fig:lines}l present the moment 0/1 maps of the SO (6$_5-$5$_4$) emission obtained from the SB-only, SBLB, and LB-only data sets, respectively.
The emission is more concentrated, i.e., within a radius of 2$\arcsec$ (312 au) from the center, compared to that from the three CO isotopologues.
Like C$^{18}$O, this emission also exhibits an S-shaped structure along the disk minor axis, which is different from the shape of the $^{12}$CO and $^{13}$CO emission tracing the bipolar outflow.
The northeastern arm is redshifted, and the southwestern arm is blueshifted.
In addition, as shown in Figure \ref{fig:lines}k, there is another arm-like feature extending toward the northwest.
The detailed spiral structure can be seen in the channel maps of the SBLB data (\revision{the upper panel of} Figure \ref{fig:chansosblb} \revision{in} Appendix).

Figures \ref{fig:lines}m, \ref{fig:lines}n, and \ref{fig:lines}o show the moment 0/1 maps of the H$_2$CO 3$_{0, 3}-$2$_{0, 2}$ emission from the SB-only, SBLB, and LB-only data, respectively.
Note that the other two observed transitions, 3$_{2, 1}-$2$_{2, 0}$ and 3$_{2, 2}-$2$_{2, 1}$, which have higher upper level energy of 68 K compared to 21 K for 3$_{0, 3}-$2$_{0, 2}$, do not show any significant features.
The emission is seen only in the SB-only image at a decent S/N and is concentrated around the protostellar position.
It appears mostly blueshifted because of the lower velocity resolution of 1.34 km s$^{-1}$, but a redshifted component is seen at the center.
These two components comprise a velocity gradient along the major axis of the dusty disk, suggesting that this molecular emission may also trace the rotational motion of the disk or inner envelope.
Furthermore, the channel maps of the SBLB data show a spiral structure, which is clearly seen in a single velocity channel at 6.08 km s$^{-1}$ and marginally in the next channel at 7.42 km s$^{-1}$ (\revision{the lower panel of} Figure \ref{fig:chansosblb} \revision{in} Appendix).

\section{Discussion}

Here, we discuss the substructure and kinematics of the disk around I04169 in Sections \ref{sec:substructures} and \ref{sec:kinematics}, respectively.
The nature and possible origins of the spiral features revealed in C$^{18}$O, SO, and H$_2$CO will be discussed in Section \ref{sec:streamer}.

\label{sec:discussion}

\subsection{Disk Substructure}
\label{sec:substructures}

\begin{figure*}[t]
    \gridline{
        \leftfig{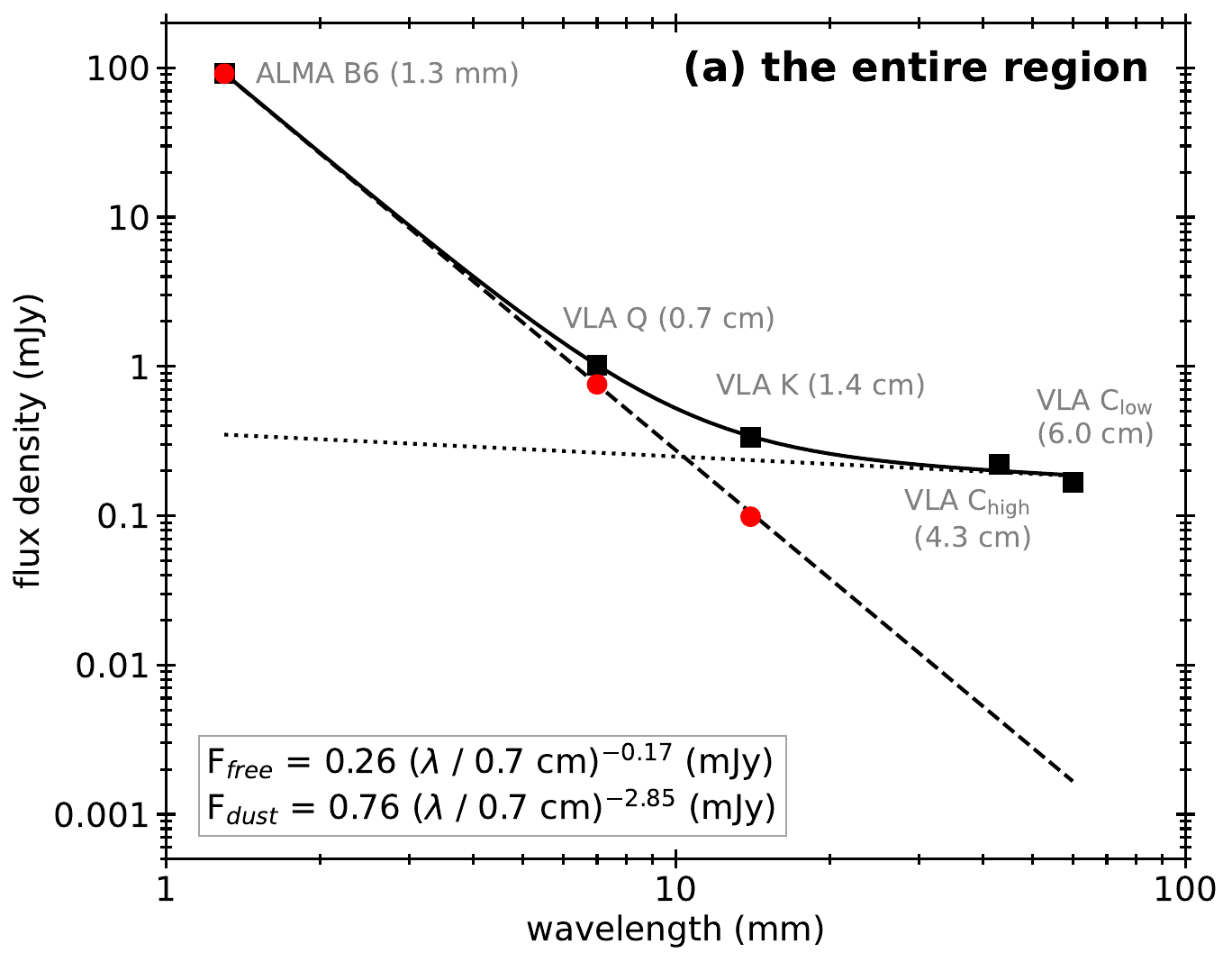}{0.45\textwidth}{}
        \rightfig{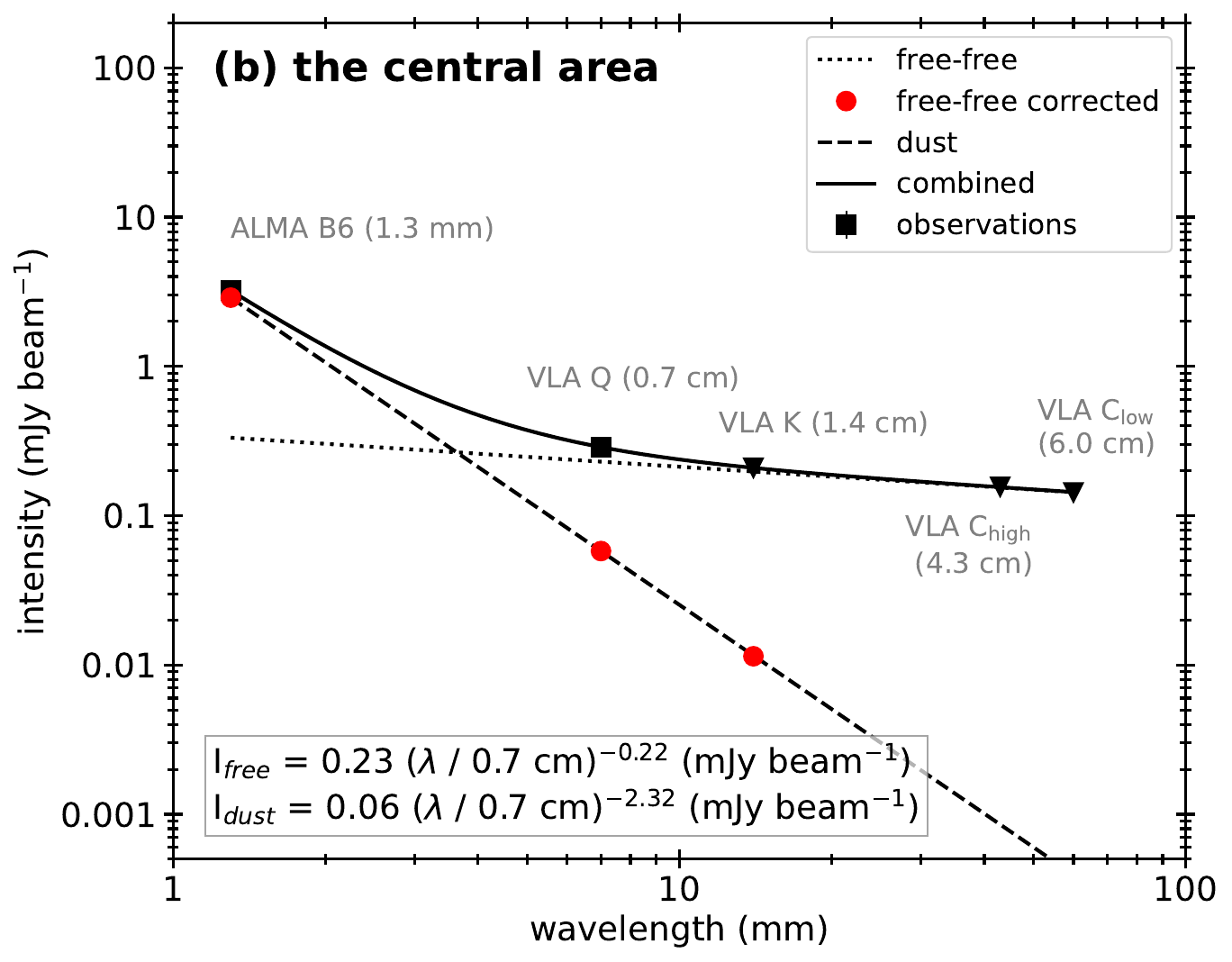}{0.45\textwidth}{}
    }
    \vspace{-0.8cm}
    \caption{Radio spectra of I04169, combining the following five observations: ALMA Band 6 (1.3 mm), VLA Q- (0.7 cm), K- (1.4 cm), C$_{\rm high}$- (4.3 cm), and C$_{\rm low}$- (6.0 cm) band data sets. \textbf{(a)} Radio spectrum over the entire emission region. The black squares indicate the flux densities measured in the 3$\sigma_I$ regions of the images. The spectra are modeled by two single power-law functions: the black dotted lines describe free-free emission, and the black dashed lines describe thermal dust emission. The thermal dust emission denoted by the red circles is obtained by subtracting the modeled free-free emission from the observations. The black solid lines are the sum of the two best-fit curves. \textbf{(b)} Radio spectrum of the central area. Each data point is likewise marked by a black symbol. The black squares represent the intensities at the single central pixels of the images where the proper motion has been corrected. The black reverse triangles indicate the upper limits of the intensities because the VLA K-, C$_{\rm high}$-, and C$_{\rm low}$-band images have coarser angular resolutions than the other two images; in particular, the central pixels of the C$_{\rm high}$- and C$_{\rm low}$-bands include nearly the whole emission of the disk.}
    \label{fig:sed}
\end{figure*}

We compare the ALMA Band 6 LB-only (1.3 mm) with the VLA Q- (7 mm), K- (1.4 cm), C$_{\rm high}$- (4.3 cm), and C$_{\rm low}$- (6.0 cm) bands.
To investigate the contribution of free-free emission in this system, we measure the flux densities of two regions: the entire and central regions of the disk.
The radio spectra of these two regions and their best-fit curves are presented in Figure \ref{fig:sed}.
We simultaneously fit each spectrum with two single power-law functions: one corresponds to free-free emission and the other thermal dust emission.
The equation used for this fitting is as follows:
\begin{equation}
    I_{\rm total} = I_{\rm dust} \left(\frac{\lambda}{7\ {\rm mm}}\right)^{\alpha_{\rm dust}} + I_{\rm free} \left(\frac{\lambda}{7\ {\rm mm}}\right)^{\alpha_{\rm free}},
\end{equation}
where $I_{\rm dust}$ is the intensity of thermal dust emission at the VLA Q band (7 mm), $\alpha_{\rm dust}$ is the spectral index of thermal dust emission, $I_{\rm free}$ is the intensity of free-free emission at the same wavelength, and $\alpha_{\rm free}$ is the spectral index of free-free emission.
Note that we only consider statistical uncertainty here.

First, Figure \ref{fig:sed}a shows the spectrum integrated over the entire disk.
The best-fit curve for thermal dust emission suggests that the flux density at 7 mm is $F_{\rm dust}$ $=$ 0.756 $\pm$ 0.051 mJy, and the spectral index is $\alpha_{\rm dust}$ $=$ 2.848 $\pm$ 0.040.
The flux density and spectral index of free-free emission are derived as $F_{\rm free}$ $=$ 0.264 $\pm$ 0.043 mJy and $\alpha_{\rm free}$ $=$ 0.166 $\pm$ 0.094, respectively.
This result indicates that thermal dust emission is more significant than free-free emission over the entire disk region at 7 mm as well as 1.3 mm.
Generally, the spectral index of thermal dust emission in the (sub)millimeter wavelength regime ($\alpha_{\rm dust}$) is closely linked to dust grain size \citep[e.g.,][]{2006ApJ...636.1114D}.
Indeed, the indices of Class II protoplanetary disks with millimeter/centimeter-sized large grains have been observed to be between 2.0 and 3.0 \citep[e.g.,][]{1991ApJ...381..250B, 2015ApJ...808..102K, 2021MNRAS.506.5117T}, which is smaller than 3.7 of the interstellar medium (ISM) consisting of (sub)micrometer-sized tiny grains \citep[e.g.,][]{2014A&A...566A..55P}.
This shallow slope of 2.848 may thus indicate that the size of the dust grains in I04169 is larger than that in the ISM\revision{, likely reaching up to millimeters or even centimeters \citep[e.g.,][]{2006ApJ...636.1114D, 2014prpl.conf..339T}}.
In addition to this source, rapid grain growth to millimeter/centimeter sizes has also been reported recently in several other protostellar disks \citep[e.g.,][]{2020ApJ...889...20L, 2021A&A...646A..18Z, 2023ApJ...956....9H, 2025ApJ...981..187R}.

Second, Figure \ref{fig:sed}b shows the spectrum at the central region.
We used flux densities of the central beams at the individual bands.
Note that the synthesized beams of the ALMA Band 6 and VLA Q-band images are comparable to each other, which are about 0$\farcs$05, while those of the VLA Q-, C$_{\rm high}$-, and C$_{\rm low}$-band images are much larger, resulting in their values being considered as upper limits.
The specific intensity of thermal dust emission at the 7 mm wavelength is derived as $I_{\rm dust}$ $=$ 0.057 $\pm$ 0.034 mJy beam$^{-1}$, and the thermal spectral index is derived as $\alpha_{\rm dust}$ $=$ 2.323 $\pm$ 0.332.
For free-free emission, the intensity and spectral index are derived as $I_{\rm free}$ $=$ 0.230 $\pm$ 0.028 mJy beam$^{-1}$ and
$\alpha_{\rm free}$ $=$ 0.220 $\pm$ 0.073, respectively.
The best-fit curves suggest that at this wavelength, free-free emission is more dominant than thermal dust emission in the central region, in contrast to the result for the whole disk region (Figure \ref{fig:sed}a).
This implies that the dust surface density of the central region is smaller than that of the outer region, which may explain the central depression observed at the 1.3-mm wavelength (Figure \ref{fig:cont}c).

A promising scenario for such a central depression is dynamical interaction with a stellar companion.
For example, \citet{1994ApJ...421..651A} demonstrated a close binary system could create a central cavity approximately three times the size of the binary semimajor axis.
Even though our data does not show any evidence for the binary with a separation of 0$\farcs$18 (28 au), as discussed in Section \ref{sec:cont}, we cannot entirely rule out a possibility of a closer binary, especially given the recent findings of highly compact Class 0/I multiples.
\citet{2022ApJ...928...61T} presented high-resolution VLA continuum observations of the Class 0 protostar L1157 MMS at the Q- and Ka- (9 mm) bands, revealing its small separation of 16 au.
High-resolution near-IR spectroscopic observations also indicated the presence of Class I binaries with very small separations, likely less than 10 au \citep[e.g.,][]{2012A&A...539A..62V, 2019AJ....158...41S}.
Hence, it would still be possible that our Q-band image does not resolve a tight binary system with a separation of less than 0$\farcs$05 (8 au), which is the synthesized beam size of the Q-band image.
In addition to a stellar companion, a giant planet can also carve such a central cavity, potentially appearing as a deficit at (sub)millimeter wavelengths \citep[e.g.,][]{2018ApJ...864L..26B}.

\subsection{Kinematics}
\label{sec:kinematics}

\begin{figure*}
    \gridline{
        \leftfig{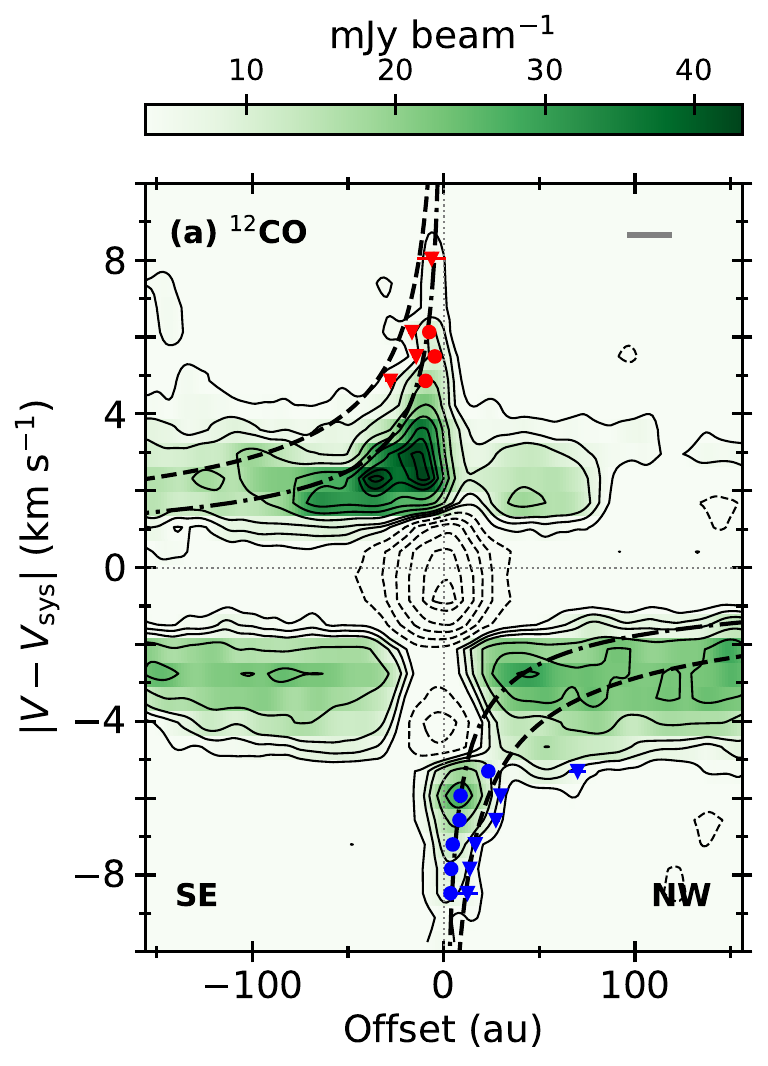}{0.30\textwidth}{}
        \fig{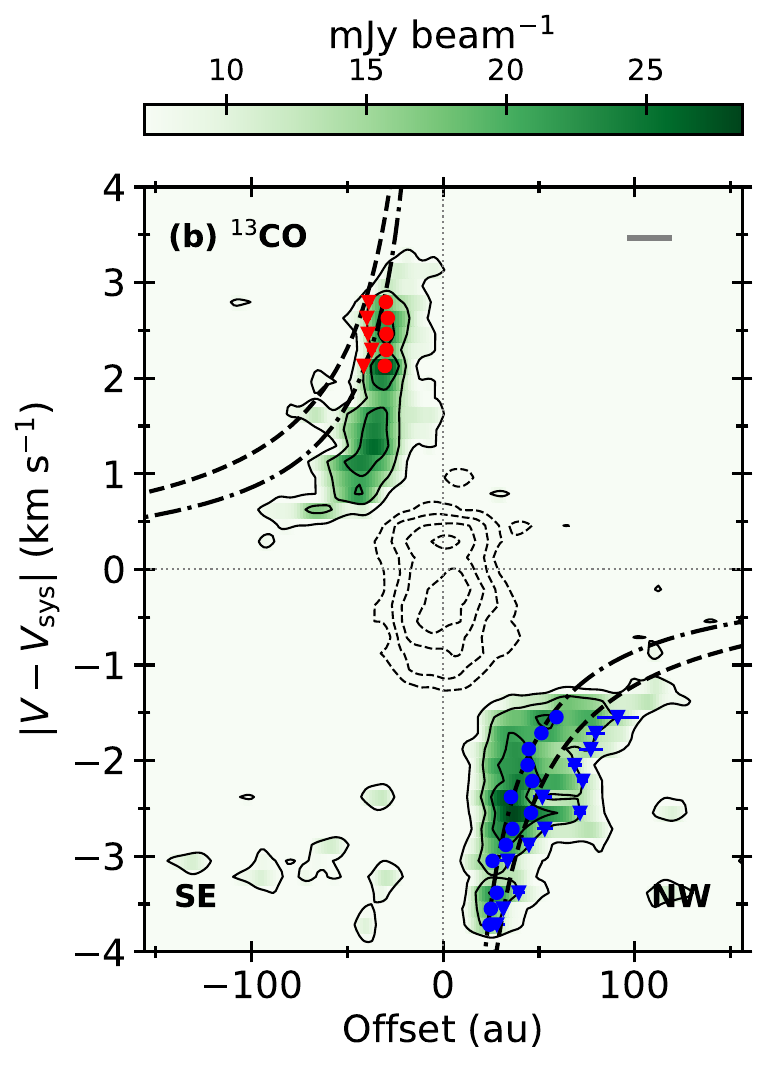}{0.30\textwidth}{}
        \rightfig{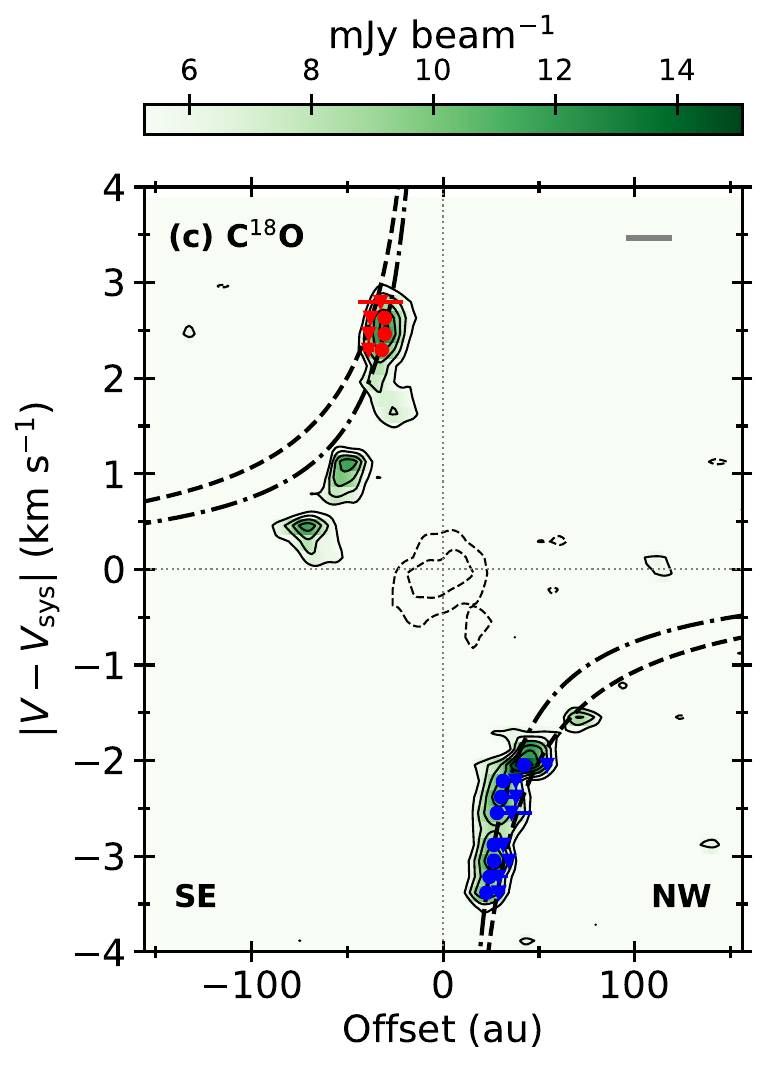}{0.30\textwidth}{}
    }
    \vspace{-0.5cm}
    \caption{Position-Velocity (PV) diagrams of the (a) $^{12}$CO (2--1), (b) $^{13}$CO (2--1), and (c) C$^{18}$O (2--1) emission along the disk major axis (P.A. $=$ 140$\arcdeg$) and their best-fit rotation curves. The LB-only data of the three emission are used to trace the embedded disk and the surrounding inner envelope, as shown in the moment 0/1 maps (Figure 2). Each panel is 2$\arcsec$ (312 au) wide: the left half represents the southeastern (SE) part along the disk major axis, while the right means the northwestern (NW). The grey horizontal bar at the upper right of each panel represents the major axis of the synthesized beam. The red and blue dots denote the red- and blue-shifted emission with respect to the identical systemic velocity of 6.9 km s$^{-1}$ for the three molecular lines. The triangles are the data points used for the edge method, while the circles are those for the ridge method. The dashed and dashed-dotted lines indicate the best-fit rotation curves of the edge and ridge methods, respectively. \textbf{(a)} PV diagram of $^{12}$CO. The velocity range is from $-$3.1 to 16.9 km s$^{-1}$, which is $\pm$10 km s$^{-1}$ with respect to the systemic velocity. The index $p$ of the two best-fit curves is fixed at 0.5, which shows that our data points follow the Keplerian rotation well. The contour levels are \{$-$30, $-$27, $-$21, $-$15, $-$9, $-$3, 3, 6, 9, 15, 21, 27, 33, 36, 39\} $\times$ $\sigma_{\rm ^{12}CO}$, where 1$\sigma_{\rm ^{12}CO}$ $=$ 1.07 mJy beam$^{-1}$. \textbf{(b)} PV diagram of $^{13}$CO. The velocity range is between $\pm$4 km s$^{-1}$ with respect to the systemic velocity: 2.9 to 10.9 km s$^{-1}$. The best-fit $p$ indices of the dashed-dotted (ridge) and dashed (edge) curves are 1.012 $\pm$ 0.058 and 0.933 $\pm$ 0.074, respectively. The contour levels are \{$-$12, $-$9, $-$6, $-$3, 3, 6, 9\} $\times$ $\sigma_{\rm ^{13}CO}$, where 1$\sigma_{\rm ^{13}CO}$ $=$ 2.35 mJy beam$^{-1}$. \textbf{(c)} PV diagram of C$^{18}$O. Its velocity range is the same as that of $^{13}$CO. The best-fit $p$ values of the ridge- and edge-method rotation curves are 1.008 $\pm$ 0.151 and 0.913 $\pm$ 0.129, respectively. The contour levels are \{$-$6, $-$3, 3, 4, 5, 6, 7, 8\} $\times$ $\sigma_{\rm C^{18}O}$, where 1$\sigma_{\rm C^{18}O}$ $=$ 1.75 mJy beam$^{-1}$. We note that the $^{13}$CO and C$^{18}$O emission traces the rotating inner envelope.}
    \label{fig:pvd}
\end{figure*}

\begin{figure*}[t]
    \gridline{
        \fig{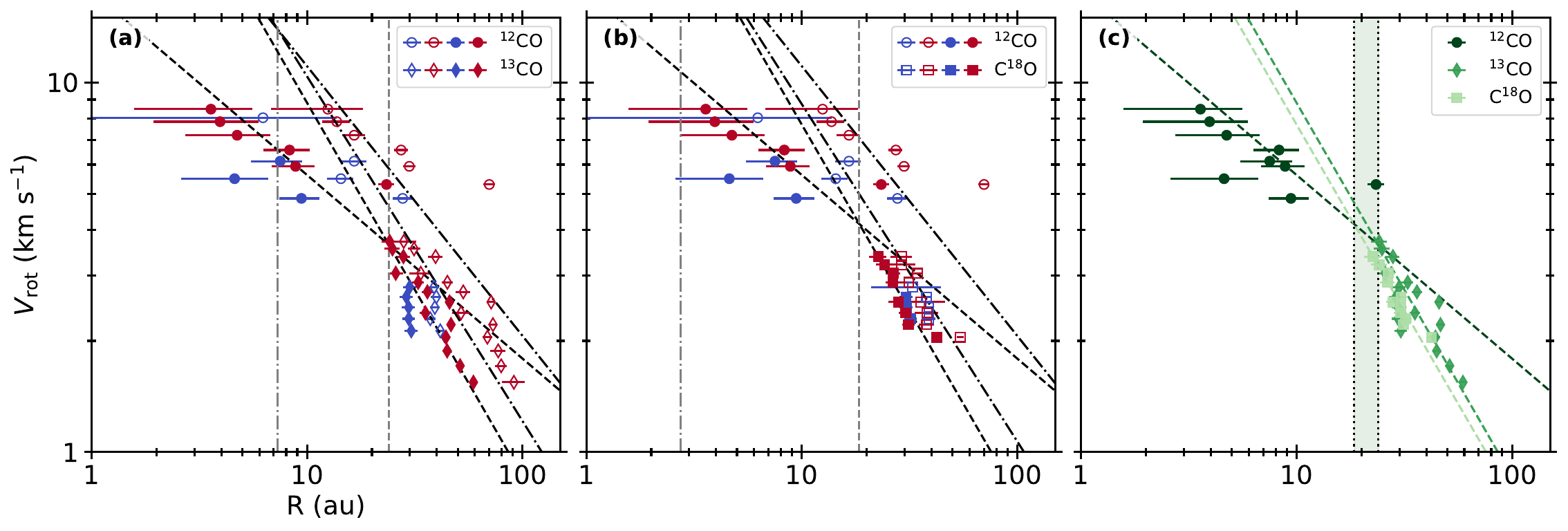}{0.95\textwidth}{}
    }
    \vspace{-0.5cm}
    \caption{Radial profiles of the rotation velocity ($V_{\rm rot}$) for the ridge and edge methods. The points are from the LB-only PV diagrams displayed in Figure \ref{fig:pvd}. \textbf{(a)} Profiles of $^{12}$CO and $^{13}$CO. The circle and diamond points indicate the $^{12}$CO and $^{13}$CO emission, respectively. The blue points represent the blueshifted components, while the red ones are the redshifted. The filled points mark the data points used in the ridge method, and the best-fit rotation curves of this method are shown by the black dashed lines. On the other hand, the open points mark the data used in the edge method, and its best-fit rotation curves are displayed by the black dashed-dotted lines. The grey vertical lines denote the cross points between the best-fit curves of $^{12}$CO and those of $^{13}$CO, which correspond to the Keplerian disk radius; likewise, the dashed-dotted line is derived from the edge method, and the dashed one is from the ridge method. \textbf{(b)} Profiles of $^{12}$CO and C$^{18}$O. The data points and best-fit curves of the $^{12}$CO emission are the same as those in the left panel, and the C$^{18}$O emission is marked by the square points. The grey vertical lines indicate the points where the best-fit curves of $^{12}$CO and C$^{18}$O cross. \textbf{(c)} Profiles of $^{12}$CO, $^{13}$CO, and C$^{18}$O, but using only the ridge method results. Note that the green-shaded region represents the finally-constrained Keplerian disk radius.}
    \label{fig:rotationprofile}
\end{figure*}

We performed a position-velocity (PV) diagram analysis to identify Keplerian rotation around I04169.
Our high-resolution observations enable us to unveil the innermost components surrounding the central protostar, particularly from the LB-only data of the three CO isotopologues.
As shown in the moment 0/1 maps of the LB-only data (the right column of Figure \ref{fig:lines}), the velocity gradients are clearly seen along the disk major axis, which appear to be consistent with that of the Keplerian disk suggested by \citet{2018ApJ...865...51T}.

We aim to estimate the dynamical mass of the central protostar and constrain the Keplerian radius of the embedded disk.
We employ the public Python package, Spectral Line Analysis/Modeling (SLAM)\footnote{\url{https://github.com/jinshisai/SLAM}} \citep{2024PKAS...39...2A}, which was also used for other eDisk targets.
The fitting is performed using two independent methods: the ridge method and the edge method.
The ridge and edge radii are defined as the intensity-mean (or peak) and outermost positions, respectively, determined by a certain threshold at a given velocity.
Both methods have been used to estimate central protostellar masses: the ridge method has widely been employed in many previous studies \citep[e.g.,][]{2009A&A...507..861J, 2012Natur.492...83T, 2013A&A...560A.103M, 2013ApJ...772...22Y, 2014A&A...562A..77H, 2014ApJ...796..131O, 2015ApJ...812...27A, 2017ApJ...849...56A, 2020A&A...635A..15M, 2020ApJ...893...51S}; on the other hand, the edge method proposed by \citet{2016MNRAS.459.1892S} has recently applied to observational data \citep[e.g.,][]{2017A&A...603L...3A, 2018ApJ...860..119G, 2020A&A...635A..15M, 2021ApJ...907L..10R}.
The overall fitting procedure is further described in \citet{2023ApJ...951....8O} and \citet{2024PKAS...39...2A}.

Our fitting was performed using the LB-only data since they include the least non-disk components compared to the SB-only and SBLB data with complicated features, particularly in the low-velocity channels mainly tracing the outflow and spirals.
However, since the LB-only data are also still contaminated by part of the outflow and spirals, we only used the high-velocity ranges of the LB-only data, thereby minimizing possible contamination by these large-scale components.
Given the narrowly limited velocity range for the fitting, we adopted a single power-law rotation curve ($V_{\rm rot}$ $\propto$ $R^{-p}$) rather than double power-law curves.

Regarding the systematic velocity $V_{\rm sys}$, when we performed fitting with it as a free parameter in the C$^{18}$O emission, which is least contaminated by the outflow and spirals compared to the other two CO emission, the best-fit values were derived to be 6.830 $\pm$ 0.056 and 6.890 $\pm$ 0.059 km s$^{-1}$ for the ridge and edge methods, respectively.
Since the mean of these two velocities is 6.858 $\pm$ 0.041 km s$^{-1}$, we use 6.9 km s$^{-1}$ as the systemic velocity in this paper.
We adopt an inclination angle of 43$\arcdeg$, which is estimated from the major and minor axis ratio of the LB-only dust continuum image in Figure \ref{fig:cont}c.

Figure \ref{fig:pvd}a shows the PV diagram of the $^{12}$CO emission along the disk major axis.
There are low-velocity components with $|V-V_{\rm sys}|$ $\lesssim$ 4 km s$^{-1}$, extending in the northwest--southeast direction of this PV diagram (see also the LB-only channel maps in Figure \ref{fig:chan12colb} of Appendix).
They are the red- and blue-shifted outflow cavity walls, part of which are also shown in the LB-only moment 0/1 maps.
We thus constrain velocity ranges of 4.5 $<$ ($V-V_{\rm sys}$) $<$ 10 km s$^{-1}$ and $-$10 $<$ ($V-V_{\rm sys}$) $<$ $-$5 km s$^{-1}$ for the red- and blue-shifted emission, respectively, to avoid the emission from the outflow.
We then use the emission above the 5$\sigma_{\rm ^{12}CO}$ level for the fitting, where $\sigma_{\rm ^{12}CO}$ indicates the rms noise level of the LB-only channel maps in $^{12}$CO.
The best-fit $p$ values derived from the ridge and edge methods are 0.497 $\pm$ 0.138 and 0.730 $\pm$ 0.136, respectively, suggesting that the \revision{high-velocity components of} $^{12}$CO trace the Keplerian rotation of the disk.

Figure \ref{fig:pvd}b shows the LB-only PV diagram of the $^{13}$CO emission and its best-fit rotation curves.
We confirmed that the low-velocity components of the SB-only and SBLB data, particularly near the systemic velocity with $|V-V_{\rm sys}|$ $\lesssim$ 1.5 km s$^{-1}$, trace the complicated outflow cavity walls, which are also marginally seen in the LB-only data (see Figures \ref{fig:chan13cosblb} and \ref{fig:chan13colb} of Appendix).
Based on this, to minimize the influence of the outflow in the fitting, we limit the velocity ranges for the red- and blue-shifted components of the LB-only data to 2 $<$ ($V-V_{\rm sys}$) $<$ 4 km s$^{-1}$ and $-$4 $<$ ($V-V_{\rm sys}$) $<$ $-$1.5 km s$^{-1}$, respectively.
We fit rotation curves to the LB-only PV diagram in this velocity range, and only the emission above the 6$\sigma_{\rm ^{13}CO}$ level is used.
The index $p$ is derived as 1.012 $\pm$ 0.058 in the ridge method and 0.933 $\pm$ 0.074 in the edge method.
These results indicate that $^{13}$CO traces the inner envelope rotating under the conservation of angular momentum.

The LB-only PV diagram of the C$^{18}$O emission and the best-fit rotation curves are presented in Figure \ref{fig:pvd}c.
The low-velocity channels with $|V - V_{\rm sys}|$ $\lesssim$ 1.9 km s$^{-1}$ of the SB-only and SBLB data detect the spiral features, and part of them are also marginally revealed in the LB-only channel maps (see Figures \ref{fig:chanc18osb}, \ref{fig:chanc18osblb}, and \ref{fig:chanc18olb} in Appendix).
We thus constrain the velocity ranges of the red- and blue-shifted components in the LB-only data to 1.9 $<$ $|V - V_{\rm sys}|$ $<$ 5 km s$^{-1}$, and only the emission above 5$\sigma_{\rm C^{18}O}$ is used for the fitting.
The best-fit $p$ index is 1.008 $\pm$ 0.151 in the ridge method and 0.913 $\pm$ 0.129 in the edge method.
In addition to $^{13}$CO, these values also indicate that the C$^{18}$O emission traces the inner rotating envelope with a conserved angular momentum.

Next, in order to estimate the dynamical mass of the central protostar, we performed the fitting again to $^{12}$CO by assuming the rotation curve index $p$ of 0.5.
The best-fit curves derived from the ridge and edge methods are presented in Figure \ref{fig:pvd}a.
The dynamical mass is computed to be 0.766 $\pm$ 0.059 $M_{\odot}$ in the ridge method and 2.007 $\pm$ 0.071 $M_{\odot}$ in the edge method.
We note that the ridge method may underestimate a central protostellar mass \citep[e.g.,][]{2015ApJ...812...27A}, while the edge method may overestimate it \citep[e.g.,][]{2020A&A...635A..15M}.
For this reason, we adopt the final mass of the protostar I04169 as the mean of the values derived from the ridge and edge methods, which is 1.312 $\pm$ 0.044 $M_{\odot}$.
We also note that \citet{2018ApJ...865...51T} previously suggested 0.1 $M_{\odot}$ for the dynamical mass of this protostar, surrounded by a Keplerian disk with a radius of 200 au, based on PV diagram analysis using the $^{13}$CO (3--2) emission.
Their observations, however, did not spatially resolve the disk, which likely introduced a large uncertainty.

Our PV diagram analysis demonstrates that $^{12}$CO traces the Keplerian disk around the 1.3-$M_{\odot}$ protostar, while $^{13}$CO and C$^{18}$O trace the inner rotating envelope.
Figure \ref{fig:rotationprofile}a and \ref{fig:rotationprofile}b present the rotation velocity profiles of $^{12}$CO and $^{13}$CO and $^{12}$CO and C$^{18}$O, respectively.
Here, we define the Keplerian radius of the disk as a point where the two rotation profiles, $^{12}$CO and $^{13}$CO or C$^{18}$O, intersect.
In the ridge method, the Keplerian radius is estimated to be 24 au for $^{12}$CO and $^{13}$CO and 18 au for $^{12}$CO and C$^{18}$O.
On the other hand, the edge method estimates the Keplerian radius to be 7 and 3 au for the two pairs of $^{12}$CO and $^{13}$CO and $^{12}$CO and C$^{18}$O, respectively.
We should note, however, the fact that the radii derived from the edge method do not lie between the data points of $^{12}$CO tracing the disk and those of $^{13}$CO or C$^{18}$O tracing the envelope.
This does not make sense because a rotation velocity curve generally exhibits a break at a point where infalling motion is changed into Keplerian motion.
This inconsistency likely results from the angular resolution of $^{12}$CO, which is insufficient to spatially resolve the compact disk, thereby overestimating its radius.
For this reason, we only rely on the two pairs derived from the ridge method, $^{12}$CO and $^{13}$CO and $^{12}$CO and C$^{18}$O, to constrain the radius of the Keplerian disk, as shown in Figure \ref{fig:rotationprofile}c.
The Keplerian radius is expected to be somewhere between 18 and 24 au.
\revision{Notably,} this range is comparable to the dusty disk radius of 29 au.
\revision{This value is derived by applying the 2$\sigma$ definition, which was proposed in \citet{2020ApJ...890..130T}, to our LB-only dust continuum image (Figure \ref{fig:cont}c)}, as described in Section \ref{sec:alma}.

\subsection{Streamers}
\label{sec:streamer}

The S-shaped spiral structures revealed in C$^{18}$O, SO, and H$_2$CO consist of a redshifted arm extending northeast and curving northwest and a blueshifted arm extending southwest and curving south.
In the following two sections, we examine the morphology and kinematics of these spiral features and explore three possible scenarios for their origin.

\subsubsection{Nature of Streamers}
\label{sec:streamernature}

\begin{figure}[t]
    \gridline{
        \fig{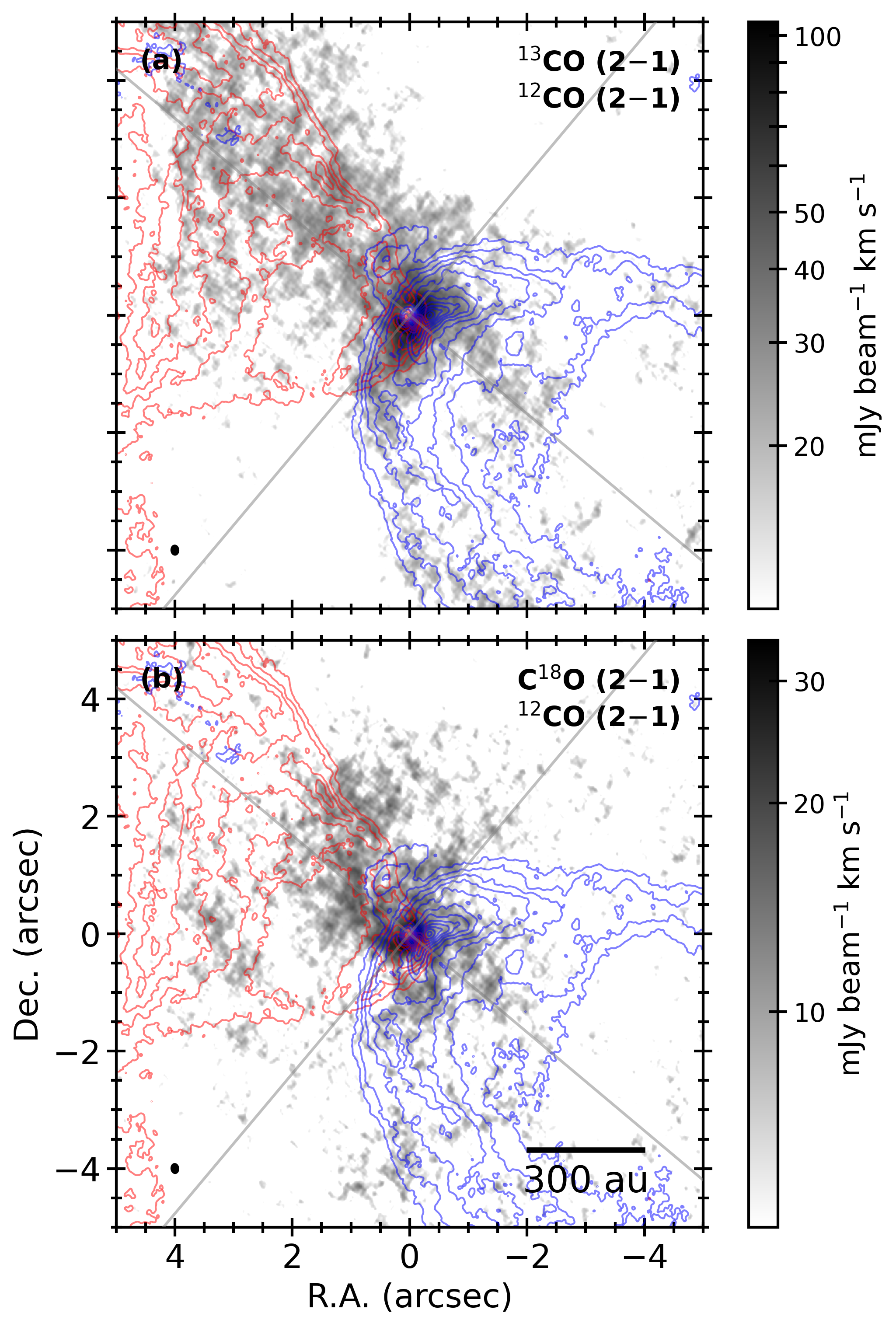}{0.4\textwidth}{}
    }
    \vspace{-0.5cm}
    \caption{Comparison of the spatial distributions between $^{12}$CO (2--1), $^{13}$CO (2--1), and C$^{18}$O (2--1) line emission of the SBLB data. In both panels, the blue and red contours indicate the integrated intensities (moment 0) of the $^{12}$CO blue- and red-shifted emission, respectively. The contour levels of the blueshifted emission are \{6, 12, 18, 24, ..., 66, 72\} $\times$ $\sigma_{\rm blue}$, where $\sigma_{\rm blue}$ $=$ 4.48 mJy beam$^{-1}$ km s$^{-1}$. The contour levels of the redshifted emission are
    \{6, 12, 18, 24, ..., 48, 54\}
    $\times$ $\sigma_{\rm red}$, where $\sigma_{\rm red}$ $=$ 3.42 mJy beam$^{-1}$ km s$^{-1}$.
    The grey maps in the background represent the moment 0 maps for $^{13}$CO in panel (a) and C$^{18}$O in panel (b), which are the same as shown in Figures \ref{fig:lines}e and \ref{fig:lines}h, respectively. The colorwedge range in both panels starts at the 2$\sigma_{M0}$ level. The black ellipses in the lower left denote the synthesized beams of these two moment 0 maps: 0$\farcs$189 $\times$ 0$\farcs$151 for $^{13}$CO and 0$\farcs$188 $\times$ 0$\farcs$149 for C$^{18}$O (see also Table \ref{tab:line}). The two grey lines denote the directions of the major (P.A. $=$ 140$\arcdeg$) and minor (P.A. $=$ 50$\arcdeg$) axes of the LB-only dust continuum image (Figure \ref{fig:cont}c).}
    \label{fig:mom0}
\end{figure}

\begin{figure*}[t]
    \gridline{
        \leftfig{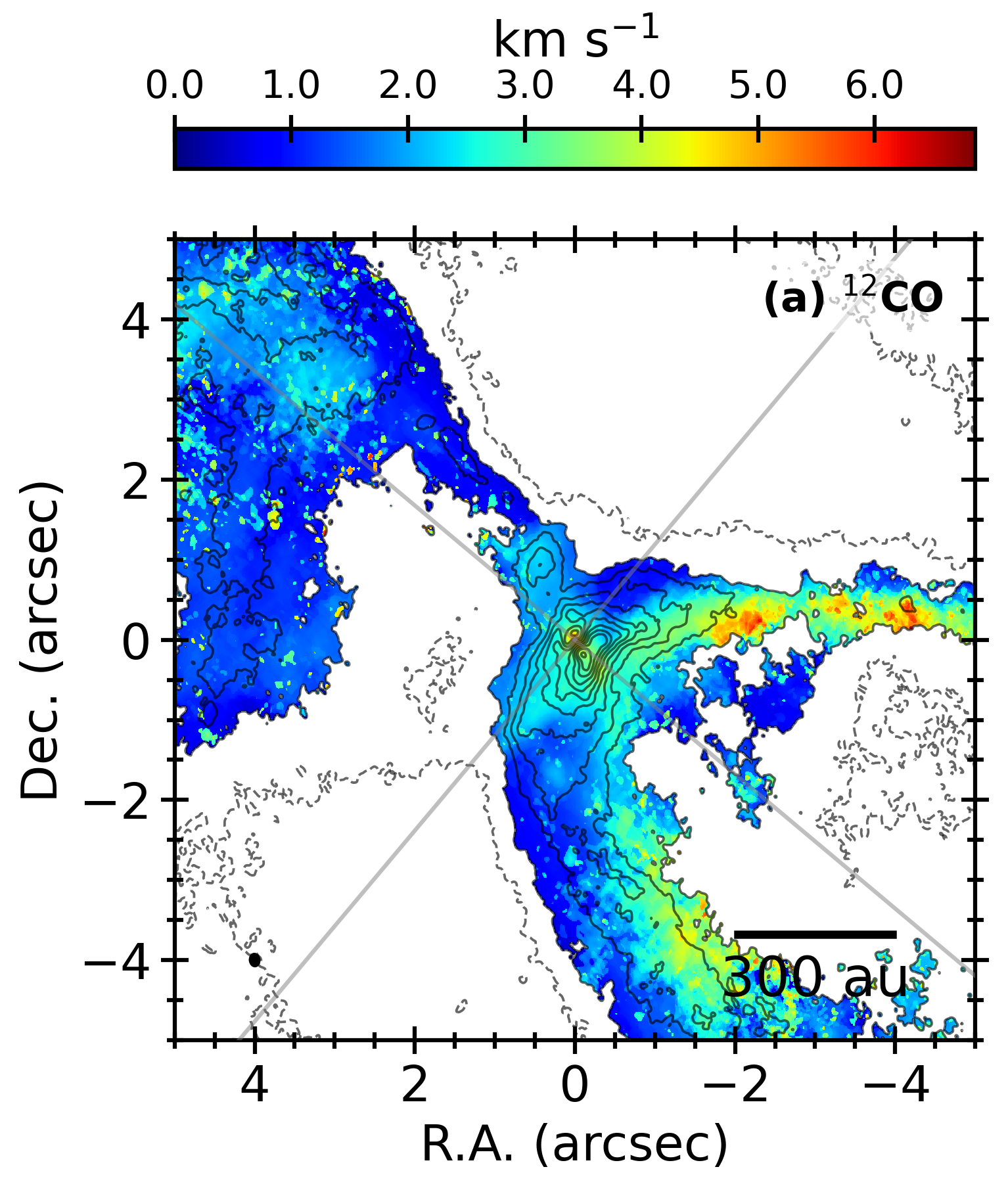}{0.3\textwidth}{}
        \fig{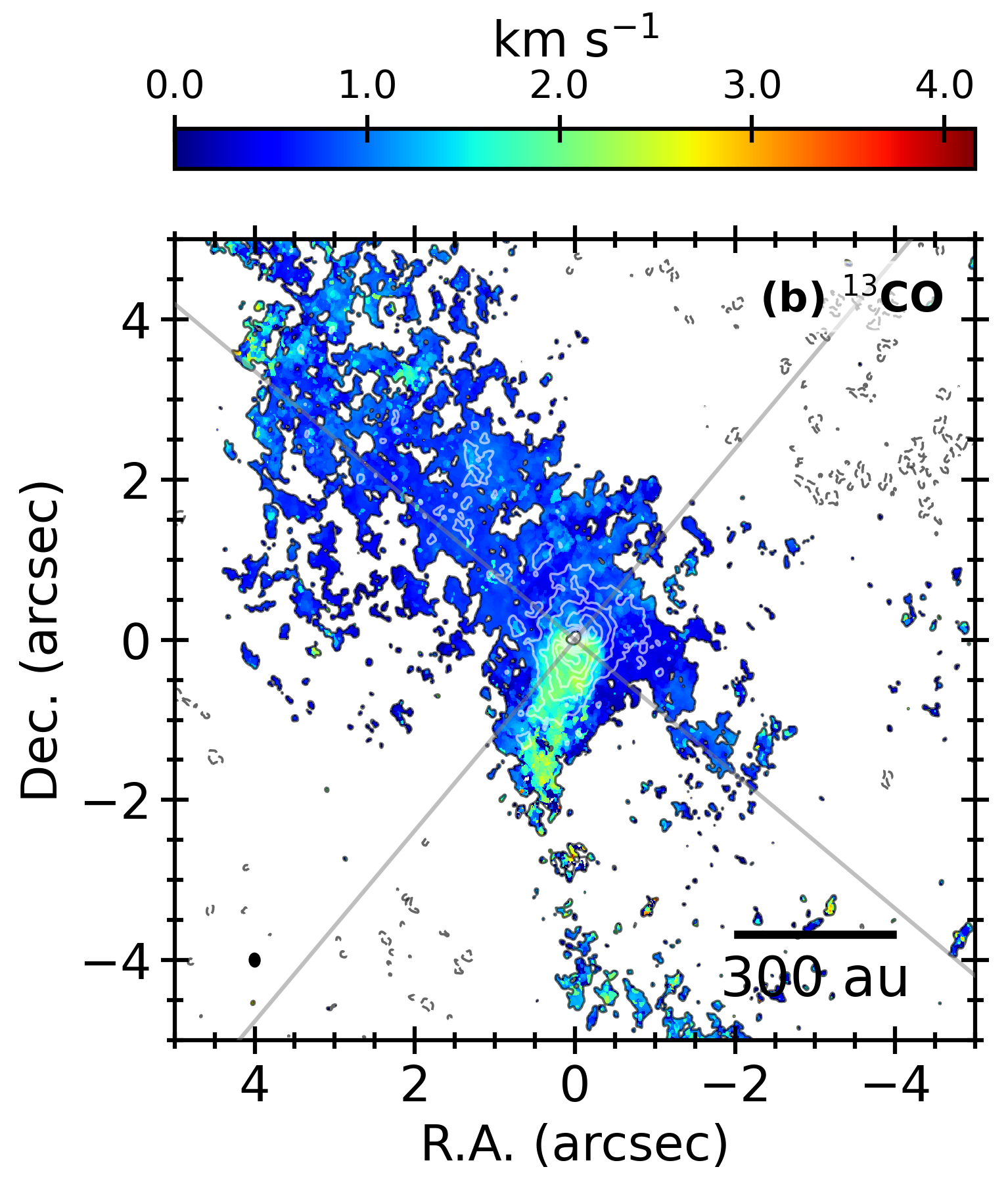}{0.3\textwidth}{}
        \rightfig{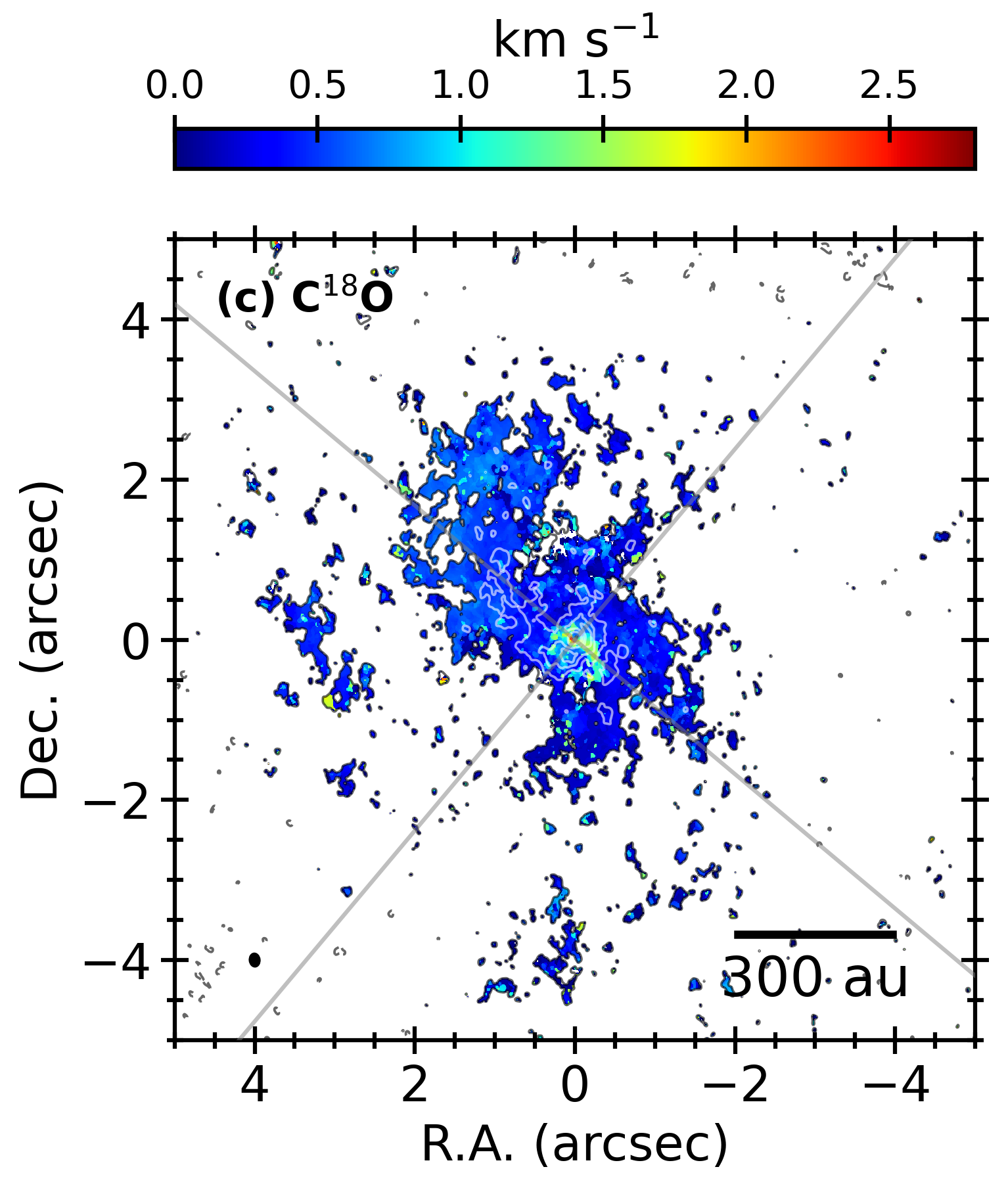}{0.3\textwidth}{}
    }
    \vspace{-0.5cm}
    \caption{Velocity dispersion (moment 2) maps for the $^{12}$CO (2--1) (from the left), $^{13}$CO (2--1), and C$^{18}$O (2--1) emission of the SBLB data. Each panel has an identical spatial scale of 10$\arcsec$ $\times$ 10$\arcsec$, allowing direct comparison of the velocity dispersions between the CO isotopologues. The contours represent the integrated intensities (moment 0) of the line emission, which are the same as the contours shown in the middle column of Figure \ref{fig:lines}. The two orthogonal grey lines indicate the directions of the disk major (P.A. $=$ 140$\arcdeg$) and minor (P.A. $=$ 50$\arcdeg$) axes of the LB-only dust continuum image (Figure \ref{fig:cont}c). The black ellipses in the lower left denote the synthesized beams of the emission: 0$\farcs$182 $\times$ 0$\farcs$152 in $^{12}$CO, 0$\farcs$189 $\times$ 0$\farcs$151 in $^{13}$CO, and 0$\farcs$188 $\times$ 0$\farcs$149 in C$^{18}$O (see also Table \ref{tab:line}).}
     \label{fig:mom2}
\end{figure*}

\begin{figure*}[t]
    \gridline{
        \leftfig{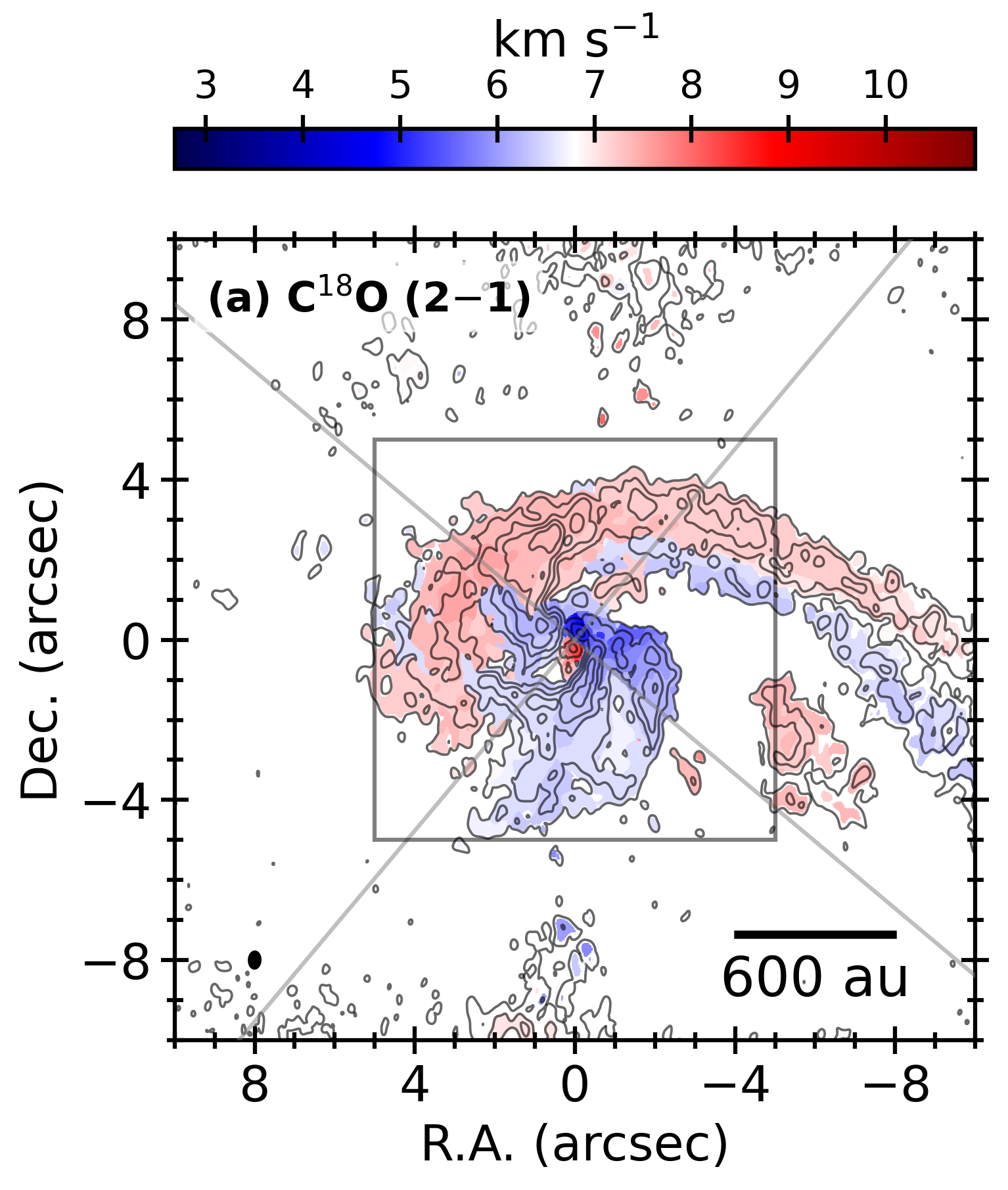}{0.3\textwidth}{}
        \fig{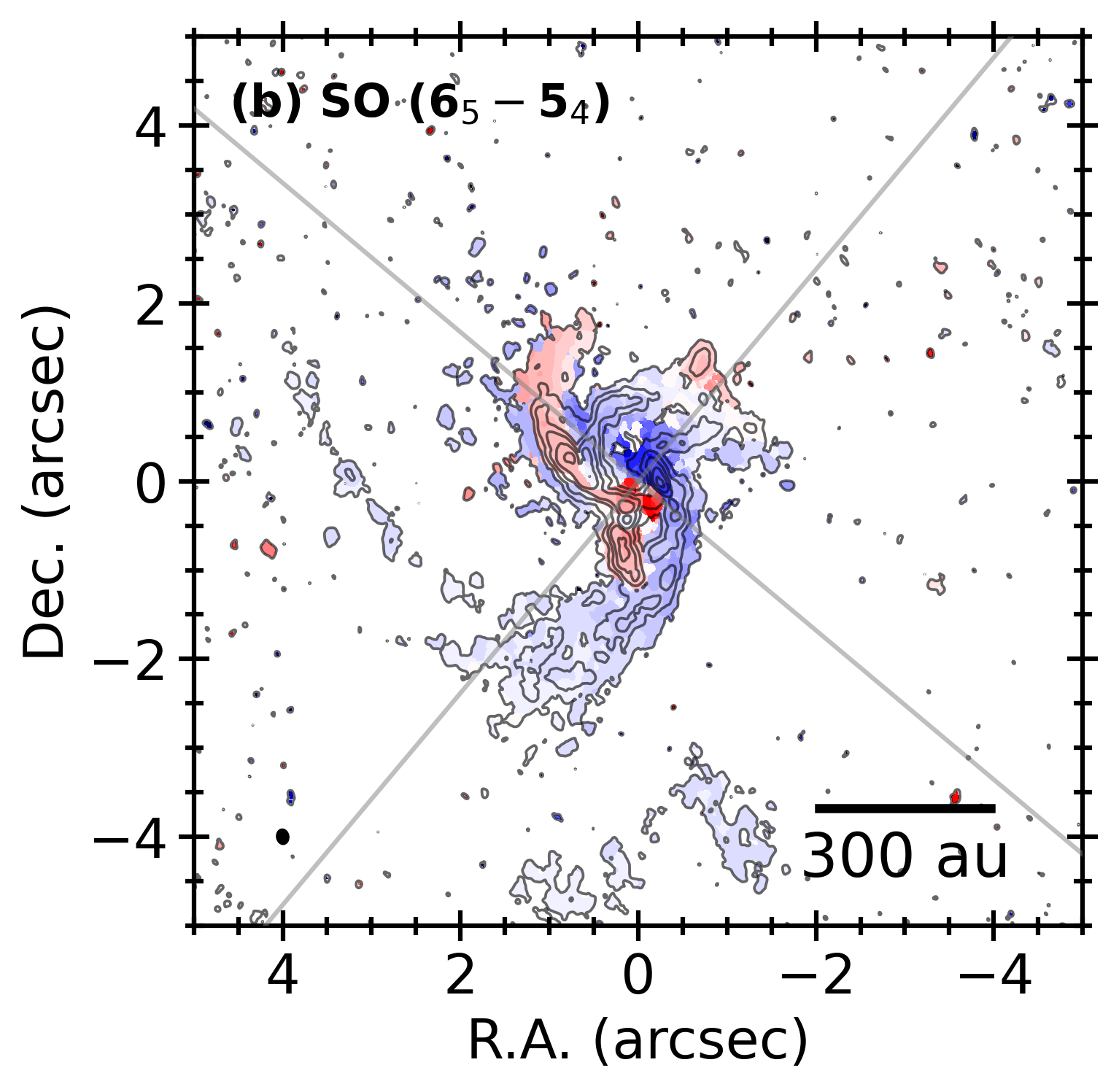}{0.3\textwidth}{}
        \rightfig{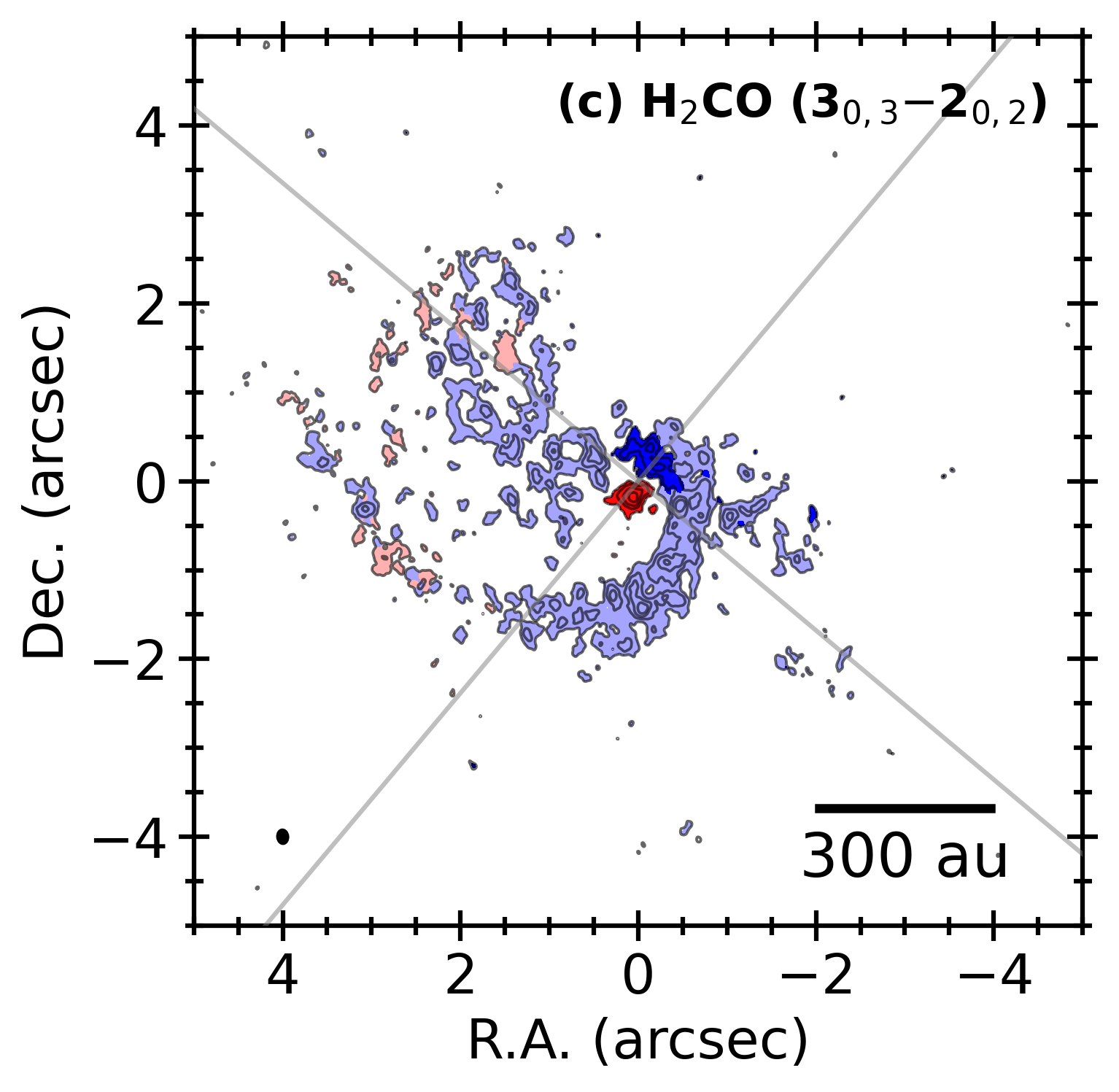}{0.3\textwidth}{}
    }
    \vspace{-0.5cm}
    \gridline{
        \leftfig{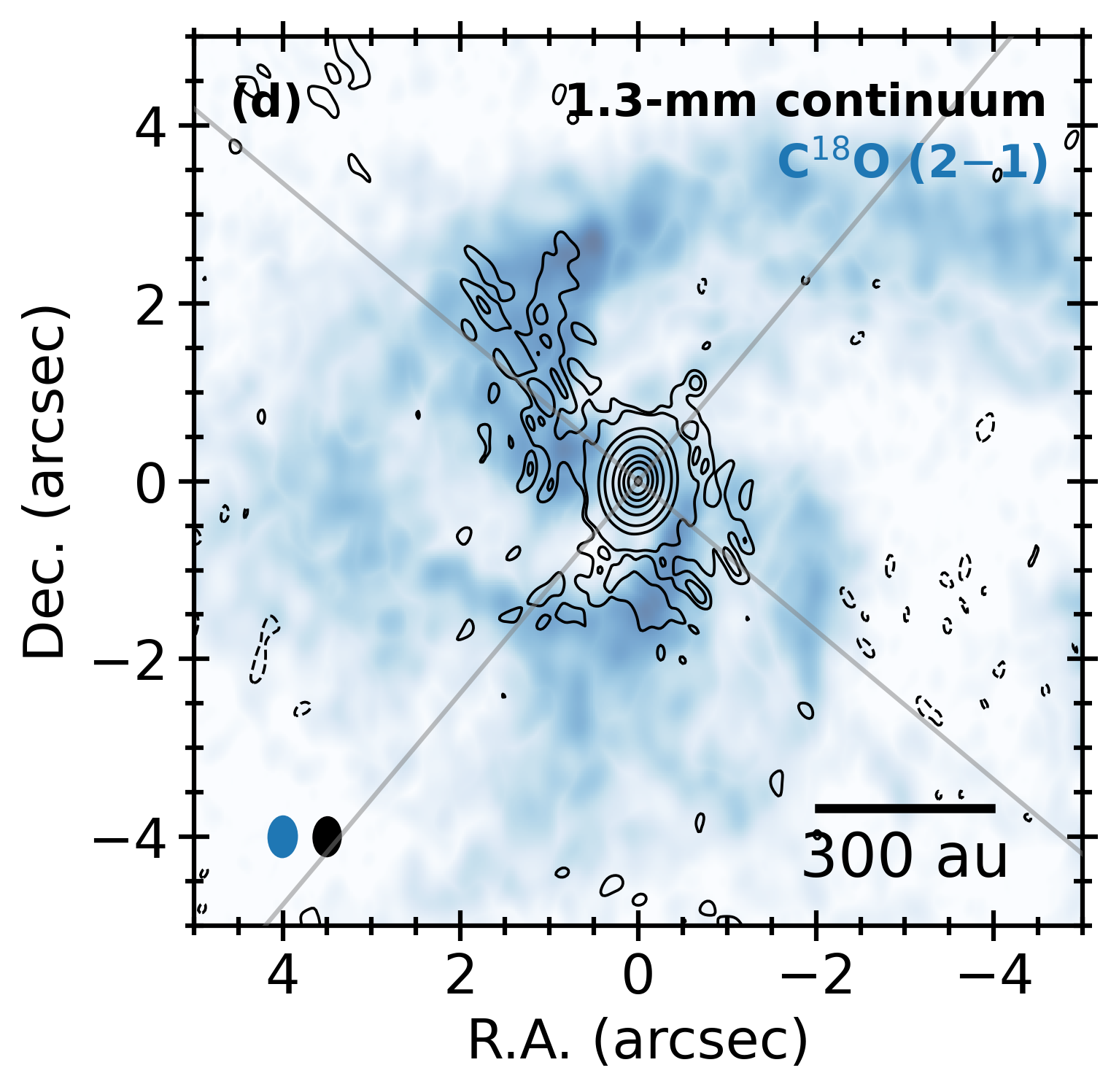}{0.3\textwidth}{}
        \fig{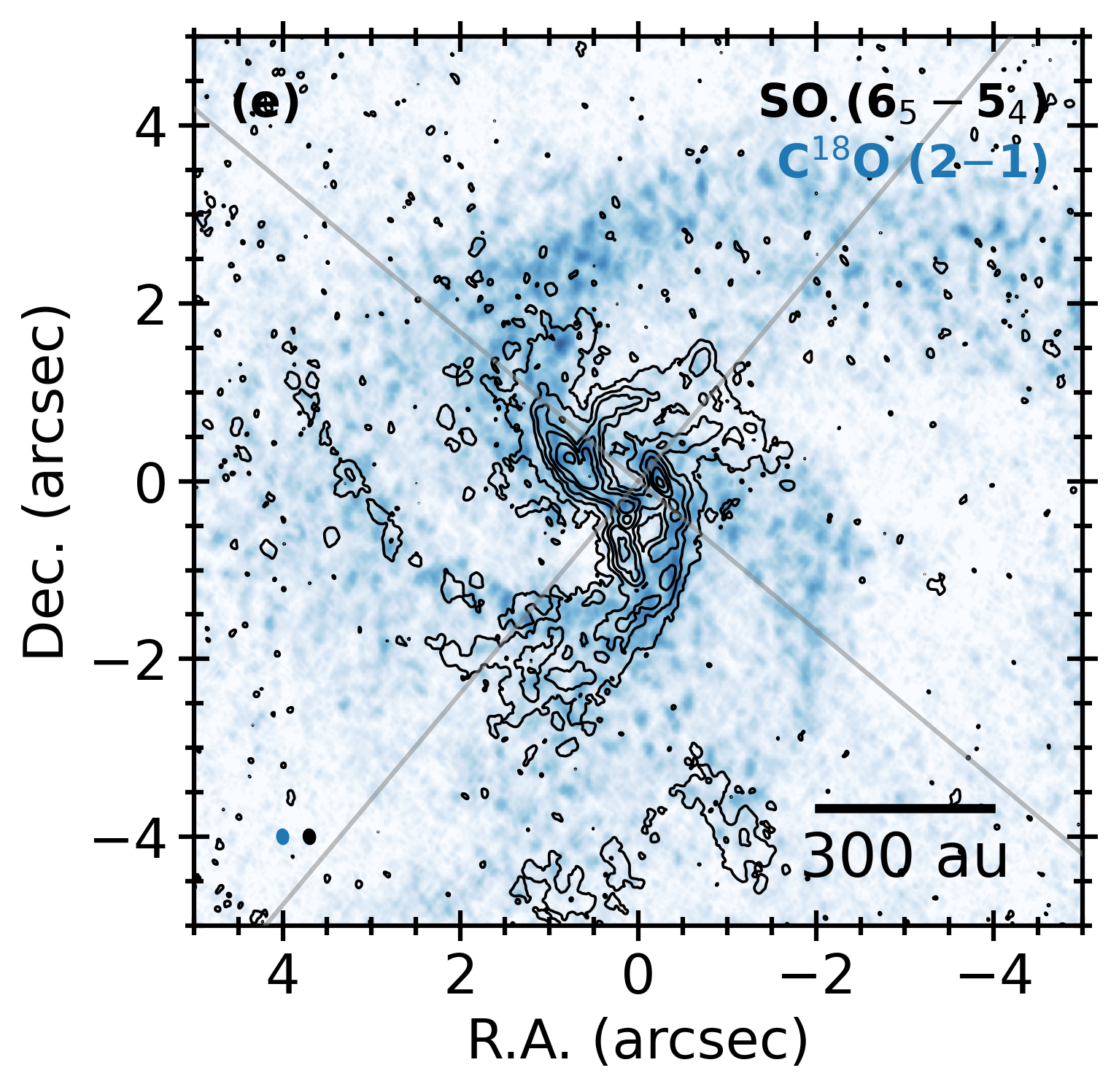}{0.3\textwidth}{}
        \rightfig{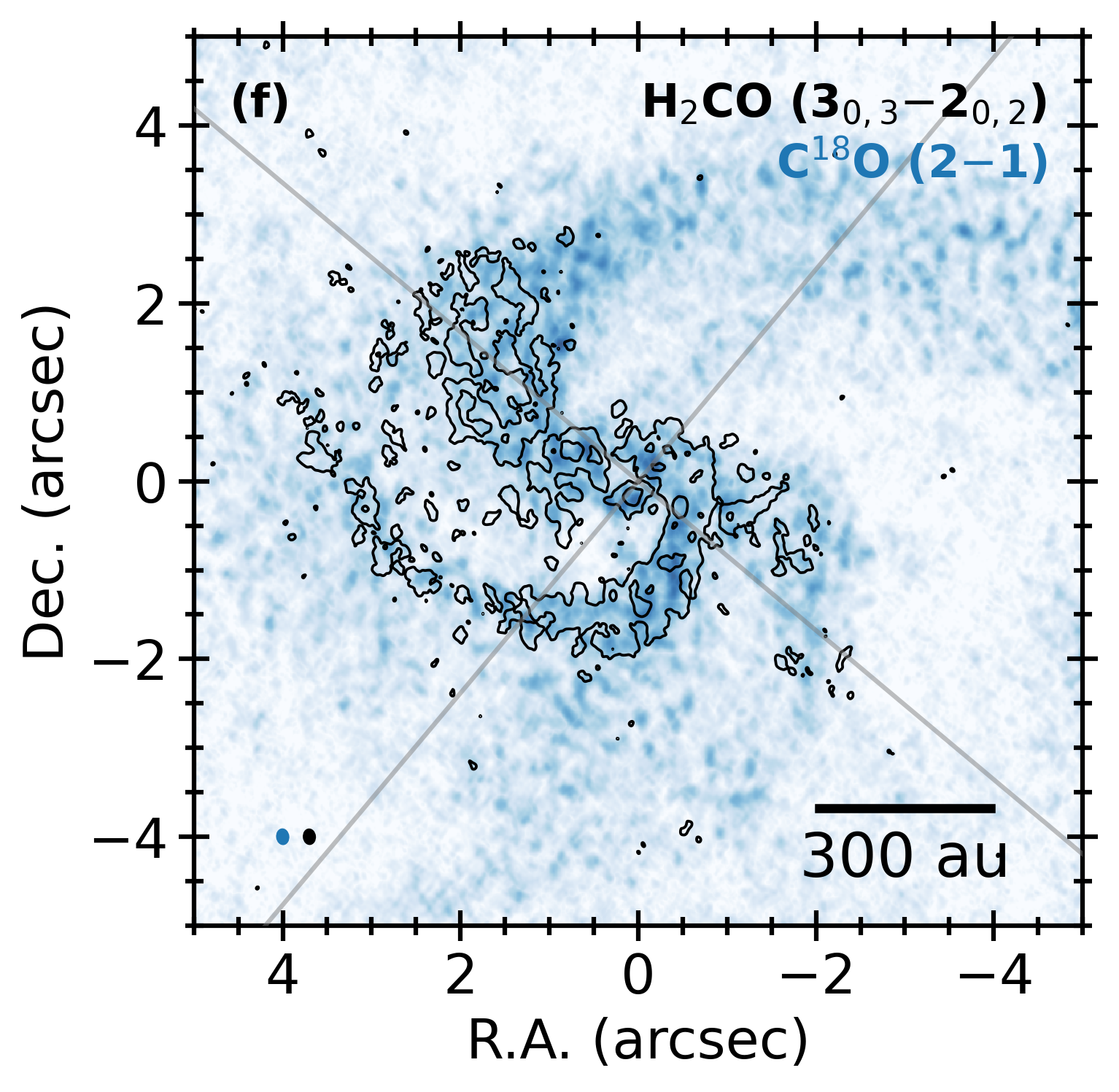}{0.3\textwidth}{}
    }
    \vspace{-0.5cm}
    \caption{Panels (a), (b), and (c) show peak intensity (moment 8) and velocity at peak intensity (moment 9) maps of C$^{18}$O (2--1), SO (6$_5$--5$_4$), and H$_2$CO (3$_{0, 3}$--2$_{0, 2}$), which are presented in contours and color, respectively. All three panels have the same color range from 2.68 to 10.92 km s$^{-1}$. Panels (a) and (d) are from the SB-only data to emphasize the large-scale spiral structure in 10$\arcsec$ $\times$ 10$\arcsec$, while the rest are from the SBLB data to examine the finer structures of these spirals in 5$\arcsec$ $\times$ 5$\arcsec$. The dark grey box in panel (a) indicates the spatial scale of the other panels. Panels (d), (e), and (f) compare the 1.3-mm dust continuum map (Figure \ref{fig:cont}a), the SO peak intensity map, and the H$_2$CO peak intensity map, which are displayed in black contours, with the C$^{18}$O peak intensity map in blue scales. In panel (d), the background color is scaled linearly from the 3$\sigma_{\rm C^{18}O, SB}$ level to a peak intensity of 65.55 mJy beam$^{-1}$. In panels (e) and (f), the color range starts at the 2$\sigma_{\rm C^{18}O, SBLB}$ level, where 1$\sigma_{\rm C^{18}O, SBLB}$ $=$ 1.57 mJy beam$^{-1}$, to a peak intensity of 20.15 mJy beam$^{-1}$. The two light grey lines mark the directions of the disk major (P.A. $=$ 140$\arcdeg$) and minor (P.A. $=$ 50$\arcdeg$) axes of the LB-only dust continuum image (Figure \ref{fig:cont}c). The ellipses shown in the lower left denote the synthesized beams of the emission, which are summarized in Tables \ref{tab:continuum} and \ref{tab:line}. The contour levels in each panel are as follows: (a) The contour levels of C$^{18}$O are \{6, 9, 12, 15, 18\} $\times$ $\sigma_{\rm C^{18}O, SB}$, where 1$\sigma_{\rm C^{18}O, SB}$ $=$ 3.17 mJy beam$^{-1}$. (b) The contour levels of SO are \{4, 8, 12, 16, 20, 24\} $\times$ $\sigma_{\rm SO}$, where 1$\sigma_{\rm SO}$ $=$ 1.88 mJy beam$^{-1}$. (c) The contour levels of H$_2$CO are \{4, 8, 12, 16, 20, 24\} $\times$ $\sigma_{\rm H_{2}CO}$, where 1$\sigma_{\rm H_{2}CO}$ $=$ 0.52 mJy beam$^{-1}$. (d) The contour levels of the dust continuum are \{$-$3, 3, 5, 50, 150, 500, 1000, 1500, 2000, 2500\} $\times$ $\sigma_{\rm 1.3 mm}$, where 1$\sigma_{\rm 1.3 mm}$ $=$ 30.0 $\mu$Jy beam$^{-1}$. The contour levels presented in panels (e) and (f) are the same as those in panels (b) and (c), respectively.}
    \label{fig:streamer}
\end{figure*}

Recently, non-axisymmetrically infalling structures, called streamers, have been discovered in several Class 0/I YSOs, including our eDisk targets, \citep[e.g.,][]{2014ApJ...793....1Y, 2019ApJ...880...69Y, 2020NatAs...4.1158P, 2022ApJ...925...32T, 2022A&A...667A..12V, 2024A&A...687A..71V, 2023ApJ...954..101A, 2023ApJ...958...98F, 2023A&A...669A.137H, 2023ApJ...953..190K, 2025arXiv250400495K, 2023ApJ...953...82L, 2023ApJ...951...11Y, 2024A&A...682A..61C}.
These streamers have a range of physical scales, from hundreds to thousands of au.
\revision{The S-shaped spiral structures of I04169, detected in C$^{18}$O, SO, and H$_2$CO, are likely streamers, given that they have morphologies similar to the previous cases.}
In addition, as briefly discussed in Section \ref{sec:c18o}, these
\revision{S-shaped structures}
seem kinematically distinct from the outflow traced in $^{12}$CO and $^{13}$CO.
In this section, we further investigate the difference between the new components and the outflow below.

Figure \ref{fig:mom0} shows the SBLB moment 0 maps of the blue- and red-shifted $^{12}$CO (2--1) emission, presented in blue and red contours, respectively, and those of the $^{13}$CO (2--1) and C$^{18}$O (2--1) emission in greyscale.
Note that C$^{18}$O shows the most prominent spiral features compared to the other two tracers, SO and H$_2$CO.
As noted in Section \ref{sec:13co}, the $^{13}$CO emission exhibits a spatial distribution quite similar to that of the $^{12}$CO emission.
It is concentrated along the northeastern wall of the redshifted outflow and is also observed along the southern wall of the blueshifted outflow.
These features indicate that $^{13}$CO traces the outflow.
In contrast, the spatial distribution of the C$^{18}$O emission is different from that of the $^{12}$CO emission.
The overall shape resembles an S-shaped spiral elongated in the northeastern--southwestern direction.
Notably, one arm-like feature extends northeast before bending northwest.

In addition to the spatial distributions of the CO isotopologues, their velocity dispersions allow us to determine whether the spirals are just part of the outflow or infalling streamers.
In general, outflows, especially their cavity walls, are expected to have higher velocity dispersions than their surroundings because they considerably perturb ambient gas \citep[e.g.,][]{2013A&A...559A..23L, 2014ApJ...784...61O, 2017ApJ...847..104O}.
Indeed, \citet{2020NatAs...4.1158P} observed a large arm-like structure of Per-emb-2 in HC$_3$N (10--9) and identified it as an infalling streamer based on the subsonic velocity dispersion of this emission.
\citet{2022ApJ...925...32T} also showed that the four streamers of Lupus 3-MMS, which were identified in C$^{18}$O (2--1), have lower velocity dispersions than its outflow cavity walls.

Figure \ref{fig:mom2} shows the velocity dispersion (moment 2) maps of the SBLB $^{12}$CO, $^{13}$CO, and C$^{18}$O emission.
The velocity dispersion of $^{12}$CO is much higher than those of the other two CO isotopologues.
Particularly, it reaches up to 7 km s$^{-1}$ along the wall carved by the western blueshifted outflow.
The overall velocity dispersion of $^{13}$CO is approximately 0.8 km s$^{-1}$, but the southern region, which coincides with the southern wall of the blueshifted outflow, has a relatively higher velocity dispersion of 2 km s$^{-1}$.
This indicates that $^{13}$CO partially traces the outflow, as expected by its spatial distribution.
In contrast, the velocity dispersion of C$^{18}$O is less than 0.5 km s$^{-1}$, except for the rotating inner envelope at the center, indicating that C$^{18}$O is not significantly tracing the outflow.

The spiral structures are more clearly revealed in the peak intensity (moment 8) maps of the C$^{18}$O, SO, and H$_2$CO emission.
The upper panels of Figure \ref{fig:streamer} show the peak intensities and the velocities at peak intensities (moment 9) together.
In Figure \ref{fig:streamer}a, the spiral structure of the SB-only C$^{18}$O emission shows that the redshifted arm extends to the northeast and then bends westwards up to around 9$\arcsec$ ($\sim$1,400 au).
There is another similarly shaped but blueshifted arm nearby, extending to the northwest and curving sharply to the west.
On the other hand, the southwestern blueshifted arm, seen in the moment 0/1 maps, is resolved into two arms: one bends sharply to the east and extends up to about 3$\arcsec$ ($\sim$500 au), and the other extends southwards by around 5$\arcsec$ ($\sim$800 au).
The velocity gradually increases along the arms toward the center, meaning that material is rotating and infalling through these arms.
Based on this feature, as well as the spatial distribution and velocity dispersion of C$^{18}$O, we therefore conclude that the spirals in C$^{18}$O are streamers, which are newly discovered in this system.
Similar features were observed in previously identified streamers, including other eDisk sources: IRAS 16253$-$2429 \citep{2023ApJ...954..101A}, IRAS 16544$-$1604 \citep{2023ApJ...953..190K, 2025arXiv250400495K}, L1489 IRS \citep{2023ApJ...951...11Y}, and Oph IRS 63 \citep{2023ApJ...958...98F}.
Furthermore, Figure \ref{fig:streamer}d shows the SB-only dust continuum map in black contours and the SB-only moment 8 map of C$^{18}$O in blue scale.
As described in Section \ref{sec:cont}, there is an arm-like structure extending to the northeast and bending to the north.
This structure coincides with the northeastern redshifted arm in C$^{18}$O, indicating that the dust continuum traces part of the streamers.

The SBLB SO maps in Figure \ref{fig:streamer}b also exhibit an S-shaped spiral structure.
It appears within an area with a radius of 4$\arcsec$ ($\sim$600 au), which is more compact than the spirals seen in C$^{18}$O.
There are two arms extending to the north from the center, one redshifted and the other blueshifted, and this blueshifted arm turns to the northwest.
On the other hand, there is a single blueshifted arm in the south, which bends sharply to the east.
These spirals coincide with part of those detected in the C$^{18}$O emission (see Figure \ref{fig:streamer}e): the two northeastern arms and one southern arm.
The velocity becomes gradually higher as the position gets closer toward the center along these three spiral arms.
Note that, in general, SO may trace accretion shocks in the region where material is infalling from an envelope onto an embedded disk \citep[e.g.,][]{2014ApJ...796..131O, 2014Natur.507...78S, 2014ApJ...793....1Y, 2022A&A...658A.104G, 2023ApJ...953...82L}.
\revision{On the other hand, the other shock tracers covered by our observations, CH$_3$OH and SiO, were not detected above a 3$\sigma_I$ threshold, with the $\sigma_I$ levels in our SBLB data of 0.52 and 0.59 mJy beam$^{-1}$ at velocity resolutions of 1.34 and 1.35 km s$^{-1}$, respectively, suggesting that the shocks occurring within these arms likely have relatively low velocities and cool temperatures \citep[e.g.,][]{1997A&A...321..293S, 2016ApJ...824...88O, 2017ApJ...837..174O}.}
\revision{Further detailed chemical modeling analysis and/or higher-sensitivity follow-up observations of this source would be necessary to distinguish whether these inferred shock characteristics are intrinsic or result from current observational sensitivity limits.}

H$_2$CO (3$_{0, 3}-$2$_{0, 2}$) also traces the S-shaped structure, and its SBLB moment 8/9 maps are presented in Figure \ref{fig:streamer}c.
Two prominent spiral arms are seen: one extends to the northeast from the center and then turns to the north, while the other extends to the southwest and then curves sharply to the southeast.
In addition, there is a weak arm-like emission in the southwest.
As shown in Figure \ref{fig:streamer}f, these three arms coincide with those of C$^{18}$O.
However, note that the northwestern arm detected in C$^{18}$O and SO is not seen in this emission.
All three arms detected in H$_2$CO appear to be blueshifted with the same velocity of 6.08 km s$^{-1}$ due to the lower velocity resolution of 1.34 km s$^{-1}$.

Indeed, \citet{2018ApJ...865...51T} also marginally detected a similar large-scale structure in the C$^{18}$O (2--1) emission using the SMA with a lower angular resolution.
They interpreted this structure as an envelope counter-rotating against an inner disk detected in $^{13}$CO (3--2) and SO (6$_5$--5$_4$).
However, our higher-resolution observations show that this structure is part of the accretion streamers, not a counter-rotating envelope.

\subsubsection{Origins of Streamers}
\label{sec:streamerorigins}

Several scenarios have been proposed to explain the origins of streamers.
Here, we discuss three possibilities of the streamers discovered in I04169 and then explore the implications for these streamers.
The first is misalignment between the magnetic field and the rotation axis of a protostellar system.
For example, \citet{2020ApJ...898..118H} magnetohydrodynamically simulated how a protostellar system evolves depending on the angle between the global magnetic field and the rotation axis of the system by considering a broad physical scale from a large spherical core to an embedded Keplerian disk.
In the misaligned case, a flattened envelope develops two prominent spiral arms, and material is accreted along these spirals.
Particularly, when the magnetic field of a core is misaligned by around 45$\arcdeg$ with the rotation axis of the core, an outflow has a more complicated shape;
\revision{for example, it loses its axisymmetry and develops a bent or even warped structure.}
\revision{Furthermore,} a disk becomes more compact.
The observed features of I04169, described in the previous sections, are consistent with these theoretical predictions.
This suggests that the streamers may be caused by core-scale magnetic misalignment.
Indeed, previous JCMT 850-$\mu$m polarimetric observations of I04169 showed that its outflow axis is misaligned by approximately between 40$\arcdeg$ and 60$\arcdeg$ with respect to the mean magnetic field orientation of the core in which this source is embedded \citep{2021ApJ...912L..27E, 2021ApJ...907...33Y, 2024ApJ...969..125Y}.
Notably, \citet{2021ApJ...907...33Y} confirmed that this misalignment remains highly probable in three-dimensional space, even when accounting for projection effects.
Assuming that the outflow axis is the same as the disk's rotation axis, these results imply that I04169 is magnetically misaligned on the core scale, supporting the scenario of magnetic field misalignment.
\citet{2020ApJ...898..118H} also demonstrated magnetic fields to become quite well-aligned with streamers on smaller physical scales, from tens to hundreds of au.
Future ALMA polarimetric observations of I04169 at a higher angular resolution are necessary to investigate such localized alignment.

Secondly, small molecular cloud fragments, so-called cloudlets, can evolve into streamers when they are captured by a protostellar system \citep{2019A&A...628A..20D}.
According to several previous observations \citep[e.g.,][]{1995ApJ...453..293L, 2012ApJ...754...95T}, cloudlets have a mass of around 0.01 to 1 $M_{J}$, which is much less than a solar mass, and range in size from hundreds to thousands of au.
\citet{2019A&A...628A..20D} proposed that when cloudlets encounter a nearby circumstellar disk, part of them can be accreted in an arc shape onto the disk, and \citet{2020A&A...633A...3K} explored a wider span of parameter space through hydrodynamic simulations.
Recently, \citet{2022ApJ...932..122H} demonstrated that such a cloudlet capture event could reproduce the arc-shaped asymmetric emission of CS (5--4) in the Class I protostellar system TMC-1A, which was observed by \citet{2016ApJ...820L..34S}.
Furthermore, \citet{2022ApJ...941..154U} investigated the effect of magnetic fields; the accretion of cloudlets depends on magnetic field strength and cloudlet size.
The resultant streamers produced by the simulations always have a single arc shape, which implies that more than two cloudlets seem to be rarely captured at once by the same protostellar system.
As shown in Figure \ref{fig:streamer}, I04169 has S-shaped streamers consisting of at least two arms.
It is thus less likely that multiple cloudlets were captured by this system simultaneously.

The third possibility is gravitational interactions within multiple stellar systems.
\citet{2015MNRAS.449L.123M} simulated that arc-shaped streamers hundreds of au long can form through interactions between multiple protostars at close separations between 10 and 200 au.
Moreover, accretion streamers can also be developed during the forming process of binary stars with much wider initial separations of more than 500 au \citep[e.g.,][]{2019A&A...628A.112K}, and such wide binaries can rapidly migrate inward to separations of less than 100 au within a few hundred kyr, which is comparable to the duration of the protostellar phase \citep[e.g.,][]{2010ApJ...725.1485O, 2016ApJ...827L..11O, 2019ApJ...887..232L, 2019A&A...628A.112K}.
Indeed, arc-shaped streamers were recently observed in several Class 0/I close multiples: e.g., Per-emb-2 \citep{2020NatAs...4.1158P}, NGC 1333-IRAS2A \citep{2022Natur.606..272J}, and IRAS 04239$+$2436 \citep{2023ApJ...953...82L}.
Particularly, the latter two demonstrated through magnetohydrodynamic simulations that their streamers were driven by gravitational interactions within the systems.
In contrast to these cases, the binarity of I04169, suggested by \citet{2008AJ....135.2496C}, is not substantiated by our higher-resolution observations; however, as discussed in Section \ref{sec:substructures}, it is also still possible that the central inner region hosts a closer binary with a separation of less than 8 au.
\citet{2017MNRAS.470.1626K} simulated that accretion streamers can also form even in a binary system separated by only 2.5 au.

Such streamers may imply development of substructures and growth of dust grains therein.
\citet{2015ApJ...805...15B} and \citet{2022ApJ...928...92K} demonstrated that the Rossby-wave instability (RWI) can be triggered when material from an envelope lands onto an embedded disk through streamers.
This instability generates vortices and pressure bumps, efficiently trapping inward-moving grains and facilitating their rapid growth to millimeter/centimeter sizes.
As a result, a compact yet ring-structured disk can form.

\section{Conclusions}
\label{sec:conclusions}

We have observed the Class I protostellar system IRAS 04169$+$2702 in the 1.3-mm dust continuum and various molecular lines, as part of the ALMA Large Program eDisk.
We also have carried out VLA continuum observations toward this target in Q (0.7 cm), K (1.4 cm), and C (5.0 cm) bands.
The main results are summarized as follows:

\begin{enumerate}
    \item
        The 1.3-mm dust continuum emission exhibits a compact circumstellar disk with a deconvolved FWHM of 0$\farcs$22 (34 au) and an arm-like feature extending toward the northeast.
        A bean-shaped asymmetric brightness distribution is revealed at the center of the disk, which is further resolved into two local peaks and a central depression between them in the higher-resolution image.
        The two peaks are located along the disk major axis with a projected separation of 0$\farcs$11 (17 au).
        The dust mass of the disk is estimated to be between 18 and 67 $M_{\oplus}$.
    \item
        The disk is further investigated by comparing the ALMA Band 6 observations with the VLA observations at multiple wavelengths.
        The Q-band image reveals a single central peak located between the two peaks seen in the ALMA image.
        The two C-band images trace the protostellar jet.
        The K-band image captures both the central peak and jet features.
        Thermal dust emission is more dominant over the full disk region, while free-free emission is more dominant in the center, supporting the central depression.
        The mean spectral index of dust emission, less than 2.9, may imply the presence of large grains that have already grown up to millimeters or centimeters in size.
    \item
        PV diagrams of the three CO (2--1) isotopologues, $^{12}$CO, $^{13}$CO, and C$^{18}$O, exhibit consistent velocity gradients along the disk major axis: whereas the northwestern part is blueshifted, the southeastern part is redshifted.
        High-velocity components in $^{12}$CO trace a Keplerian disk with a radius of $\sim$21 au surrounding a 1.3-$M_{\odot}$ central protostar.
        \revision{This radius is comparable to the radius of the dusty disk, which is 29 au.}
        On the other hand, the low-velocity components in $^{13}$CO and C$^{18}$O indicate an infalling and rotating inner envelope, conserving angular momentum.
    \item
        In addition to the disk and envelope components, the $^{12}$CO emission reveals a bipolar outflow, perpendicular to the disk major axis.
        The northeastern redshifted lobe shows a complicated shape with multiple asymmetric shells, while the southwestern blueshifted lobe displays a U-shaped symmetric cavity wall.
        Part of these outflow features are observed in the $^{13}$CO emission as well.
    \item
        A large spiral structure, which comprises two main arms along the disk minor axis, has been discovered in C$^{18}$O, SO (6$_5$--5$_4$), and H$_2$CO (3$_{0, 3}$--2$_{0, 2}$): one extends from the northeast and sharply bends west, while the other curves from the southwest to the southeast.
        This northeastern arm also corresponds to the one detected in the dust continuum emission.
        They are infalling streamers distinct from the outflow.
\end{enumerate}

\section*{Acknowledgements}
    \revision{We are grateful to the anonymous reviewer for fruitful comments.}
        This paper makes use of the following ALMA data: ADS/JAO.ALMA$\#$2019.1.00261.L. ALMA is a partnership of ESO (representing its member states), NSF (USA) and NINS (Japan), together with NRC (Canada), MOST and ASIAA (Taiwan), and KASI (Republic of Korea), in cooperation with the Republic of Chile. The Joint ALMA Observatory is operated by ESO, AUI/NRAO and NAOJ.
        The National Radio Astronomy Observatory is a facility of the National Science Foundation operated under cooperative agreement by Associated Universities, Inc.
        I. Han acknowledges support the funding from the European Research Council (ERC) under the European Union's Horizon 2020 research and innovation programme (Grant agreement No. 101098309 - PEBBLES).
        W.K. is supported by the National Research Foundation of Korea (NRF) grant funded by the Korea government (MSIT) (RS-2024-00342488 and RS-2024-00416859).
        N.O. and C.F. acknowledge support from National Science and Technology Council (NSTC) in Taiwan through the grants NSTC 109-2112-M-001-051, 110-2112-M-001-031, and 113-2112-M-001-037, and also from Academia Sinica Investigator Project Grant (AS-IV-114-M02).
        J.J.T. acknowledges support from NASA XRP 80NSSC22K1159, The National Radio Astronomy Observatory is a facility of the National Science Foundation operated under cooperative agreement by Associated Universities, Inc.
        J.K.J. and S.G. acknowledge support from the Independent Research Fund Denmark (grant No. 0135-00123B).
        S.T. is supported by JSPS KAKENHI Grant Numbers 21H00048 and 21H04495. This work was supported by NAOJ ALMA Scientific Research Grant Code 2022-20A.
        Y.A. acknowledges support by NAOJ ALMA Scientific Research Grant code 2019-13B, Grant-in-Aid for Scientific Research (S) 18H05222, and Grant-in-Aid for Transformative Research Areas (A) 20H05844 and 20H05847.
        I.d.G. acknowledges support from grant PID2020-114461GB-I00, funded by MCIN/AEI/10.13039/501100011033.
        P.M.K. acknowledges support from NSTC 108-2112-M-001-012, NSTC 109-2112-M-001-022 and NSTC 110-2112-M-001-057.
        C.W.L. is supported by the Basic Science Research Program through the National Research Foundation of Korea (NRF) funded by the Ministry of Education, Science and Technology (NRF-2019R1A2C1010851), and by the Korea Astronomy and Space Science Institute grant funded by the Korea government (MSIT; Project No. 2023-1-84000).
        J.-E.L. is supported by the National Research Foundation of Korea (NRF) grant funded by the Korean government (MSIT) (grant numbers 2021R1A2C1011718 and RS-2024-00416859).
        Z.-Y.L. is supported in part by NASA 80NSSC20K0533, NSF AST-2307199, and AST-1910106.
        Z.-Y.D.L. acknowledges support from NASA 80NSSCK1095, the Jefferson Scholars Foundation, the NRAO ALMA Student Observing Support (SOS) SOSPA8-003, the Achievements Rewards for College Scientists (ARCS) Foundation Washington Chapter, the Virginia Space Grant Consortium (VSGC), and UVA research computing (RIVANNA).
        L.W.L. acknowledges support from NSF AST-2108794.
        N.T.P. acknowledges support from Vietnam Academy of Science and Technology under grand number VAST 08.02/25-26.
        S.-P.L and T.J.T. acknowledge grants from the National Science and Technology Council of Taiwan 106-2119-M-007-021-MY3 and 109-2112-M-007-010-MY3.
        J.P.W. acknowledges support from NSF AST-2107841.
        H.-W.Y. acknowledges support from the National Science and Technology Council (NSTC) in Taiwan through the grant NSTC 110-2628-M-001-003-MY3 and from the Academia Sinica Career Development Award (AS-CDA-111-M03).

%

\vspace{5mm}
\facilities{ALMA, VLA.}


\software{CASA \citep{2007ASPC..376..127M, 2022PASP..134k4501C}, SLAM \citep{2024PKAS...39...2A}.}



\appendix
\restartappendixnumbering
\section{Channel maps for the detected five molecular lines}

The channel maps of the five molecular lines detected in I04169, $^{12}$CO (2--1), $^{13}$CO (2--1), C$^{18}$O (2--1), SO (6$_5$--5$_4$), H$_2$CO (3$_{0, 3}$--2$_{0, 2}$), are shown in the following figures.
Basically, we employ the SBLB data for all the lines, enabling us to trace both large- and small-scale features.
In addition, we utilize the LB-only data of the three CO isotopologues that trace the Keplerian disk and surrounding envelope (Section \ref{sec:kinematics}) and the SB-only data for C$^{18}$O to highlight the spiral feature (Section \ref{sec:streamer}).

\begin{figure*}[h]
    \gridline{
        \fig{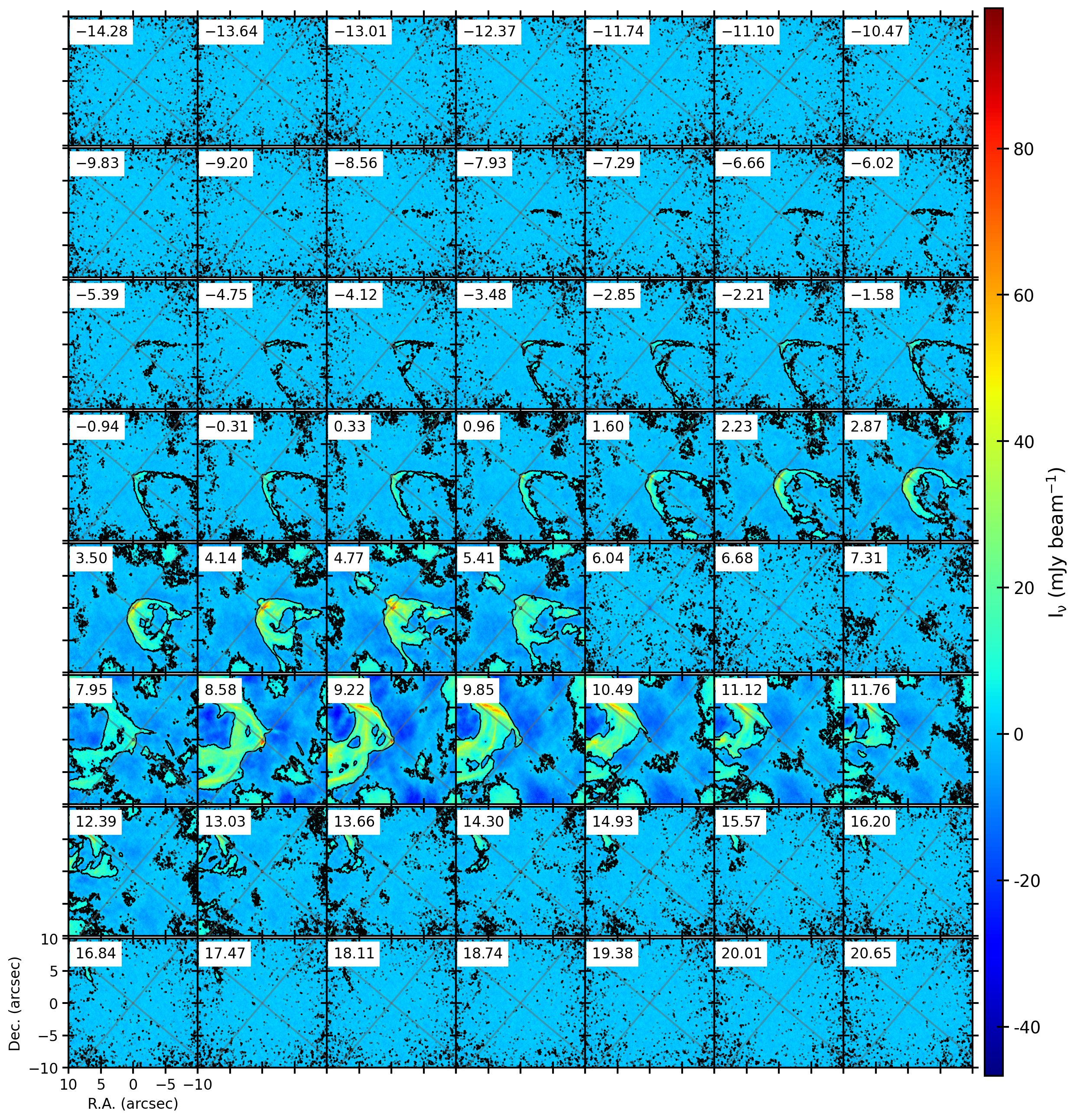}{0.9\textwidth}{}
    }
    \vspace{-0.8cm}
    \caption{Channel maps for the $^{12}$CO (2--1) emission of the SBLB data. The emission above the 3$\sigma_{\rm SBLB}$ level is detected from $-$13.01 to 18.75 km s$^{-1}$. The five channels before $-$13.01 km s$^{-1}$ and after 18.75 km s$^{-1}$ is emission-free channels. The black contours indicate the 3$\sigma_{\rm SBLB}$ level of the emission, where 1$\sigma_{\rm SBLB}$ corresponds to 0.95 mJy beam$^{-1}$. The red ellipse in the lower left denotes the synthesized beam of 0$\farcs$182 $\times$ 0$\farcs$147 (P.A. $=$ 6.5$\arcdeg$). The two perpendicular grey lines show the directions of the disk major (P.A. $=$ 140$\arcdeg$) and minor (P.A. $=$ 50$\arcdeg$) axes measured in the LB-only dust continuum image (Figure \ref{fig:cont}c). The bipolar outflow feature is clearly seen along the disk minor axis, i.e., along the northeast-southwest direction. Notably, the redshifted components exhibit complicated shapes with multiple parabolic shells and are not well aligned with the disk minor axis.}
    \label{fig:chan12cosblb}
\end{figure*}

\begin{figure*}[p]
    \gridline{
        \fig{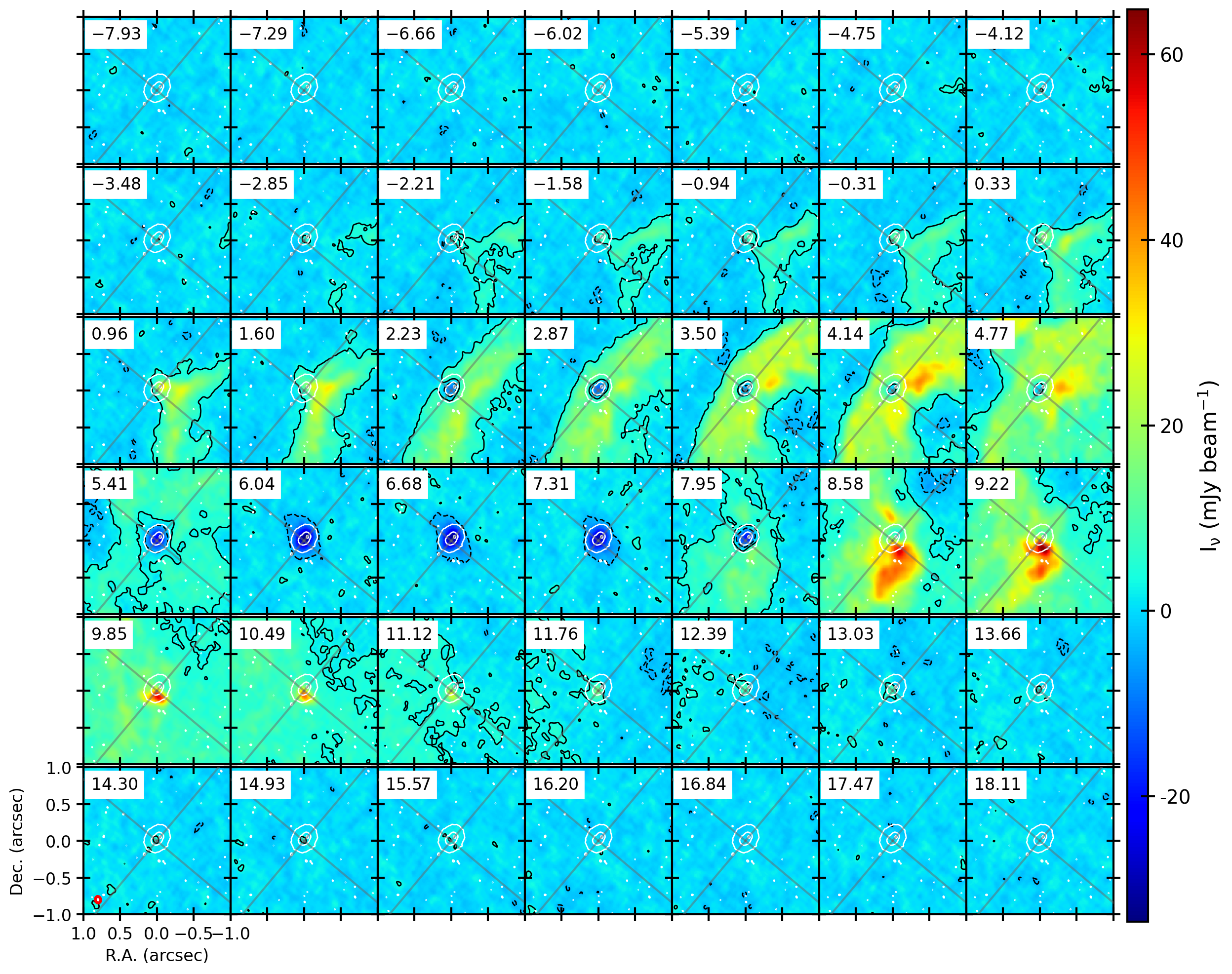}{0.9\textwidth}{}
    }
    \vspace{-0.8cm}
    \caption{Channel maps for the $^{12}$CO (2--1) emission of the LB-only data. The emission above the 3$\sigma_{\rm LB}$ emission is detected from --7.93 to 18.11 km s$^{-1}$, and the black dashed and solid contours indicate the $-$3$\sigma_{\rm LB}$ and 3$\sigma_{\rm LB}$ levels of the emission, respectively, where 1$\sigma_{\rm LB}$ corresponds to 1.07 mJy beam$^{-1}$. The red ellipse in the lower left denotes the synthesized beam of this emission, which is 0$\farcs$129 $\times$ 0$\farcs$109 (P.A. $=$ 12.9$\arcdeg$). In addition, the LB-only dust continuum emission is represented as white contours with levels of \{3, 150, 190\}, where 1$\sigma_{\rm LB}$ corresponds to 18.8 $\mu$Jy beam$^{-1}$, and its synthesized beam of 0$\farcs$044 $\times$ 0$\farcs$034 (P.A. $=$ 19.1$\arcdeg$) is denoted as a white ellipse in the lower left.}
    \label{fig:chan12colb}
\end{figure*}

\begin{figure*}[p]
    \gridline{
        \fig{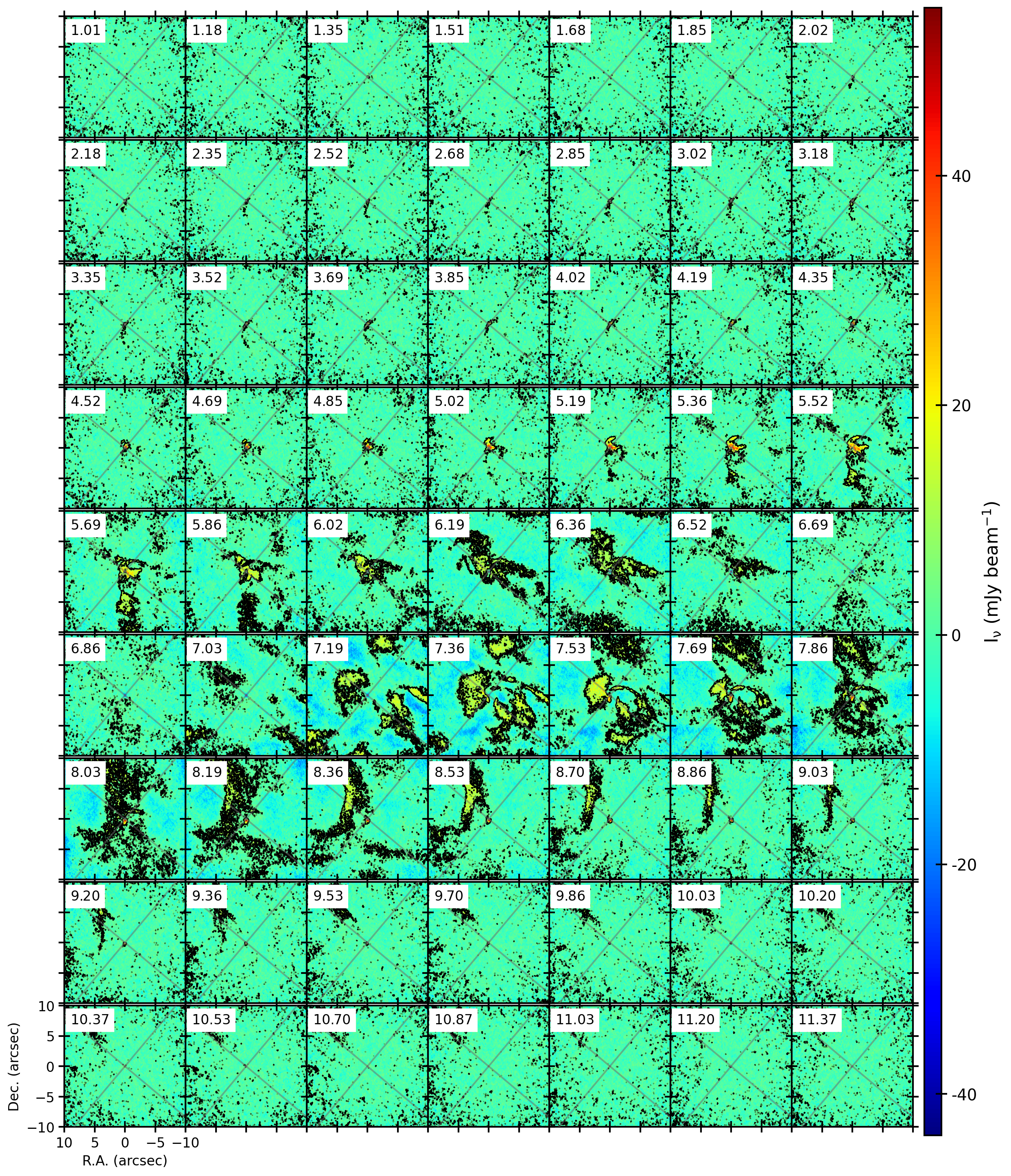}{0.9\textwidth}{}
    }
    \vspace{-0.8cm}
    \caption{Channel maps for the $^{13}$CO (2--1) emission of the SBLB data. The emission above the 3$\sigma_{\rm SBLB}$ level is detected from 1.01 to 11.37 km s$^{-1}$. The black contours indicate the 3$\sigma_{\rm SBLB}$ level of the emission, where 1$\sigma_{\rm SBLB}$ corresponds to 2.09 mJy beam$^{-1}$. The red ellipse in the lower left denotes the synthesized beam of the emission, which is 0$\farcs$189 $\times$ 0$\farcs$151 (P.A. $=$ 4.0$\arcdeg$).}
    \label{fig:chan13cosblb}
\end{figure*}

\begin{figure*}[p]
    \gridline{
        \fig{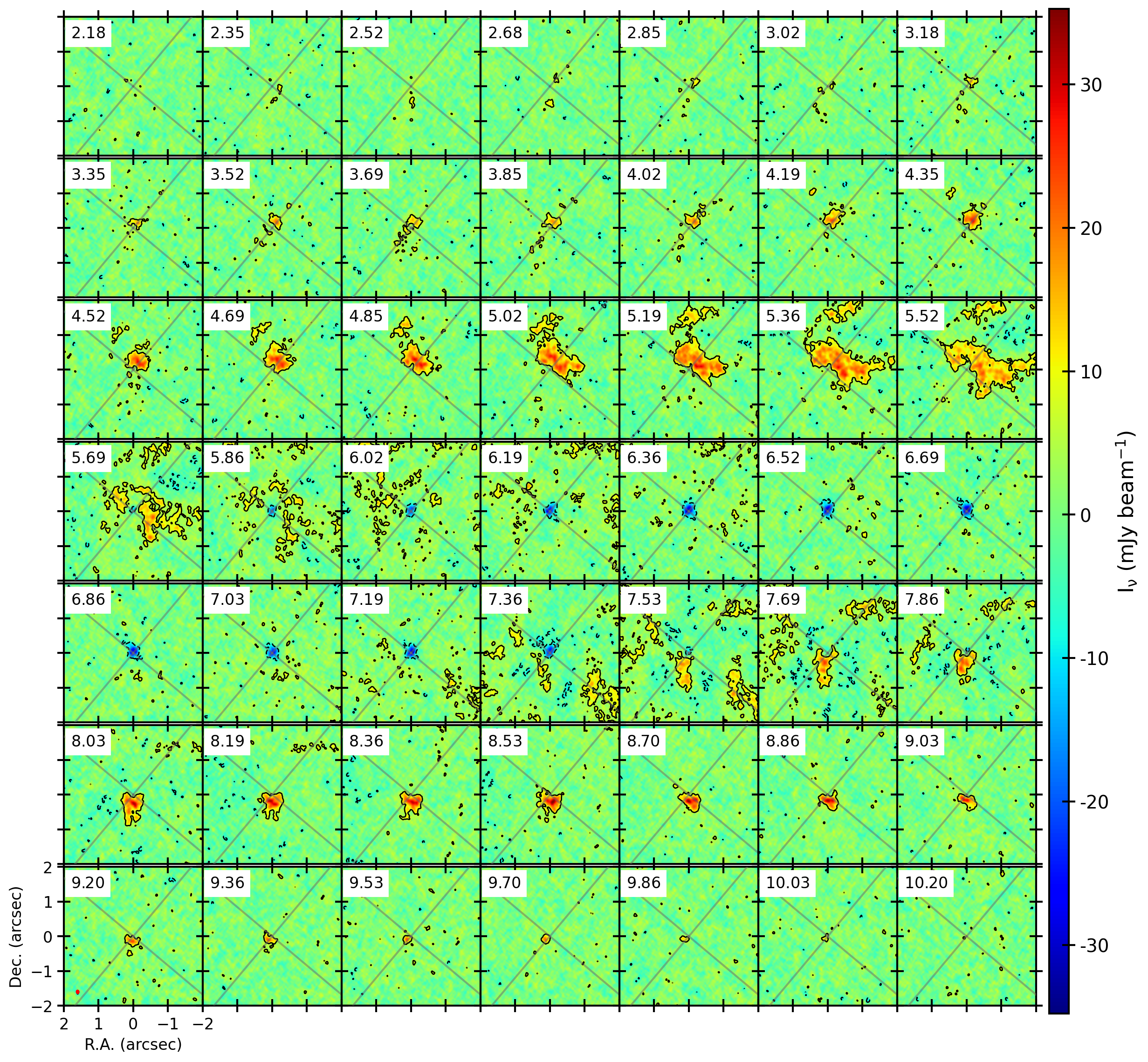}{0.9\textwidth}{}
    }
    \vspace{-0.8cm}
    \caption{Channel maps for the $^{13}$CO (2--1) emission of the LB-only data. The emission above the 3$\sigma_{\rm LB}$ level is detected from 2.18 to 10.20 km s$^{-1}$. The black dashed and solid contours indicate the $-$3$\sigma_{\rm LB}$ and 3$\sigma_{\rm LB}$ levels of the emission, respectively, where 1$\sigma_{\rm LB}$ corresponds to 2.35 mJy beam$^{-1}$. The red ellipse in the lower left denotes the synthesized beam of 0$\farcs$132 $\times$ 0$\farcs$111 (P.A. $=$ 10.8$\arcdeg$).}
    \label{fig:chan13colb}
\end{figure*}

\begin{figure*}[p]
    \gridline{
        \fig{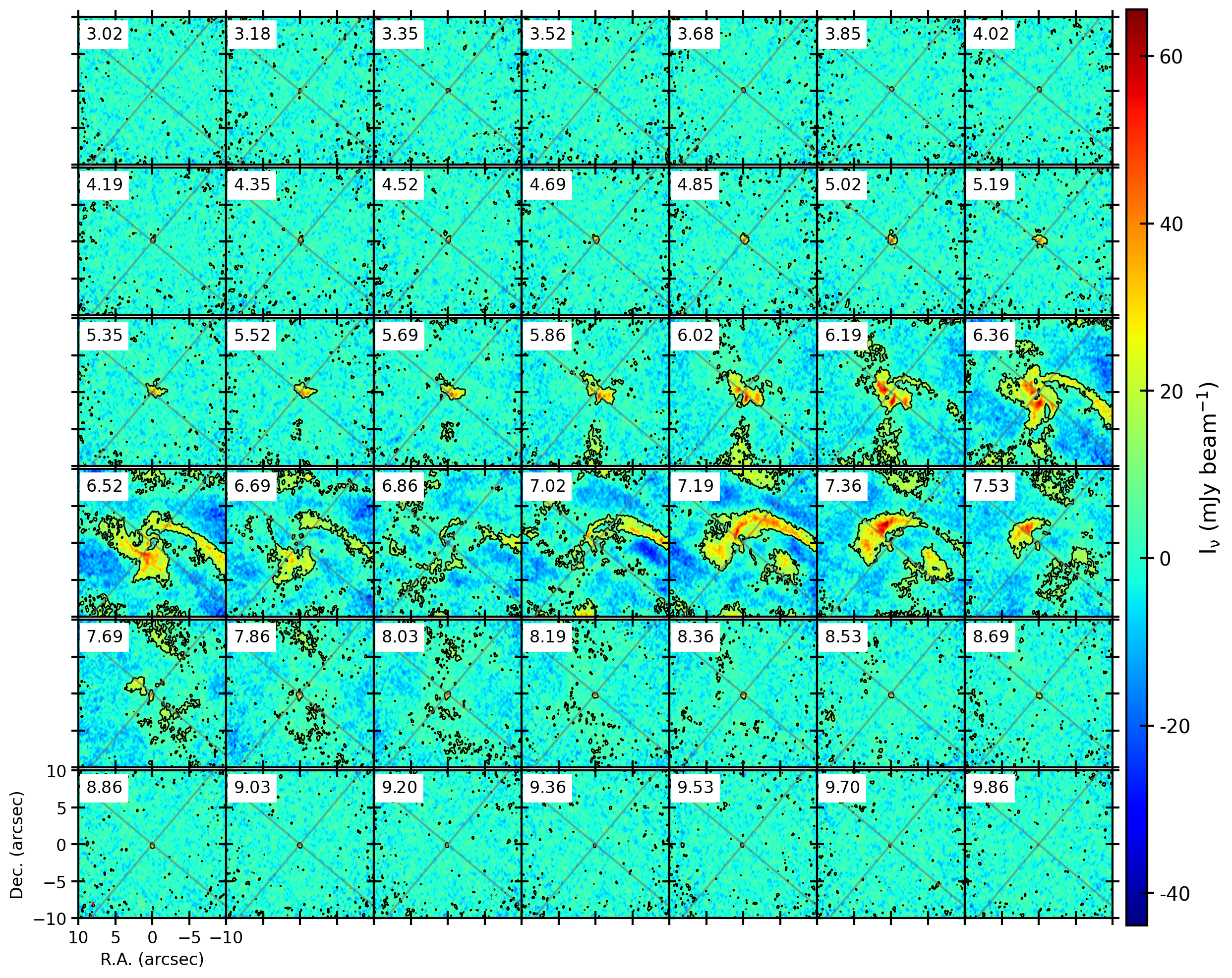}{0.9\textwidth}{}
    }
    \vspace{-0.8cm}
    \caption{Channel maps for the C$^{18}$O (2--1) emission of the SB-only data. The emission above the 3$\sigma_{\rm SB}$ level is detected from 3.18 to 9.86 km s$^{-1}$. The single channel with a velocity of 3.02 km s$^{-1}$ is an emission-free channel. The black contours indicate the 3$\sigma_{\rm SB}$ levels of the emission, where 1$\sigma_{\rm SB}$ corresponds to 3.17 mJy beam$^{-1}$. The red ellipse in the lower left denotes the synthesized beam of 0$\farcs$482 $\times$ 0$\farcs$343 (P.A. $=$ $-$2.0$\arcdeg$).}
    \label{fig:chanc18osb}
\end{figure*}

\begin{figure*}[p]
    \gridline{
        \fig{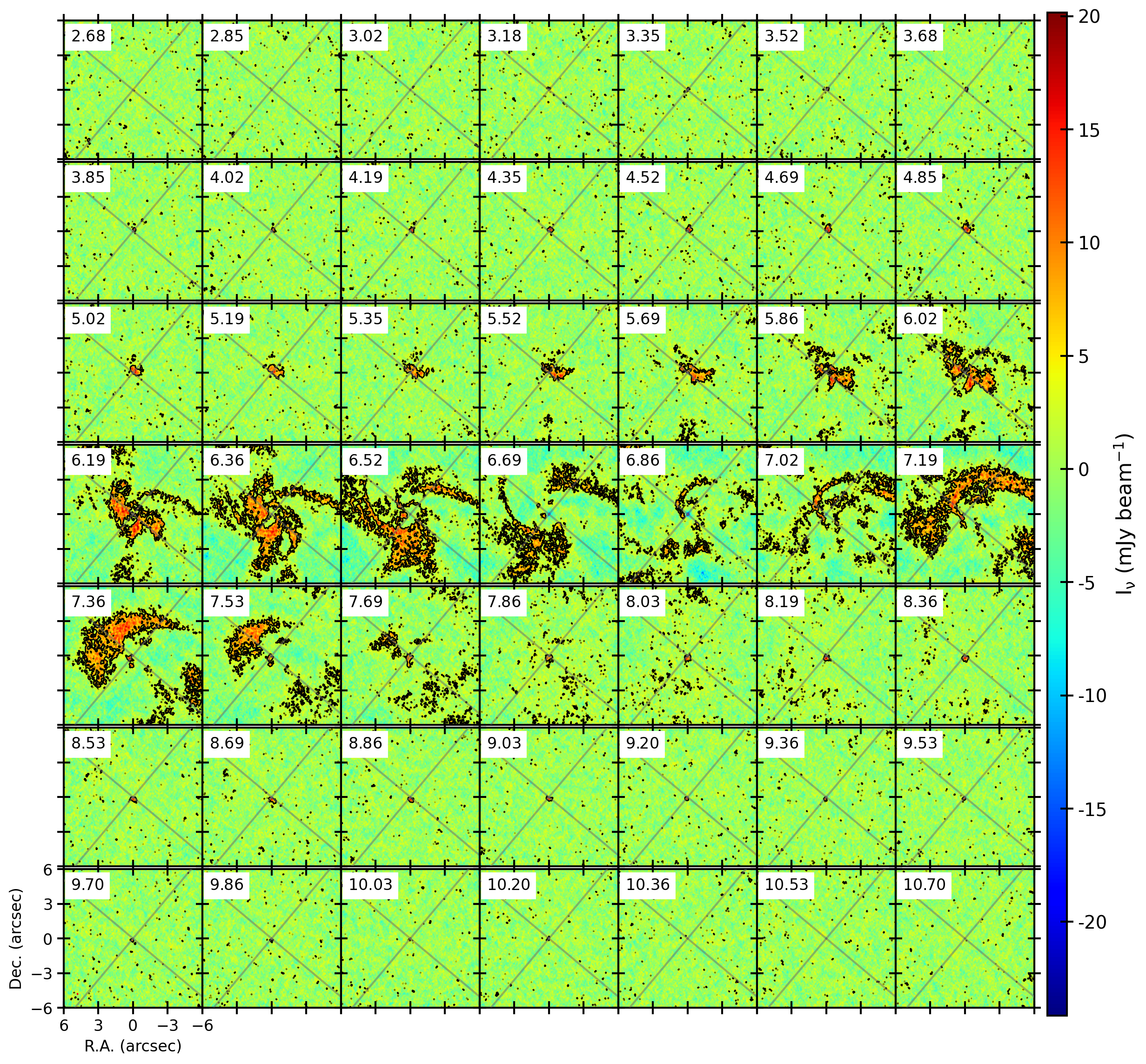}{0.9\textwidth}{}
    }
    \vspace{-0.8cm}
    \caption{Channel maps for the C$^{18}$O (2--1) emission of the SBLB data. The emission above the 3$\sigma_{\rm SBLB}$ level is detected from 3.18 to 10.20 km s$^{-1}$. The six channels before 3.18 km s$^{-1}$ and after 10.20 km s$^{-1}$ are emission-free channels. The black contours indicate the 3$\sigma_{\rm SBLB}$ levels of the emission, where 1$\sigma_{\rm SBLB}$ corresponds to 1.57 mJy beam$^{-1}$. The red ellipse in the lower left denotes the synthesized beam of 0$\farcs$188 $\times$ 0$\farcs$149 (P.A. $=$ 6.6$\arcdeg$).}
    \label{fig:chanc18osblb}
\end{figure*}

\begin{figure*}[p]
    \gridline{
        \fig{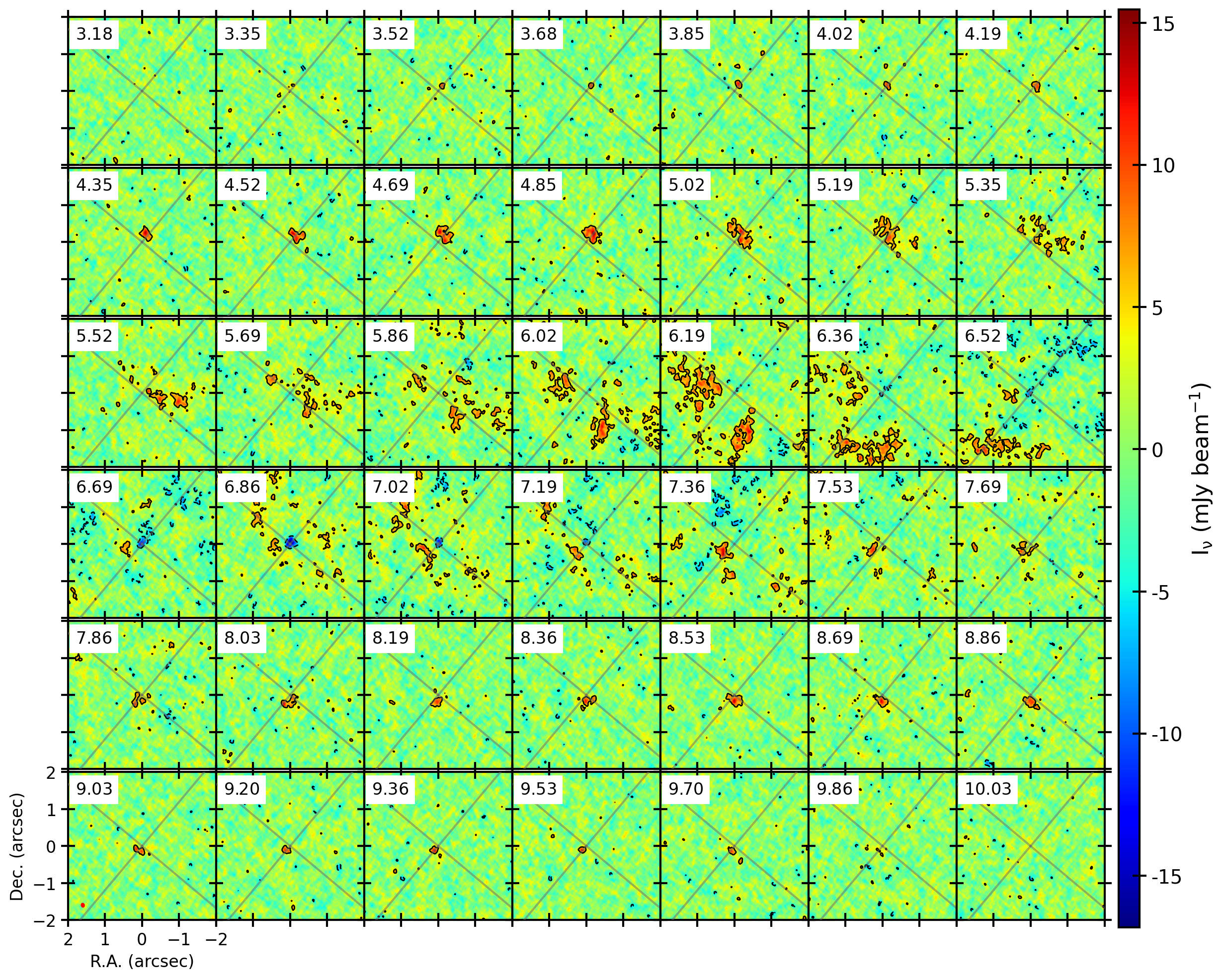}{0.9\textwidth}{}
    }
    \vspace{-0.8cm}
    \caption{Channel maps for the C$^{18}$O (2--1) emission of the LB-only data. The emission above the 3$\sigma_{\rm LB}$ level is detected from 3.35 to 9.86 km s$^{-1}$. The two channels before 3.35 km s$^{-1}$ and after 9.86 km s$^{-1}$ are emission-free channels. The black dashed and solid contours indicate the $-$3$\sigma_{\rm LB}$ and 3$\sigma_{\rm LB}$ levels of the emission, respectively, where 1$\sigma_{\rm LB}$ corresponds to 1.75 mJy beam$^{-1}$. The red ellipse in the lower left denotes the synthesized beam of 0$\farcs$133 $\times$ 0$\farcs$111 (P.A. $=$ 14.2$\arcdeg$).}
    \label{fig:chanc18olb}
\end{figure*}

\begin{figure*}[p]
    \gridline{
        \fig{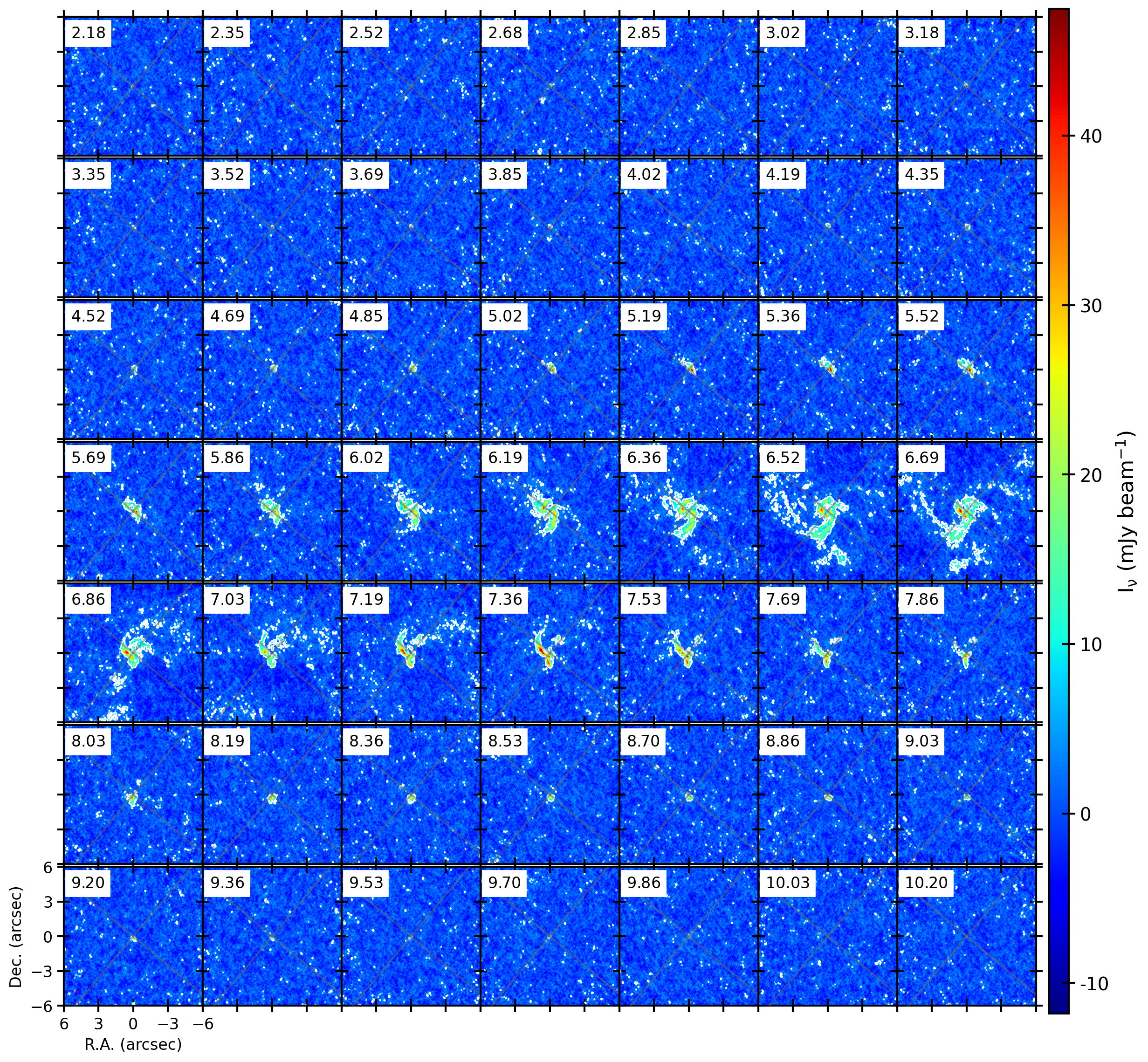}{0.9\textwidth}{}
    }
    \vspace{-0.8cm}
    \gridline{
        \fig{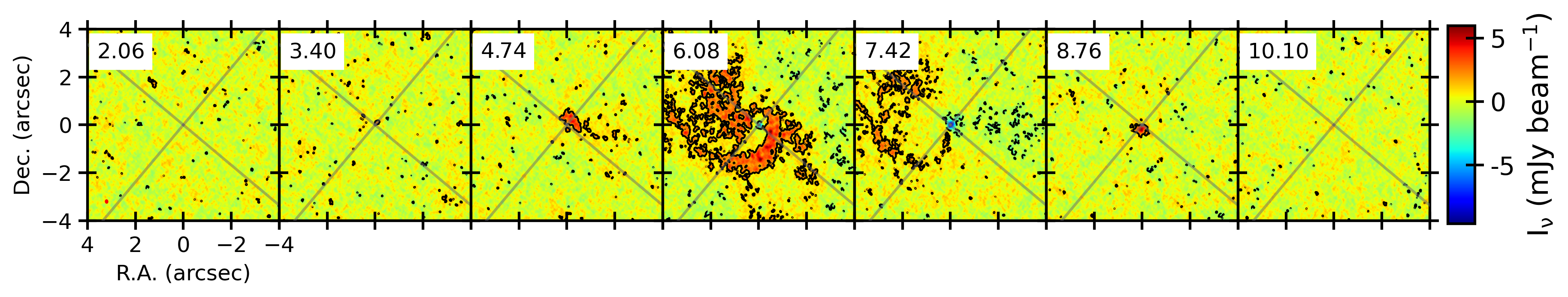}{0.9\textwidth}{}
    }
    \vspace{-0.8cm}
    \caption{\revision{\textbf{(top)}} Channel maps for the SO (6$_5-$5$_4$) emission of the SBLB data. The emission above the 3$\sigma_{\rm SBLB}$ level is detected from 2.68 to 9.70 km s$^{-1}$. The six channels before 2.68 km s$^{-1}$ and after 9.70 km s$^{-1}$ are emission-free channels. The white contours indicate the 3$\sigma_{\rm SBLB}$ levels of the emission, where 1$\sigma_{\rm SBLB}$ corresponds to 1.88 mJy beam$^{-1}$. The red ellipse in the lower left denotes the synthesized beam of 0$\farcs$188 $\times$ 0$\farcs$152 (P.A. $=$ 8.6$\arcdeg$). Note that this velocity range overlaps with the range where the spiral arms are detected in C$^{18}$O (Figures \ref{fig:chanc18osb} and \ref{fig:chanc18osblb}). \revision{\textbf{(bottom)} Channel maps for the H$_2$CO (3$_{0, 3}-$2$_{0, 2}$) emission of the SBLB data. The emission above the 3$\sigma_{\rm SBLB}$ level is detected from 3.4 to 8.76 km s$^{-1}$. The two channels before 3.4 km s$^{-1}$ and after 8.76 km s$^{-1}$ are emission-free channels. The black dashed and solid contours indicate the $-$3$\sigma_{\rm SBLB}$ and 3$\sigma_{\rm SBLB}$ levels of the emission, respectively, where 1$\sigma_{\rm SBLB}$ corresponds to 0.52 mJy beam$^{-1}$. The red ellipse in the lower left denotes the synthesized beam of 0$\farcs$186 $\times$ 0$\farcs$148 (P.A. $=$ 5.5$\arcdeg$). Note that the spiral structure is seen in the two channels from 6.08 to 7.42 km s$^{-1}$.}}
    \label{fig:chansosblb}
\end{figure*}

\clearpage


\bibliography{sample631}{}

\begin{thebibliography}{}
\expandafter\ifx\csname natexlab\endcsname\relax\def\natexlab#1{#1}\fi
\providecommand{\url}[1]{\href{#1}{#1}}
\providecommand{\dodoi}[1]{doi:~\href{http://doi.org/#1}{\nolinkurl{#1}}}
\providecommand{\doeprint}[1]{\href{http://ascl.net/#1}{\nolinkurl{http://ascl.net/#1}}}
\providecommand{\doarXiv}[1]{\href{https://arxiv.org/abs/#1}{\nolinkurl{https://arxiv.org/abs/#1}}}

\bibitem[{{Alves} {et~al.}(2017){Alves}, {Girart}, {Caselli}, {Franco}, {Zhao},
  {Vlemmings}, {Evans}, \& {Ricci}}]{2017A&A...603L...3A}
{Alves}, F.~O., {Girart}, J.~M., {Caselli}, P., {et~al.} 2017, \aap, 603, L3,
  \dodoi{10.1051/0004-6361/201731077}

\bibitem[{{Andrews} \& {Williams}(2005)}]{2005ApJ...631.1134A}
{Andrews}, S.~M., \& {Williams}, J.~P. 2005, \apj, 631, 1134,
  \dodoi{10.1086/432712}

\bibitem[{{Andrews} {et~al.}(2011){Andrews}, {Wilner}, {Espaillat}, {Hughes},
  {Dullemond}, {McClure}, {Qi}, \& {Brown}}]{2011ApJ...732...42A}
{Andrews}, S.~M., {Wilner}, D.~J., {Espaillat}, C., {et~al.} 2011, \apj, 732,
  42, \dodoi{10.1088/0004-637X/732/1/42}

\bibitem[{{Andrews} {et~al.}(2009){Andrews}, {Wilner}, {Hughes}, {Qi}, \&
  {Dullemond}}]{2009ApJ...700.1502A}
{Andrews}, S.~M., {Wilner}, D.~J., {Hughes}, A.~M., {Qi}, C., \& {Dullemond},
  C.~P. 2009, \apj, 700, 1502, \dodoi{10.1088/0004-637X/700/2/1502}

\bibitem[{{Andrews} {et~al.}(2018){Andrews}, {Huang}, {P{\'e}rez}, {Isella},
  {Dullemond}, {Kurtovic}, {Guzm{\'a}n}, {Carpenter}, {Wilner}, {Zhang}, {Zhu},
  {Birnstiel}, {Bai}, {Benisty}, {Hughes}, {{\"O}berg}, \&
  {Ricci}}]{2018ApJ...869L..41A}
{Andrews}, S.~M., {Huang}, J., {P{\'e}rez}, L.~M., {et~al.} 2018, \apjl, 869,
  L41, \dodoi{10.3847/2041-8213/aaf741}

\bibitem[{{Artymowicz} \& {Lubow}(1994)}]{1994ApJ...421..651A}
{Artymowicz}, P., \& {Lubow}, S.~H. 1994, \apj, 421, 651,
  \dodoi{10.1086/173679}

\bibitem[{{Aso} \& {Sai}(2024)}]{2024PKAS...39...2A}
{Aso}, Y., \& {Sai}, J. 2024, {PKAS}, {39}, 27,
  \dodoi{10.5303/PKAS.2024.39.2.027}

\bibitem[{{Aso} {et~al.}(2015){Aso}, {Ohashi}, {Saigo}, {Koyamatsu}, {Aikawa},
  {Hayashi}, {Machida}, {Saito}, {Takakuwa}, {Tomida}, {Tomisaka}, \&
  {Yen}}]{2015ApJ...812...27A}
{Aso}, Y., {Ohashi}, N., {Saigo}, K., {et~al.} 2015, \apj, 812, 27,
  \dodoi{10.1088/0004-637X/812/1/27}

\bibitem[{{Aso} {et~al.}(2017){Aso}, {Ohashi}, {Aikawa}, {Machida}, {Saigo},
  {Saito}, {Takakuwa}, {Tomida}, {Tomisaka}, \& {Yen}}]{2017ApJ...849...56A}
{Aso}, Y., {Ohashi}, N., {Aikawa}, Y., {et~al.} 2017, \apj, 849, 56,
  \dodoi{10.3847/1538-4357/aa8264}

\bibitem[{{Aso} {et~al.}(2023){Aso}, {Kwon}, {Ohashi}, {J{\o}rgensen}, {Tobin},
  {Aikawa}, {de Gregorio-Monsalvo}, {Han}, {Kido}, {Koch}, {Lai}, {Lee}, {Lee},
  {Li}, {Lin}, {Looney}, {Narayanan}, {Phuong}, {(Insa Choi)}, {Saigo},
  {Santamar{\'\i}a-Miranda}, {Sharma}, {Takakuwa}, {Thieme}, {Tomida},
  {Williams}, \& {Yen}}]{2023ApJ...954..101A}
{Aso}, Y., {Kwon}, W., {Ohashi}, N., {et~al.} 2023, \apj, 954, 101,
  \dodoi{10.3847/1538-4357/ace624}

\bibitem[{{Bae} {et~al.}(2015){Bae}, {Hartmann}, \&
  {Zhu}}]{2015ApJ...805...15B}
{Bae}, J., {Hartmann}, L., \& {Zhu}, Z. 2015, \apj, 805, 15,
  \dodoi{10.1088/0004-637X/805/1/15}

\bibitem[{{Bae} {et~al.}(2023){Bae}, {Isella}, {Zhu}, {Martin}, {Okuzumi}, \&
  {Suriano}}]{2023ASPC..534..423B}
{Bae}, J., {Isella}, A., {Zhu}, Z., {et~al.} 2023, in Astronomical Society of
  the Pacific Conference Series, Vol. 534, Protostars and Planets VII, ed.
  S.~{Inutsuka}, Y.~{Aikawa}, T.~{Muto}, K.~{Tomida}, \& M.~{Tamura}, 423,
  \dodoi{10.48550/arXiv.2210.13314}

\bibitem[{{Bae} {et~al.}(2018){Bae}, {Pinilla}, \&
  {Birnstiel}}]{2018ApJ...864L..26B}
{Bae}, J., {Pinilla}, P., \& {Birnstiel}, T. 2018, \apjl, 864, L26,
  \dodoi{10.3847/2041-8213/aadd51}

\bibitem[{{Beckwith} \& {Sargent}(1991)}]{1991ApJ...381..250B}
{Beckwith}, S. V.~W., \& {Sargent}, A.~I. 1991, \apj, 381, 250,
  \dodoi{10.1086/170646}

\bibitem[{{Beckwith} {et~al.}(1990){Beckwith}, {Sargent}, {Chini}, \&
  {Guesten}}]{1990AJ.....99..924B}
{Beckwith}, S. V.~W., {Sargent}, A.~I., {Chini}, R.~S., \& {Guesten}, R. 1990,
  \aj, 99, 924, \dodoi{10.1086/115385}

\bibitem[{{Benisty} {et~al.}(2021){Benisty}, {Bae}, {Facchini}, {Keppler},
  {Teague}, {Isella}, {Kurtovic}, {P{\'e}rez}, {Sierra}, {Andrews},
  {Carpenter}, {Czekala}, {Dominik}, {Henning}, {Menard}, {Pinilla}, \&
  {Zurlo}}]{2021ApJ...916L...2B}
{Benisty}, M., {Bae}, J., {Facchini}, S., {et~al.} 2021, \apjl, 916, L2,
  \dodoi{10.3847/2041-8213/ac0f83}

\bibitem[{{Benson} \& {Myers}(1989)}]{1989ApJS...71...89B}
{Benson}, P.~J., \& {Myers}, P.~C. 1989, \apjs, 71, 89, \dodoi{10.1086/191365}

\bibitem[{{Birnstiel} {et~al.}(2018){Birnstiel}, {Dullemond}, {Zhu}, {Andrews},
  {Bai}, {Wilner}, {Carpenter}, {Huang}, {Isella}, {Benisty}, {P{\'e}rez}, \&
  {Zhang}}]{2018ApJ...869L..45B}
{Birnstiel}, T., {Dullemond}, C.~P., {Zhu}, Z., {et~al.} 2018, \apjl, 869, L45,
  \dodoi{10.3847/2041-8213/aaf743}

\bibitem[{{Bontemps} {et~al.}(1996){Bontemps}, {Andre}, {Terebey}, \&
  {Cabrit}}]{1996A&A...311..858B}
{Bontemps}, S., {Andre}, P., {Terebey}, S., \& {Cabrit}, S. 1996, \aap, 311,
  858

\bibitem[{{Cacciapuoti} {et~al.}(2024){Cacciapuoti}, {Macias}, {Gupta},
  {Testi}, {Miotello}, {Espaillat}, {K{\"u}ffmeier}, {van Terwisga}, {Tobin},
  {Grant}, {Manara}, {Segura-Cox}, {Wendeborn}, {Klessen}, {Maury},
  {Lebreuilly}, {Hennebelle}, \& {Molinari}}]{2024A&A...682A..61C}
{Cacciapuoti}, L., {Macias}, E., {Gupta}, A., {et~al.} 2024, \aap, 682, A61,
  \dodoi{10.1051/0004-6361/202347486}

\bibitem[{{CASA Team} {et~al.}(2022){CASA Team}, {Bean}, {Bhatnagar}, {Castro},
  {Donovan Meyer}, {Emonts}, {Garcia}, {Garwood}, {Golap}, {Gonzalez Villalba},
  {Harris}, {Hayashi}, {Hoskins}, {Hsieh}, {Jagannathan}, {Kawasaki},
  {Keimpema}, {Kettenis}, {Lopez}, {Marvil}, {Masters}, {McNichols},
  {Mehringer}, {Miel}, {Moellenbrock}, {Montesino}, {Nakazato}, {Ott}, {Petry},
  {Pokorny}, {Raba}, {Rau}, {Schiebel}, {Schweighart}, {Sekhar}, {Shimada},
  {Small}, {Steeb}, {Sugimoto}, {Suoranta}, {Tsutsumi}, {van Bemmel},
  {Verkouter}, {Wells}, {Xiong}, {Szomoru}, {Griffith}, {Glendenning}, \&
  {Kern}}]{2022PASP..134k4501C}
{CASA Team}, {Bean}, B., {Bhatnagar}, S., {et~al.} 2022, \pasp, 134, 114501,
  \dodoi{10.1088/1538-3873/ac9642}

\bibitem[{{Cassen} \& {Moosman}(1981)}]{1981Icar...48..353C}
{Cassen}, P., \& {Moosman}, A. 1981, \icarus, 48, 353,
  \dodoi{10.1016/0019-1035(81)90051-8}

\bibitem[{{Chiang} {et~al.}(2012){Chiang}, {Looney}, \&
  {Tobin}}]{2012ApJ...756..168C}
{Chiang}, H.-F., {Looney}, L.~W., \& {Tobin}, J.~J. 2012, \apj, 756, 168,
  \dodoi{10.1088/0004-637X/756/2/168}

\bibitem[{{Cieza} {et~al.}(2021){Cieza}, {Gonz{\'a}lez-Ruilova}, {Hales},
  {Pinilla}, {Ru{\'\i}z-Rodr{\'\i}guez}, {Zurlo}, {Casassus}, {P{\'e}rez},
  {C{\'a}novas}, {Arce-Tord}, {Flock}, {Kurtovic}, {Marino}, {Nogueira},
  {Perez}, {Price}, {Principe}, \& {Williams}}]{2021MNRAS.501.2934C}
{Cieza}, L.~A., {Gonz{\'a}lez-Ruilova}, C., {Hales}, A.~S., {et~al.} 2021,
  \mnras, 501, 2934, \dodoi{10.1093/mnras/staa3787}

\bibitem[{{Connelley} {et~al.}(2007){Connelley}, {Reipurth}, \&
  {Tokunaga}}]{2007AJ....133.1528C}
{Connelley}, M.~S., {Reipurth}, B., \& {Tokunaga}, A.~T. 2007, \aj, 133, 1528,
  \dodoi{10.1086/511745}

\bibitem[{{Connelley} {et~al.}(2008){Connelley}, {Reipurth}, \&
  {Tokunaga}}]{2008AJ....135.2496C}
---. 2008, \aj, 135, 2496, \dodoi{10.1088/0004-6256/135/6/2496}

\bibitem[{Cortes {et~al.}(2025)Cortes, Carpenter, Kameno, Loomis, Vila~Vilaro,
  Immer, Vlahakis, Law, Stoehr, Saini, Hales, \&
  Kneissl}]{cortes_2025_14933753}
Cortes, P., Carpenter, J., Kameno, S., {et~al.} 2025, ALMA Cycle 12 Technical
  Handbook, \dodoi{10.5281/zenodo.14933753}

\bibitem[{{de Valon} {et~al.}(2020){de Valon}, {Dougados}, {Cabrit}, {Louvet},
  {Zapata}, \& {Mardones}}]{2020A&A...634L..12D}
{de Valon}, A., {Dougados}, C., {Cabrit}, S., {et~al.} 2020, \aap, 634, L12,
  \dodoi{10.1051/0004-6361/201936950}

\bibitem[{{Draine}(2006)}]{2006ApJ...636.1114D}
{Draine}, B.~T. 2006, \apj, 636, 1114, \dodoi{10.1086/498130}

\bibitem[{{Dullemond} {et~al.}(2019){Dullemond}, {K{\"u}ffmeier}, {Goicovic},
  {Fukagawa}, {Oehl}, \& {Kramer}}]{2019A&A...628A..20D}
{Dullemond}, C.~P., {K{\"u}ffmeier}, M., {Goicovic}, F., {et~al.} 2019, \aap,
  628, A20, \dodoi{10.1051/0004-6361/201832632}

\bibitem[{{Encalada} {et~al.}(2024){Encalada}, {Looney}, {Takakuwa}, {Tobin},
  {Ohashi}, {J{\o}rgensen}, {Li}, {Aikawa}, {Aso}, {Koch}, {Kwon}, {Lai},
  {Lee}, {Lin}, {Santamar{\'\i}a-Miranda}, {de Gregorio-Monsalvo}, {Phuong},
  {Plunkett}, {Sai (Insa Choi)}, {Sharma}, {Yen}, \&
  {Han}}]{2024ApJ...966...32E}
{Encalada}, F.~J., {Looney}, L.~W., {Takakuwa}, S., {et~al.} 2024, \apj, 966,
  32, \dodoi{10.3847/1538-4357/ad3442}

\bibitem[{{Eswaraiah} {et~al.}(2021){Eswaraiah}, {Li}, {Furuya}, {Hasegawa},
  {Ward-Thompson}, {Qiu}, {Ohashi}, {Pattle}, {Sadavoy}, {Hull}, {Berry},
  {Doi}, {Ching}, {Lai}, {Wang}, {Koch}, {Kwon}, {Kwon}, {Bastien},
  {Arzoumanian}, {Coud{\'e}}, {Soam}, {Fanciullo}, {Yen}, {Liu}, {Hoang}, {Ping
  Chen}, {Shimajiri}, {Liu}, {Chen}, {Li}, {Lyo}, {Hwang}, {Johnstone}, {Rao},
  {Bich Ngoc}, {Ngoc Diep}, {Mairs}, {Parsons}, {Tamura}, {Tahani}, {Vivien
  Chen}, {Nakamura}, {Shinnaga}, {Tang}, {Cho}, {Won Lee}, {Inutsuka}, {Inoue},
  {Iwasaki}, {Qian}, {Xie}, {Li}, {Liu}, {Zhang}, {Chen}, {Zhang}, {Zhu},
  {Zhou}, {Andr{\'e}}, {Liu}, {Yuan}, {Lu}, {Peretto}, {Bourke}, {Byun}, {Dai},
  {Duan}, {Duan}, {Eden}, {Matthews}, {Fiege}, {Fissel}, {Kim}, {Lee}, {Kim},
  {Pyo}, {Choi}, {Choi}, {Chrysostomou}, {Jung Chung}, {Ngoc Tram},
  {Franzmann}, {Friberg}, {Friesen}, {Fuller}, {Gledhill}, {Graves}, {Greaves},
  {Griffin}, {Gu}, {Han}, {Hatchell}, {Hayashi}, {Houde}, {Kawabata}, {Jeong},
  {Kang}, {Kang}, {Kang}, {Kataoka}, {Kemper}, {Rawlings}, {Rawlings},
  {Retter}, {Richer}, {Rigby}, {Saito}, {Savini}, {Scaife}, {Seta}, {Kim}, {Hee
  Kim}, {Kim}, {Kirchschlager}, {Kirk}, {Kobayashi}, {Konyves}, {Kusune},
  {Lacaille}, {Law}, {Lee}, {Lee}, {Matsumura}, {Moriarty-Schieven}, {Nagata},
  {Nakanishi}, {Onaka}, {Park}, {Tang}, {Tomisaka}, {Tsukamoto}, {Viti},
  {Wang}, {Whitworth}, {Yoo}, {Yun}, {Zenko}, {Zhang}, {de Looze}, {Dowell},
  {Eyres}, {Falle}, {Robitaille}, \& {van Loo}}]{2021ApJ...912L..27E}
{Eswaraiah}, C., {Li}, D., {Furuya}, R.~S., {et~al.} 2021, \apjl, 912, L27,
  \dodoi{10.3847/2041-8213/abeb1c}

\bibitem[{{Flores} {et~al.}(2023){Flores}, {Ohashi}, {Tobin}, {J{\o}rgensen},
  {Takakuwa}, {Li}, {Lin}, {van't Hoff}, {Plunkett}, {Yamato}, {Sai (Insa
  Choi)}, {Koch}, {Yen}, {Aikawa}, {Aso}, {de Gregorio-Monsalvo}, {Kido},
  {Kwon}, {Lee}, {Lee}, {Looney}, {Santamar{\'\i}a-Miranda}, {Sharma},
  {Thieme}, {Williams}, {Han}, {Narayanan}, \& {Lai}}]{2023ApJ...958...98F}
{Flores}, C., {Ohashi}, N., {Tobin}, J.~J., {et~al.} 2023, \apj, 958, 98,
  \dodoi{10.3847/1538-4357/acf7c1}

\bibitem[{{Gaia Collaboration} {et~al.}(2021){Gaia Collaboration}, {Brown},
  {Vallenari}, {Prusti}, {de Bruijne}, {Babusiaux}, {Biermann}, {Creevey},
  {Evans}, {Eyer}, {Hutton}, {Jansen}, {Jordi}, {Klioner}, {Lammers},
  {Lindegren}, {Luri}, {Mignard}, {Panem}, {Pourbaix}, {Randich}, {Sartoretti},
  {Soubiran}, {Walton}, {Arenou}, {Bailer-Jones}, {Bastian}, {Cropper},
  {Drimmel}, {Katz}, {Lattanzi}, {van Leeuwen}, {Bakker}, {Cacciari},
  {Casta{\~n}eda}, {De Angeli}, {Ducourant}, {Fabricius}, {Fouesneau},
  {Fr{\'e}mat}, {Guerra}, {Guerrier}, {Guiraud}, {Jean-Antoine Piccolo},
  {Masana}, {Messineo}, {Mowlavi}, {Nicolas}, {Nienartowicz}, {Pailler},
  {Panuzzo}, {Riclet}, {Roux}, {Seabroke}, {Sordo}, {Tanga}, {Th{\'e}venin},
  {Gracia-Abril}, {Portell}, {Teyssier}, {Altmann}, {Andrae}, {Bellas-Velidis},
  {Benson}, {Berthier}, {Blomme}, {Brugaletta}, {Burgess}, {Busso}, {Carry},
  {Cellino}, {Cheek}, {Clementini}, {Damerdji}, {Davidson}, {Delchambre},
  {Dell'Oro}, {Fern{\'a}ndez-Hern{\'a}ndez}, {Galluccio}, {Garc{\'\i}a-Lario},
  {Garcia-Reinaldos}, {Gonz{\'a}lez-N{\'u}{\~n}ez}, {Gosset}, {Haigron},
  {Halbwachs}, {Hambly}, {Harrison}, {Hatzidimitriou}, {Heiter},
  {Hern{\'a}ndez}, {Hestroffer}, {Hodgkin}, {Holl}, {Jan{\ss}en}, {Jevardat de
  Fombelle}, {Jordan}, {Krone-Martins}, {Lanzafame}, {L{\"o}ffler}, {Lorca},
  {Manteiga}, {Marchal}, {Marrese}, {Moitinho}, {Mora}, {Muinonen}, {Osborne},
  {Pancino}, {Pauwels}, {Petit}, {Recio-Blanco}, {Richards}, {Riello},
  {Rimoldini}, {Robin}, {Roegiers}, {Rybizki}, {Sarro}, {Siopis}, {Smith},
  {Sozzetti}, {Ulla}, {Utrilla}, {van Leeuwen}, {van Reeven}, {Abbas}, {Abreu
  Aramburu}, {Accart}, {Aerts}, {Aguado}, {Ajaj}, {Altavilla}, {{\'A}lvarez},
  {{\'A}lvarez Cid-Fuentes}, {Alves}, {Anderson}, {Anglada Varela}, {Antoja},
  {Audard}, {Baines}, {Baker}, {Balaguer-N{\'u}{\~n}ez}, {Balbinot}, {Balog},
  {Barache}, {Barbato}, {Barros}, {Barstow}, {Bartolom{\'e}}, {Bassilana},
  {Bauchet}, {Baudesson-Stella}, {Becciani}, {Bellazzini}, {Bernet}, {Bertone},
  {Bianchi}, {Blanco-Cuaresma}, {Boch}, {Bombrun}, {Bossini}, {Bouquillon},
  {Bragaglia}, {Bramante}, {Breedt}, {Bressan}, {Brouillet}, {Bucciarelli},
  {Burlacu}, {Busonero}, {Butkevich}, {Buzzi}, {Caffau}, {Cancelliere},
  {C{\'a}novas}, {Cantat-Gaudin}, {Carballo}, {Carlucci}, {Carnerero},
  {Carrasco}, {Casamiquela}, {Castellani}, {Castro-Ginard}, {Castro Sampol},
  {Chaoul}, {Charlot}, {Chemin}, {Chiavassa}, {Cioni}, {Comoretto}, {Cooper},
  {Cornez}, {Cowell}, {Crifo}, {Crosta}, {Crowley}, {Dafonte}, {Dapergolas},
  {David}, {David}, {de Laverny}, {De Luise}, {De March}, {De Ridder}, {de
  Souza}, {de Teodoro}, {de Torres}, {del Peloso}, {del Pozo}, {Delbo},
  {Delgado}, {Delgado}, {Delisle}, {Di Matteo}, {Diakite}, {Diener},
  {Distefano}, {Dolding}, {Eappachen}, {Edvardsson}, {Enke}, {Esquej}, {Fabre},
  {Fabrizio}, {Faigler}, {Fedorets}, {Fernique}, {Fienga}, {Figueras},
  {Fouron}, {Fragkoudi}, {Fraile}, {Franke}, {Gai}, {Garabato},
  {Garcia-Gutierrez}, {Garc{\'\i}a-Torres}, {Garofalo}, {Gavras}, {Gerlach},
  {Geyer}, {Giacobbe}, {Gilmore}, {Girona}, {Giuffrida}, {Gomel}, {Gomez},
  {Gonzalez-Santamaria}, {Gonz{\'a}lez-Vidal}, {Granvik},
  {Guti{\'e}rrez-S{\'a}nchez}, {Guy}, {Hauser}, {Haywood}, {Helmi}, {Hidalgo},
  {Hilger}, {H{\l}adczuk}, {Hobbs}, {Holland}, {Huckle}, {Jasniewicz},
  {Jonker}, {Juaristi Campillo}, {Julbe}, {Karbevska}, {Kervella}, {Khanna},
  {Kochoska}, {Kontizas}, {Kordopatis}, {Korn}, {Kostrzewa-Rutkowska},
  {Kruszy{\'n}ska}, {Lambert}, {Lanza}, {Lasne}, {Le Campion}, {Le Fustec},
  {Lebreton}, {Lebzelter}, {Leccia}, {Leclerc}, {Lecoeur-Taibi}, {Liao},
  {Licata}, {Lindstr{\o}m}, {Lister}, {Livanou}, {Lobel}, {Madrero Pardo},
  {Managau}, {Mann}, {Marchant}, {Marconi}, {Marcos Santos}, {Marinoni},
  {Marocco}, {Marshall}, {Martin Polo}, {Mart{\'\i}n-Fleitas}, {Masip},
  {Massari}, {Mastrobuono-Battisti}, {Mazeh}, {McMillan}, {Messina},
  {Michalik}, {Millar}, {Mints}, {Molina}, {Molinaro}, {Moln{\'a}r},
  {Montegriffo}, {Mor}, {Morbidelli}, {Morel}, {Morris}, {Mulone}, {Munoz},
  {Muraveva}, {Murphy}, {Musella}, {Noval}, {Ord{\'e}novic}, {Orr{\`u}},
  {Osinde}, {Pagani}, {Pagano}, {Palaversa}, {Palicio}, {Panahi}, {Pawlak},
  {Pe{\~n}alosa Esteller}, {Penttil{\"a}}, {Piersimoni}, {Pineau}, {Plachy},
  {Plum}, {Poggio}, {Poretti}, {Poujoulet}, {Pr{\v{s}}a}, {Pulone}, {Racero},
  {Ragaini}, {Rainer}, {Raiteri}, {Rambaux}, {Ramos}, {Ramos-Lerate}, {Re
  Fiorentin}, {Regibo}, {Reyl{\'e}}, {Ripepi}, {Riva}, {Rixon}, {Robichon},
  {Robin}, {Roelens}, {Rohrbasser}, {Romero-G{\'o}mez}, {Rowell}, {Royer},
  {Rybicki}, {Sadowski}, {Sagrist{\`a} Sell{\'e}s}, {Sahlmann}, {Salgado},
  {Salguero}, {Samaras}, {Sanchez Gimenez}, {Sanna}, {Santove{\~n}a},
  {Sarasso}, {Schultheis}, {Sciacca}, {Segol}, {Segovia}, {S{\'e}gransan},
  {Semeux}, {Shahaf}, {Siddiqui}, {Siebert}, {Siltala}, {Slezak}, {Smart},
  {Solano}, {Solitro}, {Souami}, {Souchay}, {Spagna}, {Spoto}, {Steele},
  {Steidelm{\"u}ller}, {Stephenson}, {S{\"u}veges}, {Szabados}, {Szegedi-Elek},
  {Taris}, {Tauran}, {Taylor}, {Teixeira}, {Thuillot}, {Tonello}, {Torra},
  {Torra}, {Turon}, {Unger}, {Vaillant}, {van Dillen}, {Vanel}, {Vecchiato},
  {Viala}, {Vicente}, {Voutsinas}, {Weiler}, {Wevers}, {Wyrzykowski}, {Yoldas},
  {Yvard}, {Zhao}, {Zorec}, {Zucker}, {Zurbach}, \&
  {Zwitter}}]{2021A&A...649A...1G}
{Gaia Collaboration}, {Brown}, A.~G.~A., {Vallenari}, A., {et~al.} 2021, \aap,
  649, A1, \dodoi{10.1051/0004-6361/202039657}

\bibitem[{{Gaia Collaboration} {et~al.}(2023){Gaia Collaboration}, {Vallenari},
  {Brown}, {Prusti}, {de Bruijne}, {Arenou}, {Babusiaux}, {Biermann},
  {Creevey}, {Ducourant}, {Evans}, {Eyer}, {Guerra}, {Hutton}, {Jordi},
  {Klioner}, {Lammers}, {Lindegren}, {Luri}, {Mignard}, {Panem}, {Pourbaix},
  {Randich}, {Sartoretti}, {Soubiran}, {Tanga}, {Walton}, {Bailer-Jones},
  {Bastian}, {Drimmel}, {Jansen}, {Katz}, {Lattanzi}, {van Leeuwen}, {Bakker},
  {Cacciari}, {Casta{\~n}eda}, {De Angeli}, {Fabricius}, {Fouesneau},
  {Fr{\'e}mat}, {Galluccio}, {Guerrier}, {Heiter}, {Masana}, {Messineo},
  {Mowlavi}, {Nicolas}, {Nienartowicz}, {Pailler}, {Panuzzo}, {Riclet}, {Roux},
  {Seabroke}, {Sordo}, {Th{\'e}venin}, {Gracia-Abril}, {Portell}, {Teyssier},
  {Altmann}, {Andrae}, {Audard}, {Bellas-Velidis}, {Benson}, {Berthier},
  {Blomme}, {Burgess}, {Busonero}, {Busso}, {C{\'a}novas}, {Carry}, {Cellino},
  {Cheek}, {Clementini}, {Damerdji}, {Davidson}, {de Teodoro}, {Nu{\~n}ez
  Campos}, {Delchambre}, {Dell'Oro}, {Esquej}, {Fern{\'a}ndez-Hern{\'a}ndez},
  {Fraile}, {Garabato}, {Garc{\'\i}a-Lario}, {Gosset}, {Haigron}, {Halbwachs},
  {Hambly}, {Harrison}, {Hern{\'a}ndez}, {Hestroffer}, {Hodgkin}, {Holl},
  {Jan{\ss}en}, {Jevardat de Fombelle}, {Jordan}, {Krone-Martins}, {Lanzafame},
  {L{\"o}ffler}, {Marchal}, {Marrese}, {Moitinho}, {Muinonen}, {Osborne},
  {Pancino}, {Pauwels}, {Recio-Blanco}, {Reyl{\'e}}, {Riello}, {Rimoldini},
  {Roegiers}, {Rybizki}, {Sarro}, {Siopis}, {Smith}, {Sozzetti}, {Utrilla},
  {van Leeuwen}, {Abbas}, {{\'A}brah{\'a}m}, {Abreu Aramburu}, {Aerts},
  {Aguado}, {Ajaj}, {Aldea-Montero}, {Altavilla}, {{\'A}lvarez}, {Alves},
  {Anders}, {Anderson}, {Anglada Varela}, {Antoja}, {Baines}, {Baker},
  {Balaguer-N{\'u}{\~n}ez}, {Balbinot}, {Balog}, {Barache}, {Barbato},
  {Barros}, {Barstow}, {Bartolom{\'e}}, {Bassilana}, {Bauchet}, {Becciani},
  {Bellazzini}, {Berihuete}, {Bernet}, {Bertone}, {Bianchi}, {Binnenfeld},
  {Blanco-Cuaresma}, {Blazere}, {Boch}, {Bombrun}, {Bossini}, {Bouquillon},
  {Bragaglia}, {Bramante}, {Breedt}, {Bressan}, {Brouillet}, {Brugaletta},
  {Bucciarelli}, {Burlacu}, {Butkevich}, {Buzzi}, {Caffau}, {Cancelliere},
  {Cantat-Gaudin}, {Carballo}, {Carlucci}, {Carnerero}, {Carrasco},
  {Casamiquela}, {Castellani}, {Castro-Ginard}, {Chaoul}, {Charlot}, {Chemin},
  {Chiaramida}, {Chiavassa}, {Chornay}, {Comoretto}, {Contursi}, {Cooper},
  {Cornez}, {Cowell}, {Crifo}, {Cropper}, {Crosta}, {Crowley}, {Dafonte},
  {Dapergolas}, {David}, {David}, {de Laverny}, {De Luise}, {De March}, {De
  Ridder}, {de Souza}, {de Torres}, {del Peloso}, {del Pozo}, {Delbo},
  {Delgado}, {Delisle}, {Demouchy}, {Dharmawardena}, {Di Matteo}, {Diakite},
  {Diener}, {Distefano}, {Dolding}, {Edvardsson}, {Enke}, {Fabre}, {Fabrizio},
  {Faigler}, {Fedorets}, {Fernique}, {Fienga}, {Figueras}, {Fournier},
  {Fouron}, {Fragkoudi}, {Gai}, {Garcia-Gutierrez}, {Garcia-Reinaldos},
  {Garc{\'\i}a-Torres}, {Garofalo}, {Gavel}, {Gavras}, {Gerlach}, {Geyer},
  {Giacobbe}, {Gilmore}, {Girona}, {Giuffrida}, {Gomel}, {Gomez},
  {Gonz{\'a}lez-N{\'u}{\~n}ez}, {Gonz{\'a}lez-Santamar{\'\i}a},
  {Gonz{\'a}lez-Vidal}, {Granvik}, {Guillout}, {Guiraud},
  {Guti{\'e}rrez-S{\'a}nchez}, {Guy}, {Hatzidimitriou}, {Hauser}, {Haywood},
  {Helmer}, {Helmi}, {Sarmiento}, {Hidalgo}, {Hilger}, {H{\l}adczuk}, {Hobbs},
  {Holland}, {Huckle}, {Jardine}, {Jasniewicz}, {Jean-Antoine Piccolo},
  {Jim{\'e}nez-Arranz}, {Jorissen}, {Juaristi Campillo}, {Julbe}, {Karbevska},
  {Kervella}, {Khanna}, {Kontizas}, {Kordopatis}, {Korn}, {K{\'o}sp{\'a}l},
  {Kostrzewa-Rutkowska}, {Kruszy{\'n}ska}, {Kun}, {Laizeau}, {Lambert},
  {Lanza}, {Lasne}, {Le Campion}, {Lebreton}, {Lebzelter}, {Leccia}, {Leclerc},
  {Lecoeur-Taibi}, {Liao}, {Licata}, {Lindstr{\o}m}, {Lister}, {Livanou},
  {Lobel}, {Lorca}, {Loup}, {Madrero Pardo}, {Magdaleno Romeo}, {Managau},
  {Mann}, {Manteiga}, {Marchant}, {Marconi}, {Marcos}, {Marcos Santos},
  {Mar{\'\i}n Pina}, {Marinoni}, {Marocco}, {Marshall}, {Martin Polo},
  {Mart{\'\i}n-Fleitas}, {Marton}, {Mary}, {Masip}, {Massari},
  {Mastrobuono-Battisti}, {Mazeh}, {McMillan}, {Messina}, {Michalik}, {Millar},
  {Mints}, {Molina}, {Molinaro}, {Moln{\'a}r}, {Monari}, {Mongui{\'o}},
  {Montegriffo}, {Montero}, {Mor}, {Mora}, {Morbidelli}, {Morel}, {Morris},
  {Muraveva}, {Murphy}, {Musella}, {Nagy}, {Noval}, {Oca{\~n}a}, {Ogden},
  {Ordenovic}, {Osinde}, {Pagani}, {Pagano}, {Palaversa}, {Palicio},
  {Pallas-Quintela}, {Panahi}, {Payne-Wardenaar}, {Pe{\~n}alosa Esteller},
  {Penttil{\"a}}, {Pichon}, {Piersimoni}, {Pineau}, {Plachy}, {Plum}, {Poggio},
  {Pr{\v{s}}a}, {Pulone}, {Racero}, {Ragaini}, {Rainer}, {Raiteri}, {Rambaux},
  {Ramos}, {Ramos-Lerate}, {Re Fiorentin}, {Regibo}, {Richards}, {Rios Diaz},
  {Ripepi}, {Riva}, {Rix}, {Rixon}, {Robichon}, {Robin}, {Robin}, {Roelens},
  {Rogues}, {Rohrbasser}, {Romero-G{\'o}mez}, {Rowell}, {Royer}, {Ruz Mieres},
  {Rybicki}, {Sadowski}, {S{\'a}ez N{\'u}{\~n}ez}, {Sagrist{\`a} Sell{\'e}s},
  {Sahlmann}, {Salguero}, {Samaras}, {Sanchez Gimenez}, {Sanna},
  {Santove{\~n}a}, {Sarasso}, {Schultheis}, {Sciacca}, {Segol}, {Segovia},
  {S{\'e}gransan}, {Semeux}, {Shahaf}, {Siddiqui}, {Siebert}, {Siltala},
  {Silvelo}, {Slezak}, {Slezak}, {Smart}, {Snaith}, {Solano}, {Solitro},
  {Souami}, {Souchay}, {Spagna}, {Spina}, {Spoto}, {Steele},
  {Steidelm{\"u}ller}, {Stephenson}, {S{\"u}veges}, {Surdej}, {Szabados},
  {Szegedi-Elek}, {Taris}, {Taylor}, {Teixeira}, {Tolomei}, {Tonello}, {Torra},
  {Torra}, {Torralba Elipe}, {Trabucchi}, {Tsounis}, {Turon}, {Ulla}, {Unger},
  {Vaillant}, {van Dillen}, {van Reeven}, {Vanel}, {Vecchiato}, {Viala},
  {Vicente}, {Voutsinas}, {Weiler}, {Wevers}, {Wyrzykowski}, {Yoldas}, {Yvard},
  {Zhao}, {Zorec}, {Zucker}, \& {Zwitter}}]{2023A&A...674A...1G}
{Gaia Collaboration}, {Vallenari}, A., {Brown}, A.~G.~A., {et~al.} 2023, \aap,
  674, A1, \dodoi{10.1051/0004-6361/202243940}

\bibitem[{{Galametz} {et~al.}(2019){Galametz}, {Maury}, {Valdivia}, {Testi},
  {Belloche}, \& {Andr{\'e}}}]{2019A&A...632A...5G}
{Galametz}, M., {Maury}, A.~J., {Valdivia}, V., {et~al.} 2019, \aap, 632, A5,
  \dodoi{10.1051/0004-6361/201936342}

\bibitem[{{Galli} {et~al.}(2019){Galli}, {Loinard}, {Bouy}, {Sarro},
  {Ortiz-Le{\'o}n}, {Dzib}, {Olivares}, {Heyer}, {Hernandez},
  {Rom{\'a}n-Z{\'u}{\~n}iga}, {Kounkel}, \& {Covey}}]{2019A&A...630A.137G}
{Galli}, P.~A.~B., {Loinard}, L., {Bouy}, H., {et~al.} 2019, \aap, 630, A137,
  \dodoi{10.1051/0004-6361/201935928}

\bibitem[{{Garufi} {et~al.}(2022){Garufi}, {Podio}, {Codella}, {Segura-Cox},
  {Vander Donckt}, {Mercimek}, {Bacciotti}, {Fedele}, {Kasper}, {Pineda},
  {Humphreys}, \& {Testi}}]{2022A&A...658A.104G}
{Garufi}, A., {Podio}, L., {Codella}, C., {et~al.} 2022, \aap, 658, A104,
  \dodoi{10.1051/0004-6361/202141264}

\bibitem[{{Ginsburg} {et~al.}(2018){Ginsburg}, {Bally}, {Goddi}, {Plambeck}, \&
  {Wright}}]{2018ApJ...860..119G}
{Ginsburg}, A., {Bally}, J., {Goddi}, C., {Plambeck}, R., \& {Wright}, M. 2018,
  \apj, 860, 119, \dodoi{10.3847/1538-4357/aac205}

\bibitem[{{Gomez} {et~al.}(1997){Gomez}, {Whitney}, \&
  {Kenyon}}]{1997AJ....114.1138G}
{Gomez}, M., {Whitney}, B.~A., \& {Kenyon}, S.~J. 1997, \aj, 114, 1138,
  \dodoi{10.1086/118545}

\bibitem[{{Hacar} {et~al.}(2013){Hacar}, {Tafalla}, {Kauffmann}, \&
  {Kov{\'a}cs}}]{2013A&A...554A..55H}
{Hacar}, A., {Tafalla}, M., {Kauffmann}, J., \& {Kov{\'a}cs}, A. 2013, \aap,
  554, A55, \dodoi{10.1051/0004-6361/201220090}

\bibitem[{{Han} {et~al.}(2023){Han}, {Kwon}, {Aso}, {Bae}, \&
  {Sheehan}}]{2023ApJ...956....9H}
{Han}, I., {Kwon}, W., {Aso}, Y., {Bae}, J., \& {Sheehan}, P. 2023, \apj, 956,
  9, \dodoi{10.3847/1538-4357/acf853}

\bibitem[{{Hanawa} {et~al.}(2022){Hanawa}, {Sakai}, \&
  {Yamamoto}}]{2022ApJ...932..122H}
{Hanawa}, T., {Sakai}, N., \& {Yamamoto}, S. 2022, \apj, 932, 122,
  \dodoi{10.3847/1538-4357/ac6e6a}

\bibitem[{{Harsono} {et~al.}(2014){Harsono}, {J{\o}rgensen}, {van Dishoeck},
  {Hogerheijde}, {Bruderer}, {Persson}, \& {Mottram}}]{2014A&A...562A..77H}
{Harsono}, D., {J{\o}rgensen}, J.~K., {van Dishoeck}, E.~F., {et~al.} 2014,
  \aap, 562, A77, \dodoi{10.1051/0004-6361/201322646}

\bibitem[{{Hildebrand}(1983)}]{1983QJRAS..24..267H}
{Hildebrand}, R.~H. 1983, \qjras, 24, 267

\bibitem[{{Hirano} {et~al.}(2020){Hirano}, {Tsukamoto}, {Basu}, \&
  {Machida}}]{2020ApJ...898..118H}
{Hirano}, S., {Tsukamoto}, Y., {Basu}, S., \& {Machida}, M.~N. 2020, \apj, 898,
  118, \dodoi{10.3847/1538-4357/ab9f9d}

\bibitem[{{Hsieh} {et~al.}(2023){Hsieh}, {Segura-Cox}, {Pineda}, {Caselli},
  {Bouscasse}, {Neri}, {Lopez-Sepulcre}, {Valdivia-Mena}, {Maureira},
  {Henning}, {Smirnov-Pinchukov}, {Semenov}, {M{\"o}ller}, {Cunningham},
  {Fuente}, {Marino}, {Dutrey}, {Tafalla}, {Chapillon}, {Ceccarelli}, \&
  {Zhao}}]{2023A&A...669A.137H}
{Hsieh}, T.~H., {Segura-Cox}, D.~M., {Pineda}, J.~E., {et~al.} 2023, \aap, 669,
  A137, \dodoi{10.1051/0004-6361/202244183}

\bibitem[{{Isella} {et~al.}(2019){Isella}, {Benisty}, {Teague}, {Bae},
  {Keppler}, {Facchini}, \& {P{\'e}rez}}]{2019ApJ...879L..25I}
{Isella}, A., {Benisty}, M., {Teague}, R., {et~al.} 2019, \apjl, 879, L25,
  \dodoi{10.3847/2041-8213/ab2a12}

\bibitem[{{J{\o}rgensen} {et~al.}(2022){J{\o}rgensen}, {Kuruwita}, {Harsono},
  {Haugb{\o}lle}, {Kristensen}, \& {Bergin}}]{2022Natur.606..272J}
{J{\o}rgensen}, J.~K., {Kuruwita}, R.~L., {Harsono}, D., {et~al.} 2022, \nat,
  606, 272, \dodoi{10.1038/s41586-022-04659-4}

\bibitem[{{J{\o}rgensen} {et~al.}(2009){J{\o}rgensen}, {van Dishoeck},
  {Visser}, {Bourke}, {Wilner}, {Lommen}, {Hogerheijde}, \&
  {Myers}}]{2009A&A...507..861J}
{J{\o}rgensen}, J.~K., {van Dishoeck}, E.~F., {Visser}, R., {et~al.} 2009,
  \aap, 507, 861, \dodoi{10.1051/0004-6361/200912325}

\bibitem[{{J{\o}rgensen} {et~al.}(2007){J{\o}rgensen}, {Bourke}, {Myers}, {Di
  Francesco}, {van Dishoeck}, {Lee}, {Ohashi}, {Sch{\"o}ier}, {Takakuwa},
  {Wilner}, \& {Zhang}}]{2007ApJ...659..479J}
{J{\o}rgensen}, J.~K., {Bourke}, T.~L., {Myers}, P.~C., {et~al.} 2007, \apj,
  659, 479, \dodoi{10.1086/512230}

\bibitem[{{Kenyon} {et~al.}(1993{\natexlab{a}}){Kenyon}, {Calvet}, \&
  {Hartmann}}]{1993ApJ...414..676K}
{Kenyon}, S.~J., {Calvet}, N., \& {Hartmann}, L. 1993{\natexlab{a}}, \apj, 414,
  676, \dodoi{10.1086/173114}

\bibitem[{{Kenyon} \& {Hartmann}(1995)}]{1995ApJS..101..117K}
{Kenyon}, S.~J., \& {Hartmann}, L. 1995, \apjs, 101, 117,
  \dodoi{10.1086/192235}

\bibitem[{{Kenyon} {et~al.}(1990){Kenyon}, {Hartmann}, {Strom}, \&
  {Strom}}]{1990AJ.....99..869K}
{Kenyon}, S.~J., {Hartmann}, L.~W., {Strom}, K.~M., \& {Strom}, S.~E. 1990,
  \aj, 99, 869, \dodoi{10.1086/115380}

\bibitem[{{Kenyon} {et~al.}(1993{\natexlab{b}}){Kenyon}, {Whitney}, {Gomez}, \&
  {Hartmann}}]{1993ApJ...414..773K}
{Kenyon}, S.~J., {Whitney}, B.~A., {Gomez}, M., \& {Hartmann}, L.
  1993{\natexlab{b}}, \apj, 414, 773, \dodoi{10.1086/173122}

\bibitem[{{Kido} {et~al.}(2023){Kido}, {Takakuwa}, {Saigo}, {Ohashi}, {Tobin},
  {J{\o}rgensen}, {Aikawa}, {Aso}, {Encalada}, {Flores}, {Gavino}, {de
  Gregorio-Monsalvo}, {Han}, {Hirano}, {Koch}, {Kwon}, {Lai}, {Lee}, {Lee},
  {Li}, {Lin}, {Looney}, {Mori}, {Narayanan}, {Plunkett}, {Phuong}, {(Insa
  Choi)}, {Santamar{\'\i}a-Miranda}, {Sharma}, {Sheehan}, {Thieme}, {Tomida},
  {van't Hoff}, {Williams}, {Yamato}, \& {Yen}}]{2023ApJ...953..190K}
{Kido}, M., {Takakuwa}, S., {Saigo}, K., {et~al.} 2023, \apj, 953, 190,
  \dodoi{10.3847/1538-4357/acdd7a}

\bibitem[{{Kido} {et~al.}(2025){Kido}, {Yen}, {Sai}, {Takakuwa}, {Ohashi},
  {Aikawa}, {Aso}, {Flores}, {Han}, {Koch}, {Kwon}, {Lee}, {Li}, {Looney},
  {Narang}, {Saigo}, {Sharma}, {Thieme}, {Tomida}, \&
  {Williams}}]{2025arXiv250400495K}
{Kido}, M., {Yen}, H.-W., {Sai}, J., {et~al.} 2025, arXiv e-prints,
  arXiv:2504.00495, \dodoi{10.48550/arXiv.2504.00495}

\bibitem[{{Krolikowski} {et~al.}(2021){Krolikowski}, {Kraus}, \&
  {Rizzuto}}]{2021AJ....162..110K}
{Krolikowski}, D.~M., {Kraus}, A.~L., \& {Rizzuto}, A.~C. 2021, \aj, 162, 110,
  \dodoi{10.3847/1538-3881/ac0632}

\bibitem[{{Kuffmeier} {et~al.}(2019){Kuffmeier}, {Calcutt}, \&
  {Kristensen}}]{2019A&A...628A.112K}
{Kuffmeier}, M., {Calcutt}, H., \& {Kristensen}, L.~E. 2019, \aap, 628, A112,
  \dodoi{10.1051/0004-6361/201935504}

\bibitem[{{Kuffmeier} {et~al.}(2020){Kuffmeier}, {Goicovic}, \&
  {Dullemond}}]{2020A&A...633A...3K}
{Kuffmeier}, M., {Goicovic}, F.~G., \& {Dullemond}, C.~P. 2020, \aap, 633, A3,
  \dodoi{10.1051/0004-6361/201936820}

\bibitem[{{Kuffmeier} {et~al.}(2017){Kuffmeier}, {Haugb{\o}lle}, \&
  {Nordlund}}]{2017ApJ...846....7K}
{Kuffmeier}, M., {Haugb{\o}lle}, T., \& {Nordlund}, {\r{A}}. 2017, \apj, 846,
  7, \dodoi{10.3847/1538-4357/aa7c64}

\bibitem[{{Kuruwita} {et~al.}(2017){Kuruwita}, {Federrath}, \&
  {Ireland}}]{2017MNRAS.470.1626K}
{Kuruwita}, R.~L., {Federrath}, C., \& {Ireland}, M. 2017, \mnras, 470, 1626,
  \dodoi{10.1093/mnras/stx1299}

\bibitem[{{Kuznetsova} {et~al.}(2022){Kuznetsova}, {Bae}, {Hartmann}, \& {Mac
  Low}}]{2022ApJ...928...92K}
{Kuznetsova}, A., {Bae}, J., {Hartmann}, L., \& {Mac Low}, M.-M. 2022, \apj,
  928, 92, \dodoi{10.3847/1538-4357/ac54a8}

\bibitem[{{Kwon} {et~al.}(2009){Kwon}, {Looney}, {Mundy}, {Chiang}, \&
  {Kemball}}]{2009ApJ...696..841K}
{Kwon}, W., {Looney}, L.~W., {Mundy}, L.~G., {Chiang}, H.-F., \& {Kemball},
  A.~J. 2009, \apj, 696, 841, \dodoi{10.1088/0004-637X/696/1/841}

\bibitem[{{Kwon} {et~al.}(2015){Kwon}, {Looney}, {Mundy}, \&
  {Welch}}]{2015ApJ...808..102K}
{Kwon}, W., {Looney}, L.~W., {Mundy}, L.~G., \& {Welch}, W.~J. 2015, \apj, 808,
  102, \dodoi{10.1088/0004-637X/808/1/102}

\bibitem[{{Kwon} {et~al.}(2019){Kwon}, {Stephens}, {Tobin}, {Looney}, {Li},
  {van der Tak}, \& {Crutcher}}]{2019ApJ...879...25K}
{Kwon}, W., {Stephens}, I.~W., {Tobin}, J.~J., {et~al.} 2019, \apj, 879, 25,
  \dodoi{10.3847/1538-4357/ab24c8}

\bibitem[{{Langer} {et~al.}(1995){Langer}, {Velusamy}, {Kuiper}, {Levin},
  {Olsen}, \& {Migenes}}]{1995ApJ...453..293L}
{Langer}, W.~D., {Velusamy}, T., {Kuiper}, T.~B.~H., {et~al.} 1995, \apj, 453,
  293, \dodoi{10.1086/176389}

\bibitem[{{Lee} {et~al.}(2019){Lee}, {Offner}, {Kratter}, {Smullen}, \&
  {Li}}]{2019ApJ...887..232L}
{Lee}, A.~T., {Offner}, S. S.~R., {Kratter}, K.~M., {Smullen}, R.~A., \& {Li},
  P.~S. 2019, \apj, 887, 232, \dodoi{10.3847/1538-4357/ab584b}

\bibitem[{{Lee} {et~al.}(2023){Lee}, {Matsumoto}, {Kim}, {Lee}, {Harsono},
  {Bae}, {Evans}, {Inutsuka}, {Choi}, {Tatematsu}, {Lee}, \&
  {Jaffe}}]{2023ApJ...953...82L}
{Lee}, J.-E., {Matsumoto}, T., {Kim}, H.-J., {et~al.} 2023, \apj, 953, 82,
  \dodoi{10.3847/1538-4357/acdd5b}

\bibitem[{{Lee} {et~al.}(2020){Lee}, {Lee}, {Aikawa}, {Herczeg}, \&
  {Johnstone}}]{2020ApJ...889...20L}
{Lee}, S., {Lee}, J.-E., {Aikawa}, Y., {Herczeg}, G., \& {Johnstone}, D. 2020,
  \apj, 889, 20, \dodoi{10.3847/1538-4357/ab5a7e}

\bibitem[{{Li} {et~al.}(2013){Li}, {Qiu}, {Wyrowski}, \&
  {Menten}}]{2013A&A...559A..23L}
{Li}, G.-X., {Qiu}, K., {Wyrowski}, F., \& {Menten}, K. 2013, \aap, 559, A23,
  \dodoi{10.1051/0004-6361/201220581}

\bibitem[{{Lin} {et~al.}(2023){Lin}, {Li}, {Tobin}, {Ohashi}, {J{\o}rgensen},
  {Looney}, {Aso}, {Takakuwa}, {Aikawa}, {van't Hoff}, {de Gregorio-Monsalvo},
  {Encalada}, {Flores}, {Gavino}, {Han}, {Kido}, {Koch}, {Kwon}, {Lai}, {Lee},
  {Lee}, {Phuong}, {Sai}, {Sharma}, {Sheehan}, {Thieme}, {Williams}, {Yamato},
  \& {Yen}}]{2023ApJ...951....9L}
{Lin}, Z.-Y.~D., {Li}, Z.-Y., {Tobin}, J.~J., {et~al.} 2023, \apj, 951, 9,
  \dodoi{10.3847/1538-4357/acd5c9}

\bibitem[{{Looney} {et~al.}(2007){Looney}, {Tobin}, \&
  {Kwon}}]{2007ApJ...670L.131L}
{Looney}, L.~W., {Tobin}, J.~J., \& {Kwon}, W. 2007, \apjl, 670, L131,
  \dodoi{10.1086/524361}

\bibitem[{{Luhman}(2023)}]{2023AJ....165...37L}
{Luhman}, K.~L. 2023, \aj, 165, 37, \dodoi{10.3847/1538-3881/ac9da3}

\bibitem[{{Maret} {et~al.}(2020){Maret}, {Maury}, {Belloche}, {Gaudel},
  {Andr{\'e}}, {Cabrit}, {Codella}, {Lef{\'e}vre}, {Podio}, {Anderl}, {Gueth},
  \& {Hennebelle}}]{2020A&A...635A..15M}
{Maret}, S., {Maury}, A.~J., {Belloche}, A., {et~al.} 2020, \aap, 635, A15,
  \dodoi{10.1051/0004-6361/201936798}

\bibitem[{{Matsumoto} {et~al.}(2015){Matsumoto}, {Onishi}, {Tokuda}, \&
  {Inutsuka}}]{2015MNRAS.449L.123M}
{Matsumoto}, T., {Onishi}, T., {Tokuda}, K., \& {Inutsuka}, S.~I. 2015, \mnras,
  449, L123, \dodoi{10.1093/mnrasl/slv031}

\bibitem[{{McMullin} {et~al.}(2007){McMullin}, {Waters}, {Schiebel}, {Young},
  \& {Golap}}]{2007ASPC..376..127M}
{McMullin}, J.~P., {Waters}, B., {Schiebel}, D., {Young}, W., \& {Golap}, K.
  2007, in Astronomical Society of the Pacific Conference Series, Vol. 376,
  Astronomical Data Analysis Software and Systems XVI, ed. R.~A. {Shaw},
  F.~{Hill}, \& D.~J. {Bell}, 127

\bibitem[{{Moriarty-Schieven} {et~al.}(1992){Moriarty-Schieven}, {Wannier},
  {Tamura}, \& {Keene}}]{1992ApJ...400..260M}
{Moriarty-Schieven}, G.~H., {Wannier}, P.~G., {Tamura}, M., \& {Keene}, J.
  1992, \apj, 400, 260, \dodoi{10.1086/171992}

\bibitem[{{Motte} \& {Andr{\'e}}(2001)}]{2001A&A...365..440M}
{Motte}, F., \& {Andr{\'e}}, P. 2001, \aap, 365, 440,
  \dodoi{10.1051/0004-6361:20000072}

\bibitem[{{Murillo} {et~al.}(2013){Murillo}, {Lai}, {Bruderer}, {Harsono}, \&
  {van Dishoeck}}]{2013A&A...560A.103M}
{Murillo}, N.~M., {Lai}, S.-P., {Bruderer}, S., {Harsono}, D., \& {van
  Dishoeck}, E.~F. 2013, \aap, 560, A103, \dodoi{10.1051/0004-6361/201322537}

\bibitem[{{Myers} {et~al.}(1991){Myers}, {Fuller}, {Goodman}, \&
  {Benson}}]{1991ApJ...376..561M}
{Myers}, P.~C., {Fuller}, G.~A., {Goodman}, A.~A., \& {Benson}, P.~J. 1991,
  \apj, 376, 561, \dodoi{10.1086/170305}

\bibitem[{{Offner} \& {Arce}(2014)}]{2014ApJ...784...61O}
{Offner}, S. S.~R., \& {Arce}, H.~G. 2014, \apj, 784, 61,
  \dodoi{10.1088/0004-637X/784/1/61}

\bibitem[{{Offner} \& {Chaban}(2017)}]{2017ApJ...847..104O}
{Offner}, S. S.~R., \& {Chaban}, J. 2017, \apj, 847, 104,
  \dodoi{10.3847/1538-4357/aa8996}

\bibitem[{{Offner} {et~al.}(2016){Offner}, {Dunham}, {Lee}, {Arce}, \&
  {Fielding}}]{2016ApJ...827L..11O}
{Offner}, S. S.~R., {Dunham}, M.~M., {Lee}, K.~I., {Arce}, H.~G., \&
  {Fielding}, D.~B. 2016, \apjl, 827, L11, \dodoi{10.3847/2041-8205/827/1/L11}

\bibitem[{{Offner} {et~al.}(2010){Offner}, {Kratter}, {Matzner}, {Krumholz}, \&
  {Klein}}]{2010ApJ...725.1485O}
{Offner}, S. S.~R., {Kratter}, K.~M., {Matzner}, C.~D., {Krumholz}, M.~R., \&
  {Klein}, R.~I. 2010, \apj, 725, 1485, \dodoi{10.1088/0004-637X/725/2/1485}

\bibitem[{{Ohashi} {et~al.}(1997){Ohashi}, {Hayashi}, {Ho}, {Momose}, {Tamura},
  {Hirano}, \& {Sargent}}]{1997ApJ...488..317O}
{Ohashi}, N., {Hayashi}, M., {Ho}, P. T.~P., {et~al.} 1997, \apj, 488, 317,
  \dodoi{10.1086/304685}

\bibitem[{{Ohashi} {et~al.}(2014){Ohashi}, {Saigo}, {Aso}, {Aikawa},
  {Koyamatsu}, {Machida}, {Saito}, {Takahashi}, {Takakuwa}, {Tomida},
  {Tomisaka}, \& {Yen}}]{2014ApJ...796..131O}
{Ohashi}, N., {Saigo}, K., {Aso}, Y., {et~al.} 2014, \apj, 796, 131,
  \dodoi{10.1088/0004-637X/796/2/131}

\bibitem[{{Ohashi} {et~al.}(2023){Ohashi}, {Tobin}, {J{\o}rgensen}, {Takakuwa},
  {Sheehan}, {Aikawa}, {Li}, {Looney}, {Williams}, {Aso}, {Sharma}, {Sai Insa
  Choi}, {Yamato}, {Lee}, {Tomida}, {Yen}, {Encalada}, {Flores}, {Gavino},
  {Kido}, {Han}, {Lin}, {Narayanan}, {Phuong}, {Santamar{\'\i}a-Miranda},
  {Thieme}, {van't Hoff}, {de Gregorio-Monsalvo}, {Koch}, {Kwon}, {Lai}, {Lee},
  {Plunkett}, {Saigo}, {Hirano}, {Lam}, \& {Mori}}]{2023ApJ...951....8O}
{Ohashi}, N., {Tobin}, J.~J., {J{\o}rgensen}, J.~K., {et~al.} 2023, \apj, 951,
  8, \dodoi{10.3847/1538-4357/acd384}

\bibitem[{{Oya} {et~al.}(2016){Oya}, {Sakai}, {L{\'o}pez-Sepulcre}, {Watanabe},
  {Ceccarelli}, {Lefloch}, {Favre}, \& {Yamamoto}}]{2016ApJ...824...88O}
{Oya}, Y., {Sakai}, N., {L{\'o}pez-Sepulcre}, A., {et~al.} 2016, \apj, 824, 88,
  \dodoi{10.3847/0004-637X/824/2/88}

\bibitem[{{Oya} {et~al.}(2017){Oya}, {Sakai}, {Watanabe}, {Higuchi}, {Hirota},
  {L{\'o}pez-Sepulcre}, {Sakai}, {Aikawa}, {Ceccarelli}, {Lefloch}, {Caux},
  {Vastel}, {Kahane}, \& {Yamamoto}}]{2017ApJ...837..174O}
{Oya}, Y., {Sakai}, N., {Watanabe}, Y., {et~al.} 2017, \apj, 837, 174,
  \dodoi{10.3847/1538-4357/aa6300}

\bibitem[{{Pineda} {et~al.}(2020){Pineda}, {Segura-Cox}, {Caselli},
  {Cunningham}, {Zhao}, {Schmiedeke}, {Maureira}, \&
  {Neri}}]{2020NatAs...4.1158P}
{Pineda}, J.~E., {Segura-Cox}, D., {Caselli}, P., {et~al.} 2020, Nature
  Astronomy, 4, 1158, \dodoi{10.1038/s41550-020-1150-z}

\bibitem[{{Planck Collaboration} {et~al.}(2014){Planck Collaboration},
  {Abergel}, {Ade}, {Aghanim}, {Alves}, {Aniano}, {Arnaud}, {Ashdown},
  {Aumont}, {Baccigalupi}, {Banday}, {Barreiro}, {Bartlett}, {Battaner},
  {Benabed}, {Benoit-L{\'e}vy}, {Bernard}, {Bersanelli}, {Bielewicz}, {Bobin},
  {Bonaldi}, {Bond}, {Bouchet}, {Boulanger}, {Burigana}, {Cardoso}, {Catalano},
  {Chamballu}, {Chiang}, {Christensen}, {Clements}, {Colombi}, {Colombo},
  {Couchot}, {Crill}, {Cuttaia}, {Danese}, {Davis}, {de Bernardis}, {de Rosa},
  {de Zotti}, {Delabrouille}, {D{\'e}sert}, {Dickinson}, {Diego}, {Dole},
  {Donzelli}, {Dor{\'e}}, {Douspis}, {Dupac}, {Efstathiou}, {En{\ss}lin},
  {Eriksen}, {Falgarone}, {Finelli}, {Forni}, {Frailis}, {Franceschi},
  {Galeotta}, {Ganga}, {Ghosh}, {Giard}, {Giraud-H{\'e}raud},
  {Gonz{\'a}lez-Nuevo}, {G{\'o}rski}, {Gregorio}, {Gruppuso}, {Guillet},
  {Hansen}, {Harrison}, {Helou}, {Henrot-Versill{\'e}},
  {Hern{\'a}ndez-Monteagudo}, {Herranz}, {Hildebrandt}, {Hivon}, {Hobson},
  {Holmes}, {Hornstrup}, {Hovest}, {Huffenberger}, {Jaffe}, {Jaffe}, {Joncas},
  {Jones}, {Jones}, {Juvela}, {Kalberla}, {Keih{\"a}nen}, {Kerp}, {Keskitalo},
  {Kisner}, {Kneissl}, {Knoche}, {Kunz}, {Kurki-Suonio}, {Lagache},
  {L{\"a}hteenm{\"a}ki}, {Lamarre}, {Lasenby}, {Lawrence}, {Leonardi},
  {Levrier}, {Liguori}, {Lilje}, {Linden-V{\o}rnle}, {L{\'o}pez-Caniego},
  {Lubin}, {Mac{\'\i}as-P{\'e}rez}, {Maffei}, {Maino}, {Mandolesi}, {Maris},
  {Marshall}, {Martin}, {Mart{\'\i}nez-Gonz{\'a}lez}, {Masi}, {Massardi},
  {Matarrese}, {Mazzotta}, {Melchiorri}, {Mendes}, {Mennella}, {Migliaccio},
  {Mitra}, {Miville-Desch{\^e}nes}, {Moneti}, {Montier}, {Morgante},
  {Mortlock}, {Munshi}, {Murphy}, {Naselsky}, {Nati}, {Natoli}, {Noviello},
  {Novikov}, {Novikov}, {Oxborrow}, {Pagano}, {Pajot}, {Paoletti}, {Pasian},
  {Perdereau}, {Perotto}, {Perrotta}, {Piacentini}, {Piat}, {Pierpaoli},
  {Pietrobon}, {Plaszczynski}, {Pointecouteau}, {Polenta}, {Ponthieu}, {Popa},
  {Pratt}, {Prunet}, {Puget}, {Rachen}, {Reach}, {Rebolo}, {Reinecke},
  {Remazeilles}, {Renault}, {Ricciardi}, {Riller}, {Ristorcelli}, {Rocha},
  {Rosset}, {Roudier}, {Rusholme}, {Sandri}, {Savini}, {Spencer}, {Starck},
  {Sureau}, {Sutton}, {Suur-Uski}, {Sygnet}, {Tauber}, {Terenzi}, {Toffolatti},
  {Tomasi}, {Tristram}, {Tucci}, {Umana}, {Valenziano}, {Valiviita}, {Van
  Tent}, {Verstraete}, {Vielva}, {Villa}, {Wade}, {Wandelt}, {Winkel}, {Yvon},
  {Zacchei}, \& {Zonca}}]{2014A&A...566A..55P}
{Planck Collaboration}, {Abergel}, A., {Ade}, P.~A.~R., {et~al.} 2014, \aap,
  566, A55, \dodoi{10.1051/0004-6361/201323270}

\bibitem[{{Radley} {et~al.}(2025){Radley}, {Busquet}, {Ilee}, {Liu}, {Pineda},
  {Pontoppidan}, {Mac{\'\i}as}, {Maureira}, {Bianchi}, {Bourke}, {Codella},
  {Forbrich}, {Girart}, {Hoare}, {Hern{\'a}ndez Garnica}, {Jim{\'e}nez-Serra},
  {Loinard}, {Ord{\'o}{\~n}ez-Toro}, \& {Podio}}]{2025ApJ...981..187R}
{Radley}, I.~C., {Busquet}, G., {Ilee}, J.~D., {et~al.} 2025, \apj, 981, 187,
  \dodoi{10.3847/1538-4357/adb0b9}

\bibitem[{{Reynolds} {et~al.}(2021){Reynolds}, {Tobin}, {Sheehan}, {Sadavoy},
  {Kratter}, {Li}, {Chandler}, {Segura-Cox}, {Looney}, \&
  {Dunham}}]{2021ApJ...907L..10R}
{Reynolds}, N.~K., {Tobin}, J.~J., {Sheehan}, P., {et~al.} 2021, \apjl, 907,
  L10, \dodoi{10.3847/2041-8213/abcc02}

\bibitem[{{Roccatagliata} {et~al.}(2020){Roccatagliata}, {Franciosini},
  {Sacco}, {Randich}, \& {Sicilia-Aguilar}}]{2020A&A...638A..85R}
{Roccatagliata}, V., {Franciosini}, E., {Sacco}, G.~G., {Randich}, S., \&
  {Sicilia-Aguilar}, A. 2020, \aap, 638, A85,
  \dodoi{10.1051/0004-6361/201936401}

\bibitem[{{Sai} {et~al.}(2020){Sai}, {Ohashi}, {Saigo}, {Matsumoto}, {Aso},
  {Takakuwa}, {Aikawa}, {Kurose}, {Yen}, {Tomisaka}, {Tomida}, \&
  {Machida}}]{2020ApJ...893...51S}
{Sai}, J., {Ohashi}, N., {Saigo}, K., {et~al.} 2020, \apj, 893, 51,
  \dodoi{10.3847/1538-4357/ab8065}

\bibitem[{{Sai} {et~al.}(2023){Sai}, {Yen}, {Ohashi}, {Tobin}, {J{\o}rgensen},
  {Takakuwa}, {Saigo}, {Aso}, {Lin}, {Koch}, {Aikawa}, {Flores}, {de
  Gregorio-Monsalvo}, {Han}, {Kido}, {Kwon}, {Lai}, {Lee}, {Lee}, {Li},
  {Looney}, {Mori}, {Phuong}, {Santamar{\'\i}a-Miranda}, {Sharma}, {Thieme},
  {Tomida}, \& {Williams}}]{2023ApJ...954...67S}
{Sai}, J., {Yen}, H.-W., {Ohashi}, N., {et~al.} 2023, \apj, 954, 67,
  \dodoi{10.3847/1538-4357/ace52d}

\bibitem[{{Saito} {et~al.}(2001){Saito}, {Kawabe}, {Kitamura}, \&
  {Sunada}}]{2001ApJ...547..840S}
{Saito}, M., {Kawabe}, R., {Kitamura}, Y., \& {Sunada}, K. 2001, \apj, 547,
  840, \dodoi{10.1086/318383}

\bibitem[{{Sakai} {et~al.}(2014){Sakai}, {Sakai}, {Hirota}, {Watanabe},
  {Ceccarelli}, {Kahane}, {Bottinelli}, {Caux}, {Demyk}, {Vastel}, {Coutens},
  {Taquet}, {Ohashi}, {Takakuwa}, {Yen}, {Aikawa}, \&
  {Yamamoto}}]{2014Natur.507...78S}
{Sakai}, N., {Sakai}, T., {Hirota}, T., {et~al.} 2014, \nat, 507, 78,
  \dodoi{10.1038/nature13000}

\bibitem[{{Sakai} {et~al.}(2016){Sakai}, {Oya}, {L{\'o}pez-Sepulcre},
  {Watanabe}, {Sakai}, {Hirota}, {Aikawa}, {Ceccarelli}, {Lefloch}, {Caux},
  {Vastel}, {Kahane}, \& {Yamamoto}}]{2016ApJ...820L..34S}
{Sakai}, N., {Oya}, Y., {L{\'o}pez-Sepulcre}, A., {et~al.} 2016, \apjl, 820,
  L34, \dodoi{10.3847/2041-8205/820/2/L34}

\bibitem[{{Schilke} {et~al.}(1997){Schilke}, {Walmsley}, {Pineau des Forets},
  \& {Flower}}]{1997A&A...321..293S}
{Schilke}, P., {Walmsley}, C.~M., {Pineau des Forets}, G., \& {Flower}, D.~R.
  1997, \aap, 321, 293

\bibitem[{{Segura-Cox} {et~al.}(2020){Segura-Cox}, {Schmiedeke}, {Pineda},
  {Stephens}, {Fern{\'a}ndez-L{\'o}pez}, {Looney}, {Caselli}, {Li}, {Mundy},
  {Kwon}, \& {Harris}}]{2020Natur.586..228S}
{Segura-Cox}, D.~M., {Schmiedeke}, A., {Pineda}, J.~E., {et~al.} 2020, \nat,
  586, 228, \dodoi{10.1038/s41586-020-2779-6}

\bibitem[{{Seifried} {et~al.}(2015){Seifried}, {Banerjee}, {Pudritz}, \&
  {Klessen}}]{2015MNRAS.446.2776S}
{Seifried}, D., {Banerjee}, R., {Pudritz}, R.~E., \& {Klessen}, R.~S. 2015,
  \mnras, 446, 2776, \dodoi{10.1093/mnras/stu2282}

\bibitem[{{Seifried} {et~al.}(2016){Seifried}, {S{\'a}nchez-Monge}, {Walch}, \&
  {Banerjee}}]{2016MNRAS.459.1892S}
{Seifried}, D., {S{\'a}nchez-Monge}, {\'A}., {Walch}, S., \& {Banerjee}, R.
  2016, \mnras, 459, 1892, \dodoi{10.1093/mnras/stw785}

\bibitem[{{Sheehan} \& {Eisner}(2017)}]{2017ApJ...851...45S}
{Sheehan}, P.~D., \& {Eisner}, J.~A. 2017, \apj, 851, 45,
  \dodoi{10.3847/1538-4357/aa9990}

\bibitem[{{Sheehan} {et~al.}(2020){Sheehan}, {Tobin}, {Federman}, {Megeath}, \&
  {Looney}}]{2020ApJ...902..141S}
{Sheehan}, P.~D., {Tobin}, J.~J., {Federman}, S., {Megeath}, S.~T., \&
  {Looney}, L.~W. 2020, \apj, 902, 141, \dodoi{10.3847/1538-4357/abbad5}

\bibitem[{{Sheehan} {et~al.}(2022{\natexlab{a}}){Sheehan}, {Tobin}, {Looney},
  \& {Megeath}}]{2022ApJ...929...76S}
{Sheehan}, P.~D., {Tobin}, J.~J., {Looney}, L.~W., \& {Megeath}, S.~T.
  2022{\natexlab{a}}, \apj, 929, 76, \dodoi{10.3847/1538-4357/ac574d}

\bibitem[{{Sheehan} {et~al.}(2022{\natexlab{b}}){Sheehan}, {Tobin}, {Li},
  {van't Hoff}, {J{\o}rgensen}, {Kwon}, {Looney}, {Ohashi}, {Takakuwa},
  {Williams}, {Aso}, {Gavino}, {de Gregorio-Monsalvo}, {Han}, {Lee},
  {Plunkett}, {Sharma}, {Aikawa}, {Lai}, {Lee}, {Lin}, {Saigo}, {Tomida}, \&
  {Yen}}]{2022ApJ...934...95S}
{Sheehan}, P.~D., {Tobin}, J.~J., {Li}, Z.-Y., {et~al.} 2022{\natexlab{b}},
  \apj, 934, 95, \dodoi{10.3847/1538-4357/ac7a3b}

\bibitem[{{Sullivan} {et~al.}(2019){Sullivan}, {Wilking}, {Greene}, {Lisalda},
  {Gibb}, \& {Ejeta}}]{2019AJ....158...41S}
{Sullivan}, T., {Wilking}, B.~A., {Greene}, T.~P., {et~al.} 2019, \aj, 158, 41,
  \dodoi{10.3847/1538-3881/ab24c0}

\bibitem[{{Tachihara} {et~al.}(2012){Tachihara}, {Saigo}, {Higuchi}, {Inoue},
  {Inutsuka}, {Hackstein}, {Haas}, \& {Mugrauer}}]{2012ApJ...754...95T}
{Tachihara}, K., {Saigo}, K., {Higuchi}, A.~E., {et~al.} 2012, \apj, 754, 95,
  \dodoi{10.1088/0004-637X/754/2/95}

\bibitem[{{Takakuwa} {et~al.}(2018){Takakuwa}, {Tsukamoto}, {Saigo}, \&
  {Saito}}]{2018ApJ...865...51T}
{Takakuwa}, S., {Tsukamoto}, Y., {Saigo}, K., \& {Saito}, M. 2018, \apj, 865,
  51, \dodoi{10.3847/1538-4357/aadb93}

\bibitem[{{Tamura} {et~al.}(1991){Tamura}, {Gatley}, {Waller}, \&
  {Werner}}]{1991ApJ...374L..25T}
{Tamura}, M., {Gatley}, I., {Waller}, W., \& {Werner}, M.~W. 1991, \apjl, 374,
  L25, \dodoi{10.1086/186063}

\bibitem[{{Tazzari} {et~al.}(2021){Tazzari}, {Testi}, {Natta}, {Williams},
  {Ansdell}, {Carpenter}, {Facchini}, {Guidi}, {Hogherheijde}, {Manara},
  {Miotello}, \& {van der Marel}}]{2021MNRAS.506.5117T}
{Tazzari}, M., {Testi}, L., {Natta}, A., {et~al.} 2021, \mnras, 506, 5117,
  \dodoi{10.1093/mnras/stab1912}

\bibitem[{{Terebey} {et~al.}(1984){Terebey}, {Shu}, \&
  {Cassen}}]{1984ApJ...286..529T}
{Terebey}, S., {Shu}, F.~H., \& {Cassen}, P. 1984, \apj, 286, 529,
  \dodoi{10.1086/162628}

\bibitem[{{Testi} {et~al.}(2014){Testi}, {Birnstiel}, {Ricci}, {Andrews},
  {Blum}, {Carpenter}, {Dominik}, {Isella}, {Natta}, {Williams}, \&
  {Wilner}}]{2014prpl.conf..339T}
{Testi}, L., {Birnstiel}, T., {Ricci}, L., {et~al.} 2014, in Protostars and
  Planets VI, ed. H.~{Beuther}, R.~S. {Klessen}, C.~P. {Dullemond}, \&
  T.~{Henning}, 339--361, \dodoi{10.2458/azu_uapress_9780816531240-ch015}

\bibitem[{{Thieme} {et~al.}(2022){Thieme}, {Lai}, {Lin}, {Cheong}, {Lee},
  {Yen}, {Li}, {Lam}, \& {Zhao}}]{2022ApJ...925...32T}
{Thieme}, T.~J., {Lai}, S.-P., {Lin}, S.-J., {et~al.} 2022, \apj, 925, 32,
  \dodoi{10.3847/1538-4357/ac382b}

\bibitem[{{Tobin}(2023)}]{2023zndo...7986682T}
{Tobin}, J. 2023, {eDisk data reduction scripts}, 1.0.0, Zenodo,  Zenodo,
  \dodoi{10.5281/zenodo.7986682}

\bibitem[{{Tobin} {et~al.}(2022){Tobin}, {Cox}, \&
  {Looney}}]{2022ApJ...928...61T}
{Tobin}, J.~J., {Cox}, E.~G., \& {Looney}, L.~W. 2022, \apj, 928, 61,
  \dodoi{10.3847/1538-4357/ac5594}

\bibitem[{{Tobin} {et~al.}(2012){Tobin}, {Hartmann}, {Chiang}, {Wilner},
  {Looney}, {Loinard}, {Calvet}, \& {D'Alessio}}]{2012Natur.492...83T}
{Tobin}, J.~J., {Hartmann}, L., {Chiang}, H.-F., {et~al.} 2012, \nat, 492, 83,
  \dodoi{10.1038/nature11610}

\bibitem[{{Tobin} {et~al.}(2013){Tobin}, {Hartmann}, {Chiang}, {Wilner},
  {Looney}, {Loinard}, {Calvet}, \& {D'Alessio}}]{2013ApJ...771...48T}
---. 2013, \apj, 771, 48, \dodoi{10.1088/0004-637X/771/1/48}

\bibitem[{{Tobin} {et~al.}(2010){Tobin}, {Hartmann}, {Looney}, \&
  {Chiang}}]{2010ApJ...712.1010T}
{Tobin}, J.~J., {Hartmann}, L., {Looney}, L.~W., \& {Chiang}, H.-F. 2010, \apj,
  712, 1010, \dodoi{10.1088/0004-637X/712/2/1010}

\bibitem[{{Tobin} \& {Sheehan}(2024)}]{2024ARA&A..62..203T}
{Tobin}, J.~J., \& {Sheehan}, P.~D. 2024, \araa, 62, 203,
  \dodoi{10.1146/annurev-astro-052920-103752}

\bibitem[{{Tobin} {et~al.}(2011){Tobin}, {Hartmann}, {Chiang}, {Looney},
  {Bergin}, {Chandler}, {Masqu{\'e}}, {Maret}, \&
  {Heitsch}}]{2011ApJ...740...45T}
{Tobin}, J.~J., {Hartmann}, L., {Chiang}, H.-F., {et~al.} 2011, \apj, 740, 45,
  \dodoi{10.1088/0004-637X/740/1/45}

\bibitem[{{Tobin} {et~al.}(2020){Tobin}, {Sheehan}, {Megeath},
  {D{\'\i}az-Rodr{\'\i}guez}, {Offner}, {Murillo}, {van 't Hoff}, {van
  Dishoeck}, {Osorio}, {Anglada}, {Furlan}, {Stutz}, {Reynolds}, {Karnath},
  {Fischer}, {Persson}, {Looney}, {Li}, {Stephens}, {Chandler}, {Cox},
  {Dunham}, {Tychoniec}, {Kama}, {Kratter}, {Kounkel}, {Mazur}, {Maud},
  {Patel}, {Perez}, {Sadavoy}, {Segura-Cox}, {Sharma}, {Stephenson}, {Watson},
  \& {Wyrowski}}]{2020ApJ...890..130T}
{Tobin}, J.~J., {Sheehan}, P.~D., {Megeath}, S.~T., {et~al.} 2020, \apj, 890,
  130, \dodoi{10.3847/1538-4357/ab6f64}

\bibitem[{{Ulrich}(1976)}]{1976ApJ...210..377U}
{Ulrich}, R.~K. 1976, \apj, 210, 377, \dodoi{10.1086/154840}

\bibitem[{{Unno} {et~al.}(2022){Unno}, {Hanawa}, \&
  {Takasao}}]{2022ApJ...941..154U}
{Unno}, M., {Hanawa}, T., \& {Takasao}, S. 2022, \apj, 941, 154,
  \dodoi{10.3847/1538-4357/aca410}

\bibitem[{{Valdivia-Mena} {et~al.}(2022){Valdivia-Mena}, {Pineda},
  {Segura-Cox}, {Caselli}, {Neri}, {L{\'o}pez-Sepulcre}, {Cunningham},
  {Bouscasse}, {Semenov}, {Henning}, {Pi{\'e}tu}, {Chapillon}, {Dutrey},
  {Fuente}, {Guilloteau}, {Hsieh}, {Jim{\'e}nez-Serra}, {Marino}, {Maureira},
  {Smirnov-Pinchukov}, {Tafalla}, \& {Zhao}}]{2022A&A...667A..12V}
{Valdivia-Mena}, M.~T., {Pineda}, J.~E., {Segura-Cox}, D.~M., {et~al.} 2022,
  \aap, 667, A12, \dodoi{10.1051/0004-6361/202243310}

\bibitem[{{Valdivia-Mena} {et~al.}(2024){Valdivia-Mena}, {Pineda}, {Caselli},
  {Segura-Cox}, {Schmiedeke}, {Spezzano}, {Offner}, {Ivlev}, {Kuffmeier},
  {Cunningham}, {Neri}, \& {Maureira}}]{2024A&A...687A..71V}
{Valdivia-Mena}, M.~T., {Pineda}, J.~E., {Caselli}, P., {et~al.} 2024, \aap,
  687, A71, \dodoi{10.1051/0004-6361/202449395}

\bibitem[{{Viana Almeida} {et~al.}(2012){Viana Almeida}, {Melo}, {Santos},
  {Figueira}, {Sterzik}, \& {Gameiro}}]{2012A&A...539A..62V}
{Viana Almeida}, P., {Melo}, C., {Santos}, N.~C., {et~al.} 2012, \aap, 539,
  A62, \dodoi{10.1051/0004-6361/201117703}

\bibitem[{{Whitney} {et~al.}(1997){Whitney}, {Kenyon}, \&
  {G{\'o}mez}}]{1997ApJ...485..703W}
{Whitney}, B.~A., {Kenyon}, S.~J., \& {G{\'o}mez}, M. 1997, \apj, 485, 703,
  \dodoi{10.1086/304454}

\bibitem[{{Williams} \& {Cieza}(2011)}]{2011ARA&A..49...67W}
{Williams}, J.~P., \& {Cieza}, L.~A. 2011, \araa, 49, 67,
  \dodoi{10.1146/annurev-astro-081710-102548}

\bibitem[{{Yamato} {et~al.}(2023){Yamato}, {Aikawa}, {Ohashi}, {Tobin},
  {J{\o}rgensen}, {Takakuwa}, {Aso}, {Sai}, {Flores}, {de Gregorio-Monsalvo},
  {Hirano}, {Han}, {Kido}, {Koch}, {Kwon}, {Lai}, {Lee}, {Lee}, {Li}, {Lin},
  {Looney}, {Mori}, {Narayanan}, {Phuong}, {Saigo}, {Santamar{\'\i}a-Miranda},
  {Sharma}, {Thieme}, {Tomida}, {van't Hoff}, \& {Yen}}]{2023ApJ...951...11Y}
{Yamato}, Y., {Aikawa}, Y., {Ohashi}, N., {et~al.} 2023, \apj, 951, 11,
  \dodoi{10.3847/1538-4357/accd71}

\bibitem[{{Yen} {et~al.}(2019){Yen}, {Gu}, {Hirano}, {Koch}, {Lee}, {Liu}, \&
  {Takakuwa}}]{2019ApJ...880...69Y}
{Yen}, H.-W., {Gu}, P.-G., {Hirano}, N., {et~al.} 2019, \apj, 880, 69,
  \dodoi{10.3847/1538-4357/ab29f8}

\bibitem[{{Yen} {et~al.}(2013){Yen}, {Takakuwa}, {Ohashi}, \&
  {Ho}}]{2013ApJ...772...22Y}
{Yen}, H.-W., {Takakuwa}, S., {Ohashi}, N., \& {Ho}, P. T.~P. 2013, \apj, 772,
  22, \dodoi{10.1088/0004-637X/772/1/22}

\bibitem[{{Yen} {et~al.}(2014){Yen}, {Takakuwa}, {Ohashi}, {Aikawa}, {Aso},
  {Koyamatsu}, {Machida}, {Saigo}, {Saito}, {Tomida}, \&
  {Tomisaka}}]{2014ApJ...793....1Y}
{Yen}, H.-W., {Takakuwa}, S., {Ohashi}, N., {et~al.} 2014, \apj, 793, 1,
  \dodoi{10.1088/0004-637X/793/1/1}

\bibitem[{{Yen} {et~al.}(2021){Yen}, {Koch}, {Hull}, {Ward-Thompson},
  {Bastien}, {Hasegawa}, {Kwon}, {Lai}, {Qiu}, {Ching}, {Chung}, {Coud{\'e}},
  {Di Francesco}, {Diep}, {Doi}, {Eswaraiah}, {Falle}, {Fuller}, {Furuya},
  {Han}, {Hatchell}, {Houde}, {Inutsuka}, {Johnstone}, {Kang}, {Kang}, {Kim},
  {Kirchschlager}, {Kwon}, {Lee}, {Lee}, {Liu}, {Liu}, {Lyo}, {Ohashi},
  {Onaka}, {Pattle}, {Sadavoy}, {Saito}, {Shinnaga}, {Soam}, {Tahani},
  {Tamura}, {Tang}, {Tang}, \& {Zhang}}]{2021ApJ...907...33Y}
{Yen}, H.-W., {Koch}, P.~M., {Hull}, C. L.~H., {et~al.} 2021, \apj, 907, 33,
  \dodoi{10.3847/1538-4357/abca99}

\bibitem[{{Yen} {et~al.}(2024){Yen}, {Williams}, {Sai}, {Koch}, {Han},
  {J{\o}rgensen}, {Kwon}, {Lee}, {Li}, {Looney}, {Narang}, {Ohashi},
  {Takakuwa}, {Tobin}, {de Gregorio-Monsalvo}, {Lai}, {Lee}, \&
  {Tomida}}]{2024ApJ...969..125Y}
{Yen}, H.-W., {Williams}, J.~P., {Sai}, J., {et~al.} 2024, \apj, 969, 125,
  \dodoi{10.3847/1538-4357/ad4c6b}

\bibitem[{{Young} {et~al.}(2003){Young}, {Shirley}, {Evans}, \&
  {Rawlings}}]{2003ApJS..145..111Y}
{Young}, C.~H., {Shirley}, Y.~L., {Evans}, Neal~J., I., \& {Rawlings}, J. M.~C.
  2003, \apjs, 145, 111, \dodoi{10.1086/345341}

\bibitem[{{Zhang} {et~al.}(2021){Zhang}, {Launhardt}, {Liu}, {Tobin}, \&
  {Henning}}]{2021A&A...646A..18Z}
{Zhang}, C.-P., {Launhardt}, R., {Liu}, Y., {Tobin}, J.~J., \& {Henning}, T.
  2021, \aap, 646, A18, \dodoi{10.1051/0004-6361/202039536}

\end{thebibliography}
\bibliographystyle{aasjournal}



\end{document}